\documentclass[10pt,twocolumn,twoside]{IEEEtran}

\ifCLASSINFOpdf
\else
\fi
\usepackage{cite}
\usepackage{stfloats}
\usepackage{amsmath}
\usepackage{array}
\usepackage{amsthm}
\usepackage{amssymb}
\usepackage{amstext}
\usepackage{graphicx}
\usepackage{graphbox}
\usepackage{subfig}
\usepackage{setspace}
\usepackage{epstopdf}
\usepackage{psfrag}
\usepackage{url}

\usepackage{siunitx}

\usepackage[dvips]{epsfig}
\usepackage{epsfig}
\usepackage[latin5]{inputenc}
\usepackage{float}

\usepackage{booktabs}
\usepackage{adjustbox}

\newcommand{\be}{\begin{equation}}
\newcommand{\ee}{\end{equation}}
\newcommand{\bi}{\begin{itemize}}
\newcommand{\ei}{\end{itemize}}
\newcommand{\bn}{\begin{enumerate}}
\newcommand{\en}{\end{enumerate}}
\newcommand{\bea}{\begin{eqnarray}}
\newcommand{\eea}{\end{eqnarray}}

\newcommand{\sii}[1]{\ensuremath{f_{{#1}}}}

\newcommand{\sitilde}[1]{\ensuremath{{s}_{#1}}}

\newcommand{\tk}[1]{\ensuremath{t_{#1}}}
\newcommand{\g}[2][1]{\ensuremath{g_{{#1},{#2}}}}

\newcommand{\ssb}{\mathbf{s}}
\newcommand{\zb}{\mathbf{z}}

\newcommand{\yb}{\mathbf{y}}
\newcommand{\wb}{\mathbf{w}}
\newcommand{\Hb}{\mathbf{H}}

\newcommand{\Fb}{\mathbf{F}}

\newcommand{\Cb}{\mathbf{C}}
\newcommand{\Kb}{\mathbf{K}}
\newcommand{\Pb}{\mathbf{P}}

\newcommand{\Ib}{\mathbf{I}}
\newcommand{\ub}{\mathbf{u}}
\newcommand{\vb}{\mathbf{v}}
\newcommand{\db}{\mathbf{d}}

\newcommand{\nb}{\mathbf{n}}
\newcommand{\fb}{\mathbf{f}}

\newcommand{\tb}{\mathbf{t}}
\newcommand{\xb}{\mathbf{x}}
\newcommand{\Dwb}{\mathbf{D(x)}}
\newcommand{\Dwbn}{\mathbf{D}(\mathbf{x}^l)}

\begin{document}
%
\title{High-resolution multi-spectral imaging with diffractive lenses and learned reconstruction}

%
  \author{Figen~S.~Oktem,~\IEEEmembership{Member,~IEEE}, O\u{g}uzhan Fatih Kar, Can Deniz Bezek, Farzad Kamalabadi,~\IEEEmembership{Member,~IEEE}
      \\

\thanks{
The preliminary results of this research were presented in IEEE ICIP 2014 held in Paris, France. 
This work was supported by the Scientific and Technological Research Council of Turkey (TUBITAK) under Grant 117E160 (3501 Research Program). It was also supported in part by the National Science Foundation under AGS-ATM Grant 1936663 and by NASA under Grant 80NSSC18M0052.
}
\thanks{F.~S.~Oktem and C.~D.~Bezek are with the Department of Electrical and Electronics Engineering, Middle East Technical University (METU), Cankaya, Ankara 06800 Turkey (e-mail: figeno@metu.edu.tr, bezek@metu.edu.tr).
}
\thanks{O.~F.~Kar was with the Department of Electrical and Electronics Engineering, METU, Cankaya, Ankara 06800 Turkey. He is now with the School of Computer and Communication Sciences, EPFL, CH-1015 Lausanne, Switzerland  (e-mail: oguzhan.kar@epfl.ch).
}
\thanks{F. Kamalabadi is with the Department of Electrical and Computer Engineering and the Coordinated Science Laboratory, University of Illinois at Urbana-Champaign, Urbana,
61801 USA (e-mail: farzadk@illinois.edu).
}

\thanks{\copyright~2021 IEEE.  Personal use of this material is permitted.  Permission
from IEEE must be obtained for all other uses, in any current or future
media, including reprinting/republishing this material for advertising or
promotional purposes, creating new collective works, for resale or
redistribution to servers or lists, or reuse of any copyrighted component
of this work in other works.
}
}

\maketitle


\begin{abstract}
Spectral imaging is a fundamental diagnostic technique with widespread application. Conventional spectral imaging approaches 
have intrinsic limitations on spatial and spectral resolutions 
due to the physical components they rely on. 
To overcome these physical limitations, in this paper, we develop 
a novel multi-spectral imaging modality that enables higher spatial and spectral resolutions. In the developed computational imaging modality, we
exploit a diffractive lens, such as a photon sieve, for both dispersing and focusing the optical field, and achieve 
measurement diversity by changing the focusing behavior of this 
lens. Because the focal length of a diffractive lens is wavelength-dependent, each measurement is 
a superposition of differently blurred spectral components. 
To reconstruct the individual spectral images from these superimposed and blurred measurements, model-based fast reconstruction algorithms 
are developed with deep and analytical priors using alternating minimization and unrolling. Finally, the effectiveness and performance of the developed technique is illustrated for an application in 
astrophysical imaging under various observation scenarios in the extreme ultraviolet (EUV) regime. The results demonstrate that the technique 
provides not only diffraction-limited high spatial resolution, as enabled by diffractive lenses, but also the capability of resolving close-by spectral sources that would not otherwise be possible with the existing techniques. This work enables high resolution multi-spectral imaging with low cost designs for a variety of applications and spectral regimes.

\end{abstract}




%
\IEEEpeerreviewmaketitle

\section{Introduction} \label{sec:intro}

Spectral imaging, also known as multispectral or hyperspectral imaging, 
is a fundamental diagnostic technique in the physical sciences with widespread application in 
physics, biology, chemistry, medicine, astronomy, and remote sensing. 
In spectral imaging, images of a scene are formed as a function of wavelength.
Capturing this three-dimensional (3D) spectral data cube enables 
unique identification of the biological, chemical, and physical properties of 
the scene. As a result, spectral imaging is a useful diagnostic tool in diverse applications including remote sensing of astrophysical plasmas, environmental monitoring, surveillance, biomedical diagnostics, and industrial inspection, among many others.

The inherent challenge in spectral imaging 
is to capture the 3D spectral cube with the 2D detectors.
Conventional approaches perform scanning to 
form the spectral data cube from a series of 2D measurements. 
Generally this data cube is measured using an imager with a series of 
wavelength filters scanning the scene spectrally, or by using a spectrometer with a 
slit scanning the scene spatially. Since these  
techniques purely rely on physical components, an important 
drawback is the intrinsic physical limitations on their spatial, spectral, and temporal resolutions, as well as the inherent trade-off between spatial and spectral resolutions~\cite{oktem2014icip,davila2011b,arce2014}. 
For example, for spectral imagers employing wavelength filters, the spectral resolution is strictly limited by the bandwidth of the producible wavelength filters in the desired spectral range, and the spatial resolution is limited by the quality and cost of the used reflective/refractive imaging optics~\cite{lemen2011, davila2011b,andersen2005}. 

\begin{figure}[tbh!]
\begin{center} 
        \includegraphics[scale=0.19]{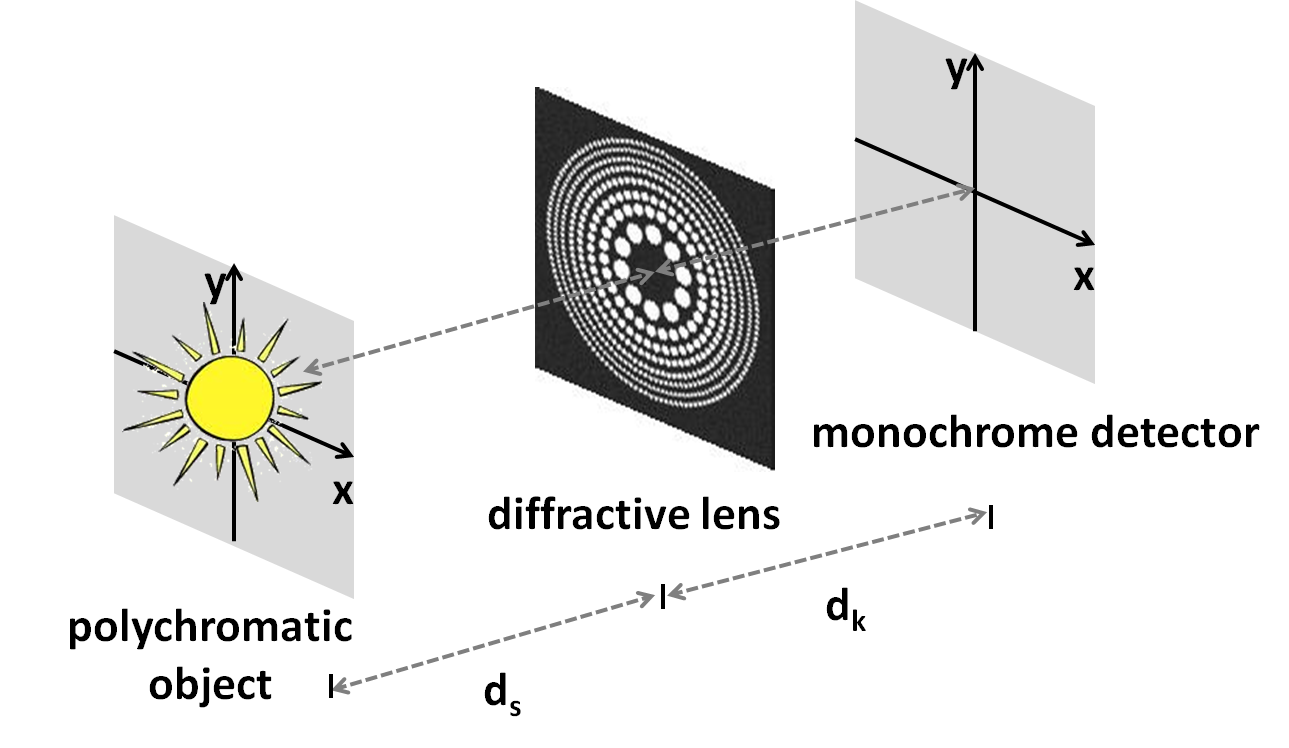}
    \caption{Schematic view of the proposed diffractive lens based spectral imaging system}  
\label{fig:system}
\end{center} \vspace{-0.0in}
\end{figure}

In this paper, we develop a computational multi-spectral imaging technique with the goal of enabling diffraction-limited high spatial resolution and 
better spectral resolution than the conventional spectral imagers 
with wavelength filters. Our imaging technique exploits a 
diffractive lens for both dispersing and focusing the optical field, allowing us to overcome these limitations in the spatial and spectral resolutions, while not creating any physical trade-off between them. Figure~\ref{fig:system} provides a schematic of the imaging
system involved. This system takes advantage of 
the wavelength-dependent behavior of a diffractive lens, i.e.
the fact that each wavelength is focused differently by a diffractive lens.
The main idea is to differently focus each spectral component 
and acquire measurements that are superpositions of these 
spectral images (with each spectral component 
exposed to different type of blur). 
Hence each measurement is the superposition of differently blurred spectral components. To obtain 
the needed information for decomposing this multiplexed data, 
multiple measurements are acquired and jointly used to reconstruct the individual spectral images by solving an inverse problem.

To solve the image reconstruction problem, we first formulate the continuous image-formation model and then obtain its discrete version. The resulting inverse problem can be viewed as a multi-frame deconvolution problem where multiple images need to be deblurred from their superimposed and blurred versions. To solve this inverse problem, we develop two fast reconstruction algorithms. The first algorithm exploits analysis priors (such as total variation) 
using the alternating direction method of multipliers (ADMM). 
As the analytical priors have recently been outperformed by deep priors~\cite{lucas2018,aggarwal2019,ongie2020}, the second algorithm learns the prior through a deep network that is end-to-end trained with an unrolling approach.
Thanks to the shift-invariance behavior of the system 
and the resulting convolutional forward model, each 
step in the developed algorithms has fast computation using FFTs. 

We illustrate the high spatial and spectral resolutions enabled by the developed spectral imaging technique for a potential application in astrophysical imaging of space plasmas in the extreme ultraviolet (EUV) regime. A detailed conditioning-based analysis is also performed for the spatial resolution of the technique. Although the presented results in this manuscript are for the UV regime, 
the developed imaging concept is equally applicable to other spectral regimes. 

To achieve measurement diversity, multiple measurements can be acquired either with a moving monochrome detector (so measurements are taken from different distances) 
or with multiple diffractive lens designs (so spectral bands focused differently onto a fixed detector). The latter may require additional optical components such as beam splitters and a larger detector. It can also be performed using a compound diffractive lens design that focus multiple wavelengths onto different locations on the fixed detector or with a variable diffractive lens through a programmable spatial light modulator (SLM) or a digital micromirror device (DMD).

Earlier approaches in computational spectral imaging have been developed with the goal of providing faster acquisition and/or reduced hardware complexity than the conventional methods. These can be regarded as \emph{compressive or snapshot} methods, and each use a different encoding scheme to acquire reduced number of measurements. Examples are coded aperture snapshot spectral imaging (CASSI)~\cite{cassi1} 
and its variants~\cite{wu2011,cao2016computational,salazar2019}, and coded aperture spectral imaging with diffractive lenses (CSID)~\cite{osa2018,kar2019}. 
However, in none of these compressive/snapshot approaches, the main 
goal was to improve the spatial and spectral resolutions
of the conventional scanning-based methods beyond their physical limitations. To the best of our knowledge, there have been no reports of higher spatial and spectral resolutions than the conventional systems, and ours is the first work that demonstrates this. 
Moreover, our system has no trade-off between spatial and spectral resolutions unlike many of the earlier systems; that is, the spatial and spectral resolutions can be increased independently. 

Note that our technique is \emph{not a compressive imaging technique}; it has indeed a different objective than the compressive/snapshot methods.
Different than these methods, we do not also solve an under-determined inverse problem for image reconstruction. Moreover, our technique is designed so as to be also applicable 
for scenes whose spectra consist of discrete spectral lines, 
or equivalently lack strong spectral correlation (i.e. has frequent peaks and fast transitions). 
Spectral sensing of such scenes are of interest in various applications including astrophysical imaging of space plasmas, atmospheric physics, and remote sensing~\cite{rybicki_book,hollas_book}.

In the earlier diffractive imaging works, diffractive lenses have been mostly used with monochromatic sources~\cite{attwood_book,kipp2001,andersen2005,menon2005} because of their wavelength-dependent focal length. When working with monochromatic illumination (i.e. single wavelength), diffractive lenses have many attractive properties such as their low weight, low cost, and flat structure. Another important advantage of diffractive imaging elements is that they can provide diffraction-limited high spatial resolution for a wide spectral range including UV and x-rays~\cite{kipp2001, andersen2005, davila2011b, kim2012,oktem2018analytical}. This is not true for the refractive or reflective imaging elements. 
For example, at short wavelengths such as UV and
x-rays, refractive lenses do not exist 
due to the strong absorption of available materials,
and mirrors are costly to fabricate to achieve diffraction-limited resolution. 

To reduce chromatic aberration and enable \emph{broadband (full-spectrum)} imaging, 
different techniques have also been developed with diffractive lenses 
to focus different wavelengths onto a common plane either  physically or computationally~\cite{andersen2007,zhou2009,wang2003,zhao2015,wang2016,mohammad2018, li2018,dun2020}. In contrast to these broadband imaging techniques, our approach
\emph{exploits chromatic aberration} to enable high-resolution multi-spectral imaging. Some preliminary results of this research were presented in \cite{oktem2014icip, davila2016spd,osa2019}. 
Other {\emph{spectral imaging}} works that take advantage of chromatic aberration focus on 
compressive or snapshot spectral imaging of spectrally correlated scenes. Different techniques and system architectures have been proposed for this purpose such as
coded aperture spectral imaging with diffractive lenses (CSID)~\cite{osa2018,kar2019}, the use of
a diffractive lens equipped with a light-field detector~\cite{hallada2017}, 
a random diffractive filter in close proximity to an image sensor~\cite{wang2018}, a diffractive optical array after a pinhole array~\cite{nimmer2018}, and a specifically designed diffractive lens attached to a DSLR camera~\cite{jeon2019}.
However, as mentioned before, in none of these compressive/snapshot imaging approaches, the main goal was to improve the spatial and spectral resolutions of the conventional systems beyond their physical limitations as in this work.
Our approach can enable high spatial and spectral
resolutions with low cost designs for a variety of applications 
and with scenes that can possibly lack spectral correlation. Moreover, the 
novel reconstruction algorithms developed in this paper can be useful
to improve the performance of the other diffractive imaging modalities.

The paper is organized as follows. We discuss the related work in spectral imaging and imaging with diffractive lenses in Section II. We introduce the forward model of the developed imaging technique in Section III. The inverse problem is discussed in Section IV. 
Section V presents the fast ADMM-based image reconstruction algorithm that exploits analysis priors (such as total variation). Section VI develops another model-based reconstruction algorithm that learns the prior through a deep network that is end-to-end trained with an unrolling approach. The performance of the developed imaging technique and the reconstruction algorithms are illustrated for EUV solar imaging in Section VII through computer simulations. 
A detailed analysis is also performed to demonstrate the resolution of the technique.
In Section VIII, conclusions and final remarks are provided.

\section{Related work}

\subsection{Multi-spectral imaging}

Over the past decades, different techniques have been used for multi-spectral imaging. These techniques can be classified into three general groups: conventional methods, spectral filter array methods, and compressive or snapshot methods.

{\bf{Conventional (scanning-based) methods}} obtain the spectral data cube from a series of 2D measurements that are acquired simultaneously, or sequentially. Generally this data cube is 
measured using an imager with a series of spectral filters (obtained through multilayer coating) scanning the scene spectrally, or by using a spectrometer that involves a dispersion element with a slit scanning the scene spatially. One of the most important drawbacks of these techniques is the intrinsic physical limitations on their spatial, spectral, and temporal resolutions, as well as the inherent trade-off between spatial and spectral resolutions~\cite{davila2011b,arce2014}. 

In fact, for multi-spectral imagers employing spectral filters, the spectral resolution is strictly limited by the bandwidth of the producible wavelength filters in that spectral range, and the spatial resolution is limited by the quality and cost of the used imaging optics~\cite{lemen2011, davila2011b,andersen2005}. 
For example, existing EUV imagers offer spatial resolutions worse than the diffraction-limit and provide spectral resolutions of 10\% of the central wavelength at best~\cite{lemen2011, davila2011b,andersen2005},
which preclude resolving important spatial and spectral features. At larger wavelengths such as visible and infrared, diffraction-limited spatial resolution is easier to attain while spectral resolution - achieved with the current multi-layer coating technology - is around 2-20\% of the central wavelength~\cite{le2017,schowengerdt2006}.
Our imaging technique enables to overcome these limitations on the spatial and spectral resolutions (without any physical trade-off between them), by using a single diffractive imaging element for both dispersing and focusing the optical field. 

In the earlier works, super-resolution methods have been proposed to improve the resolutions of the conventional systems through post-processing~\cite{yi2019}. However, these methods aim to guess information that is not captured by the observation system and rely on 
spectral correlations; as a result, their performance is limited. 
On the other hand, our approach improves the spatial and spectral resolutions by proposing a novel system architecture to capture the observation data and then reconstruct the high-resolution spectral cube without requiring spectral correlations. 

{\bf{Spectral filter array methods}} rely on recent technology developments that extend the Bayer filters used in color imaging to multi-spectral sensors~\cite{geelen2013,lapray2014,wang2014,wang2019,hahn2020}. 
These methods use tiled filter arrays and enable to measure different spectral bands with a single sensor. They have generally been realized at small spatial scales and particular spectral ranges, and have an inherent trade-off between spatial and spectral resolutions. 

{\bf{Compressive or snapshot methods}} use different encoding schemes to acquire 
reduced number of measurements than the conventional methods~\cite{cao2016computational,book_chapter,arce2014}. 
The first approach in this respect is the computed tomography imaging spectrometry (CTIS)~\cite{okamoto1991,descour1995,ford2001}, which uses a tomographic reconstruction approach with an optical system similar to a slit spectrometer, where the main difference is to remove the slit (to allow an instantaneous 2D field-of-view).
An optical system similar to CTIS is also used with a parametric reconstruction approach that requires a parametric prior to hold along the spectral dimension~\cite{oktem2014}. In the compressive spectral imaging techniques - inspired by compressed sensing, the spectral data cube is reconstructed from compressive measurements by exploiting the correlations along the spatial and spectral dimensions. 
A commonly known compressive spectral imaging technique is coded aperture snapshot spectral imaging (CASSI)~\cite{cassi1}, with many variants~\cite{wu2011,cao2016computational,salazar2019}. 
The CASSI system is similar to the CTIS system 
where the major difference is the addition of a coded aperture, and the use of a sparse recovery algorithm for reconstruction. 
To provide a better alternative to CASSI, many other compressive or snapshot spectral imaging systems have been developed such as those exploiting single-pixel compressive camera architecture~\cite{august2013}, diffractive lenses~\cite{hallada2017,osa2018,kar2019,jeon2019,baek2020}, random diffractive filter~\cite{wang2018}, diffractive optical array with a pinhole array~\cite{nimmer2018},
prism with a DSLR camera\cite{baek2017} and diffuser with a multi-spectral sensor~\cite{monakhova2020}.

All of these compressive/snapshot methods have been developed with the goal of providing faster acquisition and/or reduced hardware complexity/cost than the conventional methods. 
However, in none of these approaches, the main goal was to improve the spatial and spectral resolutions of the conventional scanning-based methods beyond their physical limitations. To the best of our knowledge, there have been no reports of \underline{higher spatial and spectral resolutions} \underline{than the conventional scanning-based systems}, and ours is the first work that demonstrates this. Moreover, our system has no trade-off between spatial and spectral resolutions unlike many of the earlier systems; that is, the spatial and spectral resolutions can be increased independently of each other.

A related work in this respect aims to overcome the physical limitation on the spectral resolution of CASSI-type architectures~\cite{parada2016}. The spectral resolution in CASSI is limited mainly by the pitch size of the detector and the spectral dispersion of the used disperser. Spectral resolution improvement by a factor of two has been demonstrated in this work by replacing the binary coded apertures in the original CASSI system with color coded ones; however there are some technical challenges in the fabrication of such apertures that may affect the resulting spatial resolution~\cite{parada2016}. There are also  superresolution works that aims to improve the spatial resolution of a CASSI system with a low resolution detector by taking additional measurements~\cite{arguello2012}. A similar superresolution problem that takes into account the decimation effect caused by a low-resolution detector has been studied in \cite{oguzhan_thesis} for the spectral imaging system developed in this paper.

\subsection{Imaging with diffractive lenses}

Diffractive lenses such as Fresnel lenses and photon sieves have been used for various imaging techniques. These techniques can be classified into three general groups: monochromatic imaging, broadband imaging, and spectral imaging. 

{\bf{Monochromatic imaging}} has been the main area of use for diffractive lenses because of their wavelength-dependent focal length. 
When working with monochromatic illumination (i.e. single wavelength), diffractive lenses have many attractive properties such as their low weight, low cost, and flat structure. Another important advantage of diffractive imaging elements is that they can provide diffraction-limited high spatial resolution for a wide spectral range including UV and x-rays~\cite{kipp2001, andersen2005, davila2011b, kim2012,oktem2018analytical}. 
This is not true for the refractive or reflective imaging elements. For example, at short wavelengths such as UV and x-rays, refractive lenses do not exist 
due to the strong absorption of available materials, and 
mirrors are costly to fabricate to achieve diffraction-limited resolution. 
In fact, for reflective optics, surface roughness and figure errors cause
substantially worse spatial resolution than the diffraction limit~\cite{attwood_book}. 
On the other hand, diffractive lenses can provide diffraction-limited high
spatial resolution with relaxed fabrication tolerances. Many monochromatic imaging systems have been developed with diffractive lenses at visible, UV, and x-ray wavelengths to approach diffraction-limited imaging performance~\cite{attwood_book,kipp2001,andersen2005,menon2005}. 

{\bf{Broadband (full-spectrum) imaging}} techniques are based on the development of new diffractive lens designs with reduced chromatic aberration than the conventional designs~\cite{andersen2007,zhou2009,wang2003,zhao2015,wang2016,mohammad2018, li2018}. 
These techniques aim to focus different wavelengths onto 
a common plane so as to operate with broad-band illumination. Recent approaches~\cite{dun2020} also exploit the computational imaging framework and neural networks to improve the broadband imaging performance.

In contrast to these broadband imaging techniques, our approach
exploits chromatic aberration to enable high-resolution multi-spectral imaging. Some preliminary results of this research were presented in \cite{oktem2014icip, oktem2014agu,davila2016spd,osa2019}.
Other {\bf{spectral imaging}} works exploiting chromatic aberration focus on compressive or snapshot spectral imaging of spectrally correlated scenes. Different techniques and system architectures have been proposed for this purpose such as
coded aperture spectral imaging with diffractive lenses (CSID)~\cite{osa2018,kar2019}, the use of
a diffractive lens equipped with a light-field detector~\cite{hallada2017}, 
a random diffractive filter in close proximity to an image sensor~\cite{wang2018}, a diffractive optical array after a pinhole array~\cite{nimmer2018}, and a specifically designed diffractive lens attached to a DSLR camera~\cite{jeon2019,dun2020}.
However, in none of these compressive/snapshot 
approaches, the main goal was to improve the spatial and spectral resolutions
of the conventional methods beyond their physical limitations as in this work.
Our approach can enable unprecedented spatial and spectral
resolutions with low cost designs for a variety of applications 
and for scenes that can possibly lack spectral correlation. Moreover, the efficient reconstruction algorithms developed in this paper can be useful to improve the performance of other diffractive imaging modalities as well.

\section{Forward problem}

\subsection{Proposed multi-spectral imaging system}

Fig.~\ref{fig:system} depicts a schematic of the proposed spectral imaging system, which contains a diffractive lens (such as a photon sieve) and a monochrome detector. Here the optical field is first passed through a diffractive lens before recording with a monochrome detector. The diffractive lens performs both focusing and dispersion of the field.  
Because the focal length of the diffractive lens varies with the wavelength, each spectral component is focused at a different distance. 
As a result of this, when one of the spectral components is in focus, on the same plane there also exists defocused images of the other spectral components.
Hence, if a measurement is taken from a plane where one spectral component is focused, the focused image of this spectral component overlaps with the defocused images of the remaining components.  A total of $K$ such measurements are taken by the spectral imaging system. For the $k$th measurement, $d_s$ and $d_k$ respectively denote the distances from the object and measurement planes to the plane where the diffractive lens lies, where $k=1,\hdots,K$.

The measurement diversity can be achieved in different ways. One possible way is to use a moving detector along the axial direction in order to take successive measurements 
at different planes. In this case, a single diffractive lens can be used by changing the distance $d_k$ between the diffractive lens and the measurement plane. 
Alternatively, one can fix the measurement plane, i.e. the distance $d_k$, and use different diffractive lens designs with a different focusing behavior to  
take multiple measurements at the fixed held detector.
Such measurements can be acquired in many different ways depending on the wavelength regime. For example, one possible way is to generate a variable diffractive lens through a programmable spatial light modulator (SLM) or a digital micromirror device (DMD). 
This is a feasible approach in visible and infrared regimes at the expense of an additional optical element. More cost-effective approaches that also enable \emph{snapshot imaging} are i)  to design a single compound diffractive lens that focus multiple wavelengths onto different locations on the fixed detector~\cite{zhou2009,chung2008dual,yontem2018imaging,ayazgok2019},  
or ii) to use multiple optical paths with different diffractive lenses by using beam-splitters. These require a larger detector (or multiple detectors), but are feasible for a wide spectral range.

From the input object, we consider a polychromatic incoherent illumination 
consisting of $P$ different wavelengths $\lambda_p$ for $p=1,\hdots,P$. In the general case, these wavelengths are uniformly sampled from the spectral range observed by the system. The spectral sampling interval should be chosen in accordance with the spectral bandwidth of the designed diffractive lens (which determines the length of the spectral range that the 
response of the lens starts to change moderately with wavelength~\cite{attwood_book}[Chap.9].)
However, it is not necessary that the sampled  wavelengths are contiguous or uniform. 
In various multi-spectral imaging applications such as astrophysical imaging, 
remote sensing, and tissue spectroscopy, the wavelengths that will be observed  
are known a priori from spectroscopic observations, and these wavelengths do not need to be contiguous, uniformly sampled, or correlated~\cite{arce2014}. 
Unlike earlier approaches, our approach is general enough to work for such cases. This will be illustrated in Section VII.

In general, the diffractive lens can form images either with spatially coherent or incoherent illumination~\cite{oktem2018analytical}. In this work, we consider the incoherent case where the diffractive lens produces images in intensity, but similar concepts 
can also be applied to the coherent or partially coherent case.

\subsection{Continuous image-formation model} 

The first task is to mathematically relate the input spectral images to the multiple measurements obtained. Each measurement is a superposition of differently blurred spectral images. 
Mathematically, this can be formulated as follows:  
\be
\tk{k}(u,v)=\sum_{p=1}^P \sitilde{p}(u,v)   \ast    \g[k]{p}(u,v). \label{convolutionModel_ch2} 
\ee
Here $\tk{k}(u,v)$ is the intensity of the $k$th measurement, $\ast$ denotes 
convolution, and $\sitilde{p}(u,v)=\frac{d_s^2}{d_k^2} \sii{p}(-\frac{d_s}{d_k} u, -\frac{d_s}{d_k} v)$ is a scaled version of the $p$th spectral image 
$\sii{p}(u,v)$ at the wavelength $\lambda_p$. 
Hence each measurement contains $P$ terms resulting from $P$ different spectral components.

At distance $d_k$ the intensity of the spectral component at 
wavelength $\lambda_p$ is convolved with the point-spread
function (PSF), $\g[k]{p}(u,v)$, of the diffractive lens,
given by~\cite{oktem2018analytical,ayazgok2020}
\be
\g[k]{p}(u,v)= \left| i\frac{\lambda_p}{\Delta_k} e^{-i \pi \frac{u^2+v^2}{\Delta_k \lambda_p d_k^2}} \ast A_k\left(\frac{u}{\lambda_p d_k},\frac{v}{\lambda_p d_k} \right) \! \right|^2. \label{exactPsf} \!\!
\ee
Here $\Delta_k=1/d_s+1/d_k$, and $A_k(f_u,f_v)$ is the Fourier transform of the aperture (transmittance) function of the diffractive lens used in the $k$th measurement. The aperture 
function of a diffractive lens is defined as the ratio of the transmitted field
amplitude to the incident field amplitude at every point on the lens.
For the commonly used diffractive lens designs with circular holes~\cite{kipp2001}, $A_k(f_u,f_v)$ is sum of $\text{jinc}$ functions resulting from the Fourier transform of the circle functions corresponding to each hole. Note that this incoherent PSF formula is derived under Fresnel approximation~\cite{oktem2018analytical}, which is valid in almost all practical imaging scenarios of interest, and also has a fast computation algorithm~\cite{ayazgok2020}. 

An approximate, but a simpler model can also be used for the PSF in Eq.~\eqref{exactPsf} when the number of zones in the diffractive lens design is large and the diffractive lens behaves like a conventional (refractive) lens in its first diffraction order. This approximate PSF is given as follows~\cite{oktem2018analytical}:
\be
\g[k]{p}(u,v)=
\begin{cases}

\frac{1}{\pi^2}(\lambda_p d_k)^4 D_k^4 \text{jinc}^2(\frac{D_k}{\lambda_p d_k}u,\frac{D_k}{\lambda_p d_k}v ),& \epsilon_k = 0 \\
\left| \frac{1}{\pi}(\lambda_p d_k)^2 D_k^2 \text{jinc}(\frac{D_k}{\lambda_p d_k}u,\frac{D_k}{\lambda_p d_k}v )\ast \right.\\ \left. i\frac{\lambda_p}{\epsilon_k}\exp(-i\pi \frac{u^2+v^2}{\epsilon_k \lambda_p d_k^2}) \right|^2,& \epsilon_k \neq 0,

\end{cases}
\label{approxPsf} \!\!
\ee
where $D_k$ is the outer diameter of the diffractive lens 
in the $k$th measurement, $f_1=\frac{D_k w_k}{\lambda_p}$ is its first-order focal distance, $w_k$ is the outer zone width of the lens, and $\epsilon_k = 1/d_k + 1/d_s - 1/f_1$ is a parameter related to the amount of defocusing. This approximate model provides easier computation as well as simpler analysis of the imaging system.

\subsection{Discrete model}

In practice, only a finite number of discrete measurements is  
available through a digital sensor such as a CCD array. 
Since image reconstruction will be performed computationally on a computer, a discrete representation of the spectral images is also necessary. Now, our goal is to obtain such a discrete-to-discrete model between the unknown spectral images and measurements.

For this, we exploit the band-limitedness of the continuous functions involved. 
First note that each PSF $\g[k]{p}$ is band-limited to a circle of diameter $2D_k/(\lambda_p d_k)$. This is because the argument inside the magnitude sign in Eq.\eqref{exactPsf} has a circular frequency support with diameter $D_k/(\lambda_p d_k)$. The incoherent PSF is the magnitude square of this function, and hence the frequency support of this PSF can be found by convolving this circular support with itself, which results in a circular support with twice diameter~\cite{blahut_book}. Due to the band-limitedness of these PSFs $\g[k]{p}$ in Eq.~\eqref{convolutionModel_ch2}, each measurement $t_k$ is bandlimited.

Secondly, note that high frequencies of the spectral images, $\sitilde{p}$, that are 
outside the frequency support of the PSF, $\g[k]{p}$, are not captured in the measurements, which is known as the inherent \emph{diffraction-limit}~\cite{attwood_book}. 
As a result, the forward operator involving convolutions with these PSFs has a non-trivial nullspace. For discretization, we restrict our attention to the band-limited version of each spectral image and aim for recovering these band-limited versions, which are simply given by 
\be
x_p(u,v) \equiv \sitilde{p}(u,v) \ast {\rm{jinc}}\left(\frac{2D_k}{\lambda_p d_k}u, \frac{2D_k}{\lambda_p d_k} v\right).
\ee 
The jinc function ${\rm{jinc}}(u,v) = \frac{{\rm{J}}_1(\pi\sqrt{u^2+v^2})}{2\sqrt{u^2+v^2}}$ where ${\rm{J}}_1(u)$ is the first-order Bessel function of the first kind.
The forward model in Eq.~\eqref{convolutionModel_ch2} is still valid when the unknown source intensities, $\sitilde{p}$, are replaced with their band-limited versions, $x_p$. 
Hence, we can treat all functions in the continuous forward model as band-limited functions and represent them using sinc basis~\cite{blahut_book}.

Now by representing each continuous band-limited function with sinc basis, 
the continuous convolution operations in Eq.~\eqref{convolutionModel_ch2} 
reduce to discrete convolutions of the form 
\be
t_k[m,n]=\sum_{p=1}^P x_p[m, n] \ast \g[k]{p}[m, n], \label{forwardModel}
\ee 
where $m,n = [0,\hdots, N-1]$. Here $t_k[m,n]$, $x_p[m, n]$, and $\g[k]{p}[m, n]$ are  
sampled versions of their continuous counterparts, e.g. $t_k[m,n]=t_k(m \Delta, n \Delta)$ for some $\Delta$ smaller than the Nyquist sampling interval (i.e. $\Delta < \frac{\lambda_p d_k}{2D_k}$ for all $k$). 
Here the uniformly sampled observations, $t_k[m,n]$, are assumed to be same as the detector measurements, i.e. the averaged intensity over detector pixels of width $\Delta$.

Let the PSF $\g[k]{p}[m, n]$ has $M \times M$ support, i.e. $\g[k]{p}[m, n] \approx 0$ for $m,n \notin [0, M-1]$.
We assume that the supports of the spectral images are slightly smaller than the detector range covered by $N$ pixels, i.e. $x_p[m, n] = 0$ for $m,n \notin [0, N-M]$. 
Then the linear convolution in Eq.~\eqref{forwardModel} can be replaced with a circular convolution of $N$ points, which will be useful to develop fast image reconstruction methods.

Due to the linearity of the convolution operator, the discrete model can be 
formulated in the following matrix-vector form using appropriate lexicographic ordering:
\be
{\tb_k}=\sum_{p=1}^P \Hb_{k,p} \xb_p.
\ee
Here $\Hb_{k,p}$ is an $N^2 \times N^2$ block circulant matrix with circular blocks corresponding to the circular convolution operation with the PSF $\g[k]{p}[m,n]$. Vectors $\tb_k$ and $\xb_p$ correspond to lexicographically ordered versions of the observation, $t_k[m,n]$, and spectral image, $x_p[m, n]$, respectively. By combining all the observation vectors into a single vector, $\tb$, and all the image vectors into the vector $\xb$, we obtain
\be
\tb=\Hb\xb,
\ee
\bea
\Hb=\left[ \begin{array}{ccc}
\Hb_{1,1} & \hdots & \Hb_{1,P} \\ 
\vdots   &            & \vdots \\
\Hb_{K,1} & \hdots & \Hb_{K,P}
\end{array}
\right], 
\eea
where $\tb=[\tb_1^T | \hdots | \tb_K^T]^T$, $\xb=[\xb_1^T| \hdots | \xb_p^T ]^T$, and $\Hb$ 
is a $KN^2 \times P N^2$ matrix. 
The final observation model can be expressed with noise as follows:
\vspace{-0.05in}
\be
\yb=\tb+\wb=\Hb\xb+\wb \label{noisyModel} \vspace{-0.05in}
\ee
where $\wb=[\wb_1^T | \hdots | \wb_K^T]^T$ is the additive noise vector. Here, white Gaussian noise is assumed, where $(\wb_k)_i \sim N(0,\sigma_k^2)$ is uncorrelated 
over different pixels $i$ and measurements $k$, with $\sigma_k$ representing the noise standard deviation for the $k$th measurement.
This noise assumption is generally valid for practical multi-spectral imaging scenarios of interest.

\section{Inverse problem}  

In the inverse problem, the goal is to recover the unknown spectral images, $\xb$, from the measurements, $\yb$, obtained with the proposed spectral imaging system. This inverse problem can be considered as a multi-frame deconvolution problem where the measurements are in the form of superimposed blurred images. That is, each measurement is a superposition of focused or defocused versions of all spectral images.
This problem is inherently ill-posed; as the PSFs, $\g[k]{p}[m, n]$, of different wavelengths, $p$, and measurements, $k$, become more similar (for example, when the difference between different wavelengths or measurement distances decreases), the conditioning of the problem gets worse due to increased dependency of the columns or rows of $\Hb$, respectively.

There are various approaches for solving ill-posed linear inverse problems. A systematic approach to regularization uses prior knowledge about the unknown solution 
and leads to the minimization of an appropriately formulated cost function~\cite{hansen_book,vogel2002computational}. The prior information can be introduced in a deterministic or a statistical setting, with the latter related to the Bayes framework~\cite{geman1987stochastic}. A general formulation of the inverse problem can be expressed as
\be
\min_{{{\xb}}} \;  || \yb-\Hb\xb ||_2^2 +\lambda \Phi(\Pb\xb).
\label{inverseProblem_pssi} 
\ee 
where the first term measures data fidelity, and the second term, $\Phi(\Pb\xb)$, controls how well the solution matches the prior knowledge, with the parameter $\lambda$ trading off between these two terms. Here $\Pb$ is a matrix representing an analysis operator. 

Equivalently, we can reformulate this problem as a constrained problem:
\be
\min_{{{\xb}}} \;  \Phi(\Pb\xb)~ {\text{subject to}}~ ||  \yb-\Hb\xb ||_2\leq\epsilon,
\label{inverseProblem2_pssi} 
\ee 
where $\epsilon\geq0$ is a parameter that depends on noise level. Note that if the problem in Eq.~\eqref{inverseProblem2_pssi} is feasible for some $\epsilon\geq0$, then it is equivalent to Eq.~\eqref{inverseProblem_pssi} for some $\lambda\geq0$. 
The formulation in Eq.~\eqref{inverseProblem2_pssi} has advantage over Eq.~\eqref{inverseProblem_pssi} due to the fact that the parameter $\epsilon$ is directly proportional to the noise standard deviation, so it is easier to choose than the parameter $\lambda$~\cite{c-salsa}.

There are popular and powerful choices for the regularizer $\Phi(.)$~\cite{c-salsa,rudin1992nonlinear,aggarwal2019,ongie2020}. 
Because we want our approach to be applicable for scenes that 
can possibly lack spectral correlation, we exploit 2D priors for each spectral image (rather than a 3D prior for the entire spectral cube). Among 2D analytical priors, total-variation (isotropic or anisotropic) has significant popularity as it provides superior results for 
images with significant structure or piecewise-constant characteristics~\cite{vogel2002computational}.   
As the analytical priors have recently been shown to be outperformed by deep priors~\cite{lucas2018,aggarwal2019,ongie2020}, 
there is also significant interest in learning the prior through a deep network that is trained offline or end-to-end with an unrolling approach. In this work, we develop two reconstruction algorithms to solve the inverse problem, one with analytical priors and another with deep priors, and comparatively evaluate their performance.

\section{Image Reconstruction with Analytical Priors}

In this work, we first develop a fast reconstruction algorithm using the ADMM framework to solve the resulting optimization problem in Eq.~\eqref{inverseProblem2_pssi} with an analytical prior (particularly, with an analysis prior). ADMM belongs to the family of augmented Lagrangian methods~\cite{nocedal2006numerical} and is used in many signal and image reconstruction problems~\cite{c-salsa,ng2010solving}. It provides a divide-and-conquer approach by splitting the minimization steps in an unconstrained multi-objective optimization problem. Its convergence is guaranteed under mild conditions.

Following the ADMM framework~\cite{c-salsa}, we first convert the problem in Eq.~\eqref{inverseProblem2_pssi} to an unconstrained problem by adding the constraint to the 
cost function as a penalty: 
\be
\min_{{{\xb}}} \;  \Phi(\Pb\xb)+\iota_{(|| \yb-\Hb\xb ||_2\leq\epsilon)}(\xb),
\label{inverseProblem3} \vspace{-0.0in}
\ee 
where the indicator function $\iota_{(|| \yb-\Hb\xb ||_2\leq\epsilon)}(\xb)$ is defined as
$$
\iota_{(|| \yb-\Hb\xb ||_2\leq\epsilon)}(\xb)=
\begin{cases}

0,~ &\text{if}~ || \yb-\Hb\xb ||_2\leq\epsilon \\
+\infty,~ &\text{if}~ ||\yb-\Hb\xb ||_2>\epsilon.

\end{cases}
$$

After variable-splitting, we arrive at the following problem:
\begin{align}
\begin{array}{cc}
\underset{\xb,\ub,\vb}{\text{minimize}} & \Phi(\ub) +\iota_{(|| {\yb} - {\vb} ||_2\leq\epsilon)}(\vb) \\
\text{subject to} & \ub = \Pb\xb ,~~ \vb = {\Hb}\xb
\end{array}  \label{eq:admm_real}
\end{align} 
where $\ub$ and $\vb$ are the auxiliary variables in the ADMM framework. After 
formulating the problem in \eqref{eq:admm_real} in augmented Lagrangian form~\cite{c-salsa}, minimization over $\xb$, $\ub$, and $\vb$ is required. 
We perform each minimization in an alternating fashion. For the $l$th iteration, 
the required minimizations are 
\begin{align}
\xb^{l+1} = \underset{\xb}{\text{arg min}}~ \frac{\mu}{2} \left\Vert \left[ \begin{array}{ccc} \Hb \\ \Pb \end{array} \right] \xb - \left[ \begin{array}{ccc} \vb^l \\ \ub^l \end{array}\right] - \left[ \begin{array}{ccc} \fb^l \\ \db^l \end{array}\right]    \right\Vert_2^2 \label{x-update_pssi}
\end{align}
\begin{align}
\begin{array}{cc}
\ub^{l+1} = \underset{\ub}{\text{arg min}}~ \Phi(\ub)+\frac{\mu}{2} \Vert\ub-(\Pb\xb^{l+1}-\db^{l}) \Vert_2^2   
\end{array}  \label{eq:prox1tv_pssi}
\end{align} 
\begin{align}
\begin{array}{cc}
\vb^{l+1} = \underset{\vb}{\text{arg min}}~ \iota_{(|| {\yb} - {\vb} ||_2\leq\epsilon)}(\vb)+
\frac{\mu}{2} \Vert\vb-(\Hb\xb^{l+1}-\fb^{l}) \Vert_2^2   
\end{array}  \label{eq:prox2proj_pssi}
\end{align} 
with $\db$ and $\fb$ denoting the dual variables in the ADMM framework, and $\mu$ is a penalty parameter. We now explain how these update steps, referred to as $\xb$-update, $\ub$-update, and $\vb$-update, are performed.

In the $\xb$-update step, we face the least squares problem in Eq.~\eqref{x-update_pssi} with the following closed-form solution:
\begin{align}
\xb^{l+1} = (\Ib + \Hb^{H}\Hb)^{-1}(\Pb^{H}(\ub^{l}+\db^{l})+\Hb^{H}(\vb^{l}+\fb^{l})).
\label{x-ls_final_pssi}
\end{align}
Here $\Pb$ is assumed to be a unitary transform resulting in $\Pb^H\Pb = \Ib$. 
This solution can be efficiently obtained through computations
in the frequency domain by exploiting the diagonalizability of the 2D
circular convolution operations~\cite{kamaci2017efficient}.

Note that each block of $\Hb$ is diagonalized by the discrete Fourier transform (DFT) matrix because $\Hb_{k,p}$ is block circulant with circular blocks (BCCB). Hence $\Hb_{k,p} = \Fb_{2D}^{H} \mathbf{\Lambda}_{k,p} \Fb_{2D}$
where $\Fb_{2D}$ is the unitary 2D DFT matrix and $\mathbf{\Lambda}_{k,p}$ is a
diagonal matrix whose diagonal consists of the 2D DFT of the
PSF $\g[k]{p}[m,n]$, for $k=1,\hdots,K$ and $p = 1,\hdots,P$. As
a result, the overall matrix $\Hb$ can be written as $\Hb = \bar{\Fb}^H \mathbf{\Lambda} \tilde{\Fb}$
where $\bar{\Fb} = \Ib_K \otimes \Fb_{2D}$ and $\tilde{\Fb} = \Ib_P \otimes \Fb_{2D}$ with $\otimes$
denoting the Kronecker product and $\Ib_n$ denoting the identity matrix of
size $n \times n$. Here $\mathbf{\Lambda}$ is a matrix of  $K \times P$ blocks with each
block given by $\mathbf{\Lambda}_{k,p}$. By inserting this expression of $\Hb$ in
Eq.~\eqref{x-ls_final_pssi}, the following form can be obtained for the efficient
computation of the image update step:
\begin{align}
\xb^{l+1} = \tilde{\Fb}^H ( \Ib + \mathbf{\Lambda}^{H}\mathbf{\Lambda})^{-1} (\tilde{\Fb}\Pb^{H}(\ub^{l}+\db^{l})+\mathbf{\Lambda}^{H}\bar{\Fb}(\vb^{l}+\fb^{l})).
\label{x-ls_final_final_final}
\end{align}

To compute the solution in Eq.~\eqref{x-ls_final_final_final}, we do not need to form huge matrices, which provides significant savings for memory and computation time. In fact, multiplication by $\tilde{\Fb}$ or $\tilde{\Fb}^H$ corresponds to taking the DFT or inverse DFT of all 2D spectral bands for $p = 1,\hdots,P$. Similarly, multiplication by $\bar{\Fb}$ corresponds to taking the DFT of the 2D measurement signals for $k=1,\hdots,K$. For example, $\bar{\Fb}(\vb^{l}+\fb^{l}) = [(\Fb_{2D}(\vb_1^{l}+\fb_1^{l}))^T | \hdots | (\Fb_{2D}(\vb_K^{l}+\fb_K^{l}))^T]^T$, 
where each term can be computed via the 2D FFT. Moreover, because $\mathbf{\Lambda}$ is a block matrix consisting of diagonal matrices, multiplication
by $\mathbf{\Lambda}^H$ corresponds to element-wise 2D multiplication with the conjugated DFTs of the underlying PSFs and then summation. Furthermore, for a unitary $\Pb$, multiplication by $\Pb^{H}$ corresponds to taking the inverse transform. Note that when the image data is correlated in all directions, this can be chosen as a 3D transform; otherwise, it can be a 2D transform applied on each spectral band separately.

Lastly, the inverse of $\mathbf{\Sigma} \triangleq (\Ib + \mathbf{\Lambda}^{H}\mathbf{\Lambda})$ needs to be computed only once, and hence does not affect the computational
cost of the iterations. Nevertheless, it is possible to reduce the
required time and memory for this pre-computation through a
recursive block matrix inversion approach~\cite{noble1988}. Note that $\mathbf{\Sigma}$ is
a block matrix of $P \times P$ blocks, where each block is a diagonal
matrix given by $\mathbf{\Sigma}_{i,j}=\sum_{k=1}^K \mathbf{\Lambda}_{k,i}^H\mathbf{\Lambda}_{k,j} + \delta_{i,j}\Ib$
with $\delta_{i,j}$ denoting the Kronecker delta function and $i,j = 1, \hdots, P$.
Hence, for P = 2, the inverse can be computed as
\vspace{-0.0in}
\bea
\left[ \begin{array}{ccc}
\mathbf{\Sigma}_{1,1} &  \mathbf{\Sigma}_{1,2} \\ 
\mathbf{\Sigma}_{2,1} &  \mathbf{\Sigma}_{2,2}
\end{array}
\right]^{-1}=\left[ \begin{array}{ccc}
 \Cb  &  \mathbf{\Sigma}_{1,1}^{-1}\mathbf{\Sigma}_{1,2}\Kb \\ 
\Kb \mathbf{\Sigma}_{2,1}\mathbf{\Sigma}_{1,1}^{-1}  &  -\Kb
\end{array}
\right] \label{recursion}
\eea
where $\Kb = -(\mathbf{\Sigma}_{2,2} - \mathbf{\Sigma}_{2,1}\mathbf{\Sigma}_{1,1}^{-1}\mathbf{\Sigma}_{1,2})^{-1}$ and $\Cb =\mathbf{\Sigma}_{1,1}^{-1} - \mathbf{\Sigma}_{1,1}^{-1}\mathbf{\Sigma}_{1,2}\Kb\mathbf{\Sigma}_{2,1}\mathbf{\Sigma}_{1,1}^{-1}$.
For $P>2$ case, the overall matrix $\mathbf{\Sigma}$ is partitioned into $2 \times 2$ blocks and each block is inverted recursively using Eq.~\eqref{recursion}. Because all the matrices involved in these computations are diagonal, computing the inverse of $\mathbf{\Sigma}$ requires element-wise multiplication and division operations.

For the $\ub$-update step, the problem in Eq.~\eqref{eq:prox1tv_pssi} needs to be solved, which corresponds to a denoising problem of the form:
\begin{align}
\begin{array}{cc}
\Psi_{f}(\zb) = \underset{\ub}{\text{arg min}}~ f(\ub)+\frac{\mu}{2} \Vert\ub-\zb \Vert_2^2.
\end{array}  \label{eq:prox_general_pssi}
\end{align} 
Here $f$ is a regularization functional and $\zb$ is a noisy observation for $\ub$. The solution is given by the Moreau proximal mapping of the regularization functional $f/\mu$
evaluated at $\zb$.  
Hence, the solution of Eq.~\eqref{eq:prox1tv_pssi} is, by definition, the Moreau proximal mapping of $\Phi(.)/\mu$ evaluated at $\Pb\xb^{l+1}-\db^{l} $. We denote this proximal mapping as $\Psi_{\Phi/\mu}(\Pb\xb^{l+1}-\db^{l})$:
\be
\ub^{l+1}= \Psi_{\Phi/\mu}(\Pb\xb^{l+1}-\db^{l}). \label{eq:u_update_final}
\ee

There are efficient computations of $\Psi_\Phi(\Pb\xb^{l+1}-\db^{l})$ for the different choices of the functional $\Phi(.)$. For example, if $\ell_1$-norm is used, i.e. $\Phi(\Pb\xb)=\Vert \Pb\xb \Vert_1$, then $\Psi_{\Phi/\mu}(\Pb\xb^{l+1}-\db^{l})$ is simply soft-thresholding, i.e. 
$\textit{soft}(\Pb\xb^{l+1}-\db^{l},1/\mu)$ with $1/\mu$ denoting the threshold.
Here, multiplication by $\Pb$ corresponds to either a single 3D transform or multiple 2D transforms along each spectral band. The soft-thresholding operation, $\textit{soft}(\nb,\tau)$, is component-wise computed as $\nb_i\rightarrow \textit{sign}(\nb_i)\max(|\nb_i |-\tau,0)$ for each voxel $i$, with $\textit{sign}(\nb_i)$ taking value $1$ if $\nb_i>0$ and $-1$ otherwise~\cite{c-salsa}. 
If $\Phi(.)$ is chosen as the 2D isotropic TV operator applied on each spectral band, $\Pb$ becomes an identity matrix, and the resulting proximal mapping, $\Psi_{\text{TV}}(\xb^{l+1}-\db^{l})$, has an efficient decoupled calculation for each spectral band using Chambolle's algorithm~\cite{chambolle,c-salsa}. Specifically, we compute the proximal mapping for each band of $\xb^{l+1}-\db^{l}$ using Chambolle's algorithm and then concatenate the updated bands along the spectral dimension.

For the $\vb$-update step given in Eq.~\eqref{eq:prox2proj_pssi}, similar to the $\ub$-update, we have the proximal mapping of $\iota_{(|| {{\yb}} -  {\vb} ||_2\leq\epsilon)/\mu}(.)$ evaluated at $\Hb\xb^{l+1}-\fb^{l}$, which we denote as $\Psi_{\iota_{(|| {\yb} - {\vb} ||_2\leq\epsilon)/\mu}}(\Hb\xb^{l+1}-\fb^{l})$.
Calculation of this proximal mapping requires a projection of $\ssb\triangleq(\Hb\xb^{l+1}-\fb^{l}) $ onto $\epsilon$-radius ball centered at ${{\yb}}$. The solution is given~\cite{c-salsa} by 
\begin{align}
\begin{array}{cc}
\Psi_{\iota_{(|| {\yb} - {\vb} ||_2\leq\epsilon)}}(\ssb)=
\begin{cases}
\yb + \epsilon\frac{\ssb-\yb}{\Vert \ssb-\yb \Vert_2},~ &\text{if}~~ \Vert \ssb-\yb \Vert_2 > \epsilon \\
\ssb,~ &\text{if}~~ \Vert \ssb-\yb \Vert_2 \leq \epsilon.
\end{cases}
 \end{array}  \label{eq:normaleq_prox_proj_pssi}
\end{align}

Finally, ADMM dual variables, $\db$ and $\fb$, are updated as 
\begin{align}
\begin{array}{cc}
{\db}^{l+1} = {\db}^l - (\Pb\xb^{l+1}-\ub^{l+1}),
\end{array}  \label{eq:d1_update_pssi}
\end{align} 
\begin{align}
\begin{array}{cc}
{\fb}^{l+1} = {\fb}^l -(\Hb\xb^{l+1}-\vb^{l+1}).
\end{array}  \label{eq:d2_update_pssi}
\end{align} 

The overall algorithm is summarized in Table~\ref{tab:pssi_ps}. In the numerical results, we choose $\Phi(.)$ as the 2D isotropic TV operator and Step 4 is solved efficiently using Chambolle's algorithm~\cite{chambolle} for each spectral band, as explained before.

\begin{table}[t]
	\centering
	\caption{Image reconstruction algorithm with analytical prior} 
	\label{tab:pssi_ps}
	\begin{tabular}{l}
		\hline
		\textbf{Input:} $\yb:$ measurements, $\Hb:$ system matrix, $\Pb:$ transform \\
		\textbf{Output:} $\xb:$ reconstructed image \\
		1. set: $l = 0$ \\ \quad choose: $\mu>0$, $\ub^{0}$, $\vb^{0}$, $\db^{0}$, $\fb^{0}$ \\
		2. \textbf{repeat} \\
		3. ~~~~update $\xb^{l+1}$ using Eq.~\eqref{x-ls_final_final_final}.\\
		4. ~~~~update $\ub^{l+1}$ using Eq.~\eqref{eq:u_update_final}. \\ 
		5. ~~~~update $\vb^{l+1}$ using Eq.~\eqref{eq:normaleq_prox_proj_pssi}.\\
		6. ~~~~update ${\db}^{l+1}$ using Eq.~\eqref{eq:d1_update_pssi}.\\
		7. ~~~~update ${\fb}^{l+1}$ using Eq.~\eqref{eq:d2_update_pssi}.\\
		9. ~~~~set $l \leftarrow l + 1$. \\
		10. \textbf{until:} some stopping criterion is satisfied. \\
		\hline
	\end{tabular} 
\end{table}

\subsection{Computational Complexity}
The computational complexity of the algorithm is dominated by the $\xb$-update, given in Step 3 in Table~\ref{tab:pssi_ps}. This requires $2P$ FFT and $P$ inverse FFT computations. Thus, its computational complexity is $O(PN^2 \log(N))$ where $N^2$ is the size of each spectral image and $P$ is the number of spectral bands. Calculation of the proximal mapping in Step 4 and ADMM dual variable update in Step 6 have $O(N^2)$ complexity if $\Pb$ is a diagonal matrix, or $O(PN^2 \log(N))$ if $\Pb$ has a fast implementation such as with FFT. Steps 5 and 7 require multiplication with $\Hb$, which are computed using FFT with $O(PN^2 \log(N))$ complexity. Thus, the overall complexity of the algorithm is $O(PN^2 \log(N))$ as determined by the complexity of the computation of the forward operator and its adjoint. It is also worth noting that the complexity of the recursive inversion in Eq.~\eqref{recursion} is $O(P^3 N^2)$. Since it is pre-calculated once, it does not affect the algorithm's overall complexity. 

\section{Learned Reconstruction with Deep Priors} 

We also develop a model-based learned reconstruction method with deep priors motivated by the idea presented in \cite{aggarwal2019}. Model-based learned reconstruction methods can simultaneously incorporate physics-based forward model and the data, and have been successfully applied to various inverse problems in imaging~\cite{lucas2018,aggarwal2019,ongie2020,jeon2019}. 

Our approach learns a 2D prior through a deep network that is end-to-end trained through unrolling. For this, we apply half-quadratic splitting to Eq.~\eqref{inverseProblem_pssi} with $\Pb = \Ib$, resulting in 
\be
\min_{{{\xb, \zb}}} \; || \yb-\Hb\xb ||_2^2 + \lambda \Phi(\zb) + \nu || \xb-\zb ||_2^2.
\label{inverseProblem_modl} \vspace{-0.0in}
\ee 
Here $\nu$ is a learned penalty parameter. 
This problem can be alternatingly solved as follows:
\begin{subequations}
\begin{align}
\zb^l && = &&\Dwbn,  \hspace{1.6in}\label{inverseProblem_modl:a}  \\
\xb^{l+1} &&= &&\underset{\xb}{\text{arg min}}~ || \yb-\Hb\xb ||_2^2 +\nu || \xb-\zb^l ||_2^2,  \vspace{-0.0in}\label{inverseProblem_modl:b} 
\end{align}
\end{subequations}
These can be regarded as denoising and data fidelity updates, respectively. Here $\Dwb$ is the denoised and artifact free version of $\xb$ that is obtained by passing $\xb$ through a convolutional neural network (CNN).

For the data-fidelity update, the least-squares problem in Eq. \eqref{inverseProblem_modl:b} can be solved in closed-form as follows:
\be
\xb^{l+1} = (\nu \Ib + \Hb^{H}\Hb)^{-1}(\Hb^{H}\yb+ \nu \zb^l).
\label{modl_normal_eq} \vspace{-0.0in}
\ee 
This solution can be efficiently obtained through computations in the frequency domain by exploiting the diagonalizability of the convolution operations involved.  Remember that $\Hb$ can be written as $\Hb= \bar{\Fb}^H \mathbf{\Lambda} \tilde{\Fb}$. By inserting this expression in Eq.~\eqref{modl_normal_eq}, we obtain the following form for the efficient data fidelity update step:
\begin{align}
\xb^{l+1} = \tilde{\Fb}^H ( \nu \Ib + \mathbf{\Lambda}^{H}\mathbf{\Lambda})^{-1} (\mathbf{\Lambda}^{H} \bar{\Fb} \yb + \nu \tilde{\Fb} \zb^{l}).
\label{data_consistency_update}
\end{align}
Here, multiplication by $\tilde{\Fb}$ or $\tilde{\Fb}^H$ corresponds to taking the FFT or inverse FFT of all 2D spectral bands for $p = 1,\hdots,P$. Similarly, multiplication by $\bar{\Fb}$ corresponds to taking the FFT of the 2D measurements for $k=1,\hdots,K$. Moreover, because $\mathbf{\Lambda}$ is a block matrix consisting of diagonal matrices, multiplication
by $\mathbf{\Lambda}^H$ corresponds to element-wise 2D multiplication with the conjugated FFTs of the PSFs and then summation. The inverse of $\nu  \Ib + \mathbf{\Lambda}^{H}\mathbf{\Lambda}$ can be computed through a recursive block matrix inversion approach~\cite{noble1988}, as before. 

\begin{figure}[htp]
	\begin{center} 
		\subfloat[]{
		\includegraphics[width=3.0in]{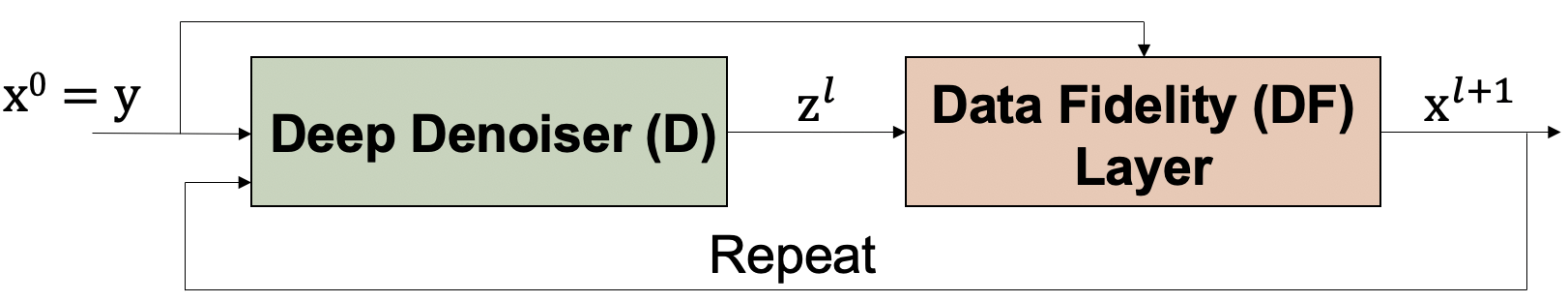}
		}\\
		\subfloat[]{
		\includegraphics[width=3.5in]{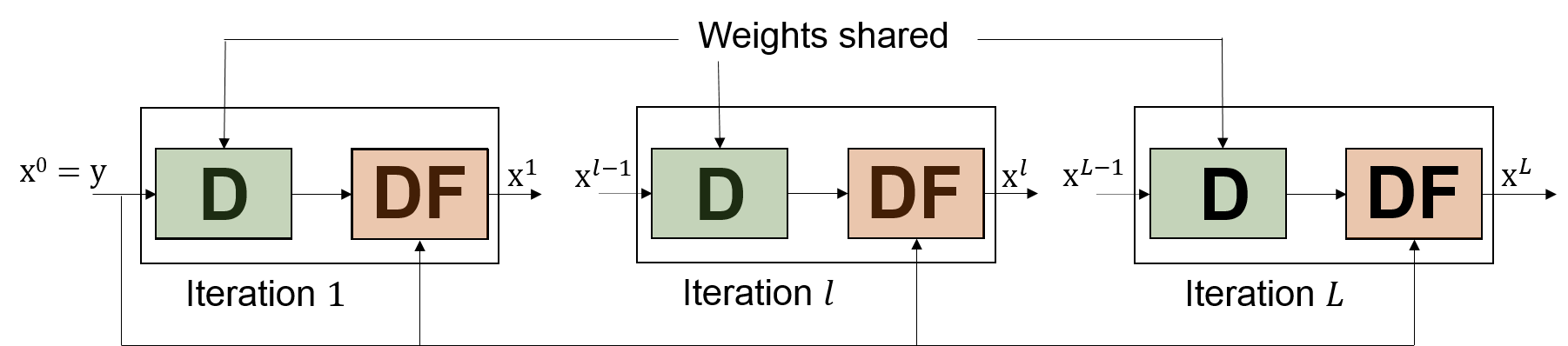}
		}\\
		\caption{Model-based learned reconstruction framework. (a) The iterative algorithm described by \eqref{inverseProblem_modl:a} and \eqref{inverseProblem_modl:b}, (b) the resulting architecture after unrolling.		
		}
		\label{fig-modl_structure}
	\end{center} 
\end{figure}

For the denoising update in Eq. \eqref{inverseProblem_modl:a}, the CNN architecture proposed in \cite{aggarwal2019} is used. This CNN consists of $N$ layers, with each layer having 64 filters of size $3\times 3$. First $N-1$ layer has convolution, batch normalization, and non-linear activation function ReLU units. The last layer, i.e., $N^{th}$ layer, does not have ReLU so that the negative part of the learned noise is not truncated. Residual learning strategy is used to speed up the learning. After fixing the number of iterations to $L$ for the alternating updates, the steps in Eq. \eqref{inverseProblem_modl:a} and \eqref{inverseProblem_modl:b} can be unrolled to an end-to-end trainable network as shown in Fig.~\ref{fig-modl_structure}. Sharing weights are used between different iterations to decrease the number of trainable parameters. 
After training the network, the images are reconstructed by simply feeding the network with the available measurements.

\section{Numerical Results} 

In this section, we demonstrate the high spatial and spectral resolutions enabled by the developed multi-spectral imaging technique in solar imaging~\cite{davila2011b}. For this, we consider three nearby solar emission lines in the EUV regime, with central wavelengths of $\lambda_1=33.28$ nm, $\lambda_2=33.42$ nm and $\lambda_3=33.54$ nm (i.e., $P=3$)~\cite{brekke2000extreme}.
Our goal is to obtain the spectral images for these three nearby emission lines with diffraction-limited spatial resolution. 
Such high spatial and spectral resolutions are not possible to achieve with the conventional spectral imagers involving wavelength filters~\cite{lemen2011, davila2011b,andersen2005,oktem2014icip}. Obtaining these spectral images will provide information about the important physical parameters of the Sun's extended atmosphere, and hence will enable the investigation of the dynamic plasma behavior~\cite{young2007euv, brekke2000extreme}.

For the diffractive lens, a sample photon sieve (PS) design developed in \cite{davila2011b} for the considered application is used, where the outer diameter is chosen as $D=25$ mm, and the diameter of the smallest hole as $\Delta=$ \SI{5}{\micro\metre}. 
The resulting photon sieve has first-order focal lengths of $f_1=3.756$ m, $f_2=3.740$ m, and $f_3=3.727$ m at the corresponding wavelengths (i.e. $f=D\Delta / \lambda$)~\cite{davila2011b}. As determined by the smallest hole diameter $\Delta$,
the resulting (Abbe's) diffraction-limited spatial resolution is \SI{5}{\micro\metre}~\cite{attwood_book}. 
The detector pixel size is chosen as \SI{2.5}{\micro\metre} to match the diffraction-limited resolution with two pixels on the detector (i.e. the Nyquist rate sampling). Moreover, the expected spectral resolution is $4 \Delta \lambda  / D  \approx 0.03$ nm near a wavelength of $33$ nm, as given by the spectral bandwidth of the diffractive lens~\cite{attwood_book}[Chap.9]. This bandwidth is slightly larger than the typical width of the considered emission lines and hence the entire spectral line will be in good focus at each focal plane~\cite{davila2011b}.
Moreover, the distance between the wavelength sources is $\sim 0.1$ nm, which is larger than the spectral resolution supported by the system. Note that the spatial and spectral resolutions of the system can be increased independently of each other, as the former is determined by the minimum hole diameter $\Delta$ while the latter is by $4 \Delta \lambda  / D$.

For the imaging system, two different measurement settings are considered. In the $\textit{moving detector}$ (MD) case, the system measures the intensities at the three focal planes, $f_1$, $f_2$ and $f_3$ (i.e., $K=3$). Hence at the first focal plane, the measurement contains a focused image of the first source and the defocused images of the second and third sources, and vice versa at the other focal planes. Note that the spectral bandwidth of the PS design is equal to $4 \Delta \lambda  / D \approx 0.03$ nm near a wavelength of $33$ nm, which is slightly larger than the typical width of an emission line in this application. Hence, in each measurement, one of the spectral (emission) lines will be entirely in focus, while the other two will be defocused.

Alternatively, in the $\textit{fixed detector}$ (FD) case, we fix the measurement plane at the distance $f_2$ and obtain the measurements by changing the PS design at each shot. For each measurement, same design parameters are used as before by only changing the diameter $D$ to focus $\lambda_1$ or $\lambda_3$ onto the distance $f_2$. Because the measurement plane is fixed to $f_2 = 3.740$ m, the outer diameter is decreased by $104.6$ $\mu$m for $\lambda_1$ and is increased by $89.9$ $\mu$m for $\lambda_3$. These modifications on the PS design are well within the limits of modern lithography techniques~\cite{menon2005}. 
FD measurement setting obtains similar measurements with the MD setting without moving the detector in the axial direction, but at the expense of additional optical components (SLM or beamsplitters), larger detector array, and/or lower SNR due to the reduced diffraction efficiency of the compound lens designs. Replacing the diffractive lens may also be considered for the FD setting using a mechanical component, but this may be more challenging in practice.

\subsection{Performance Analysis}

We present numerical simulations to demonstrate the performance of the developed imaging technique and algorithms under different imaging scenarios with MD and FD settings. For this, solar EUV images of size $256\times 256$ pixels are used as the input images. However, because the spatial resolution of the existing solar imagers are worse than the diffraction-limited resolution 
enabled by the proposed technique, realistic (high-resolution) solar images are not available
for the simulations. Here we use these images as if they were images of some other sun-like object, and demonstrate the diffraction-limited resolution for this. Hence this experiment will illustrate that objects with similar characteristics can be observed with diffraction-limited high spatial resolution.  

\begin{figure*}[tbh!]
\begin{center} 
\subfloat[]{\hspace{-0.0in} 
    \includegraphics[scale=0.21]{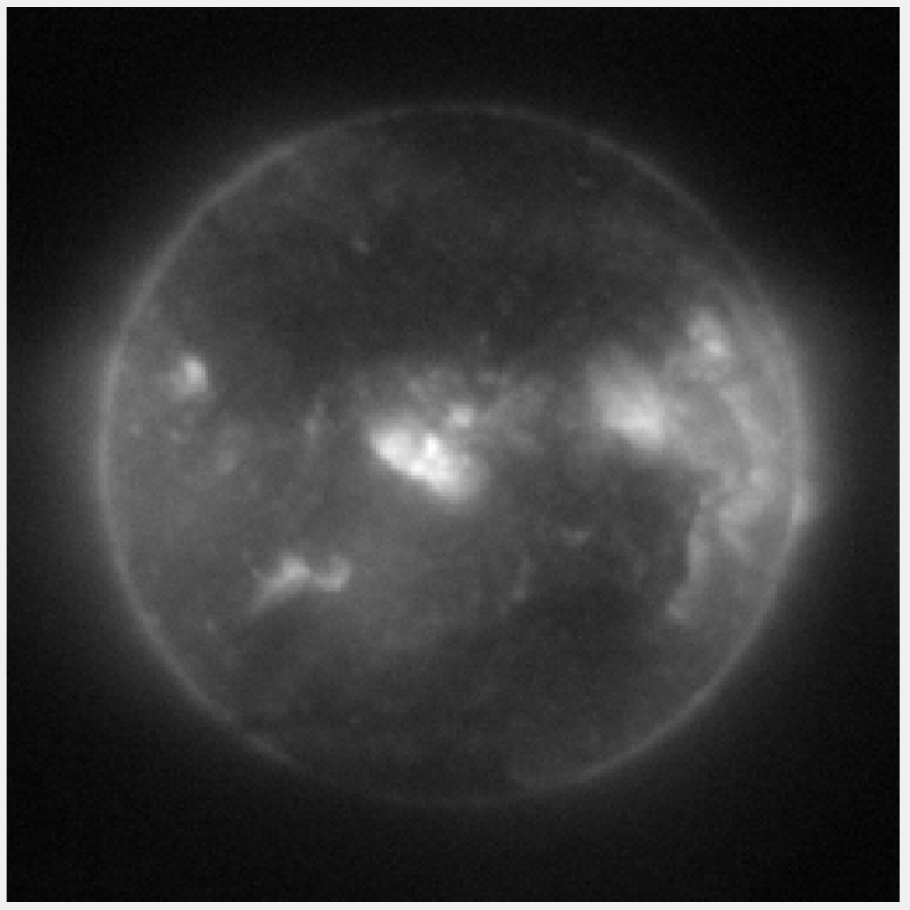}
    } 
\subfloat[]{\hspace{-0.0in}
    \includegraphics[scale=0.21]{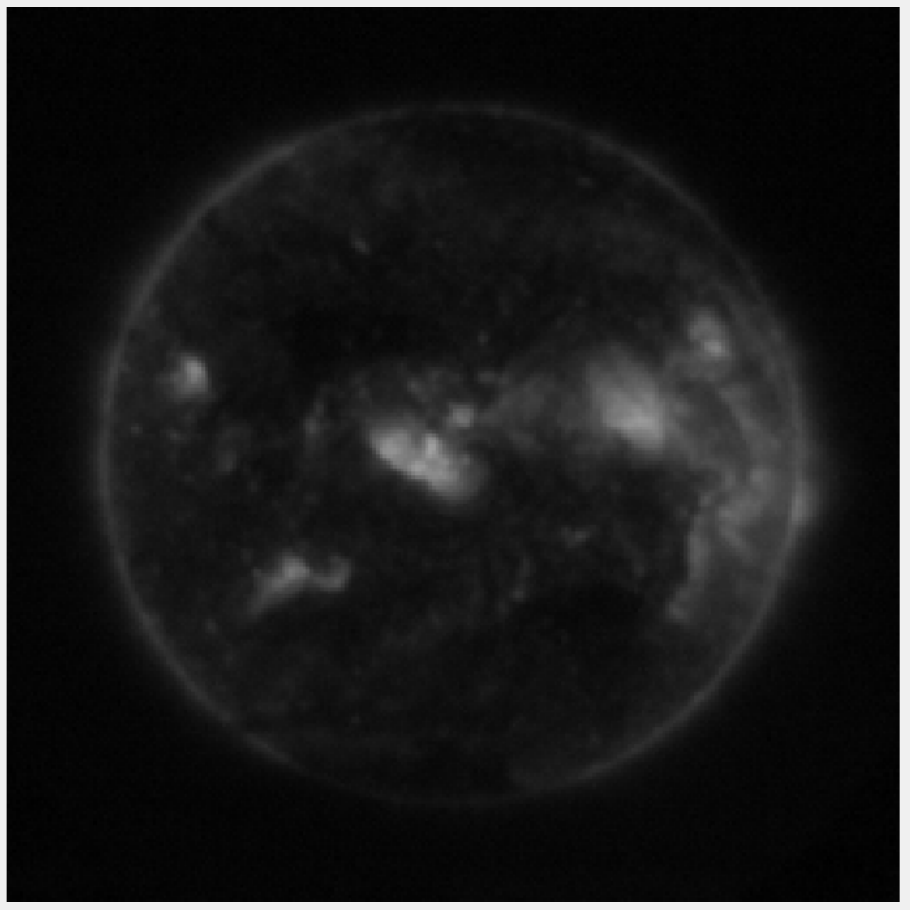}
    }
    \subfloat[]{\hspace{-0.0in}
    \includegraphics[scale=0.21]{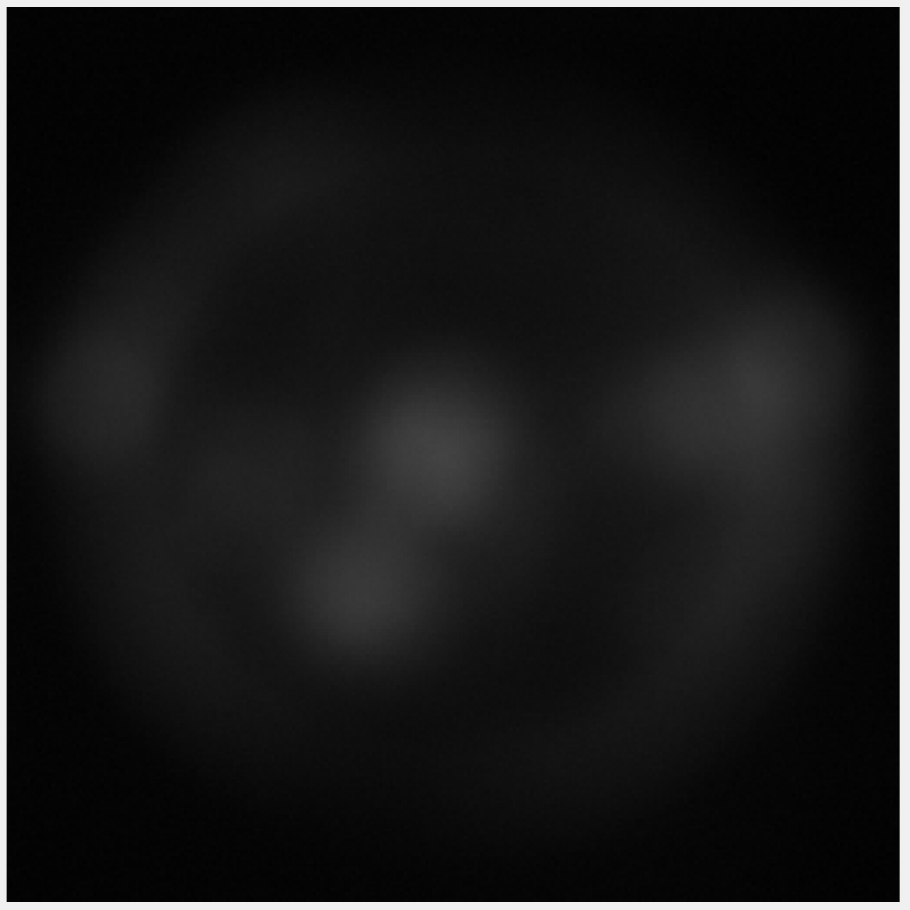}
    }
    \subfloat[]{\hspace{-0.0in}
    \includegraphics[scale=0.21]{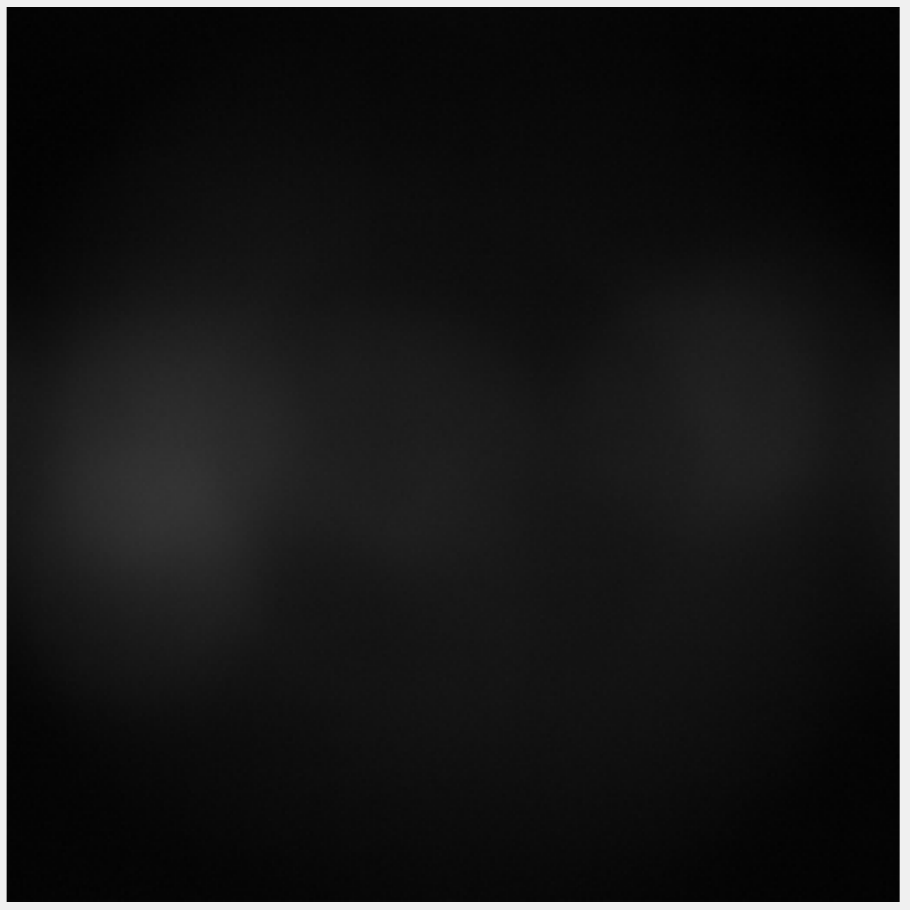}
    }
    \subfloat[]{\hspace{-0.0in}
    \includegraphics[scale=0.21]{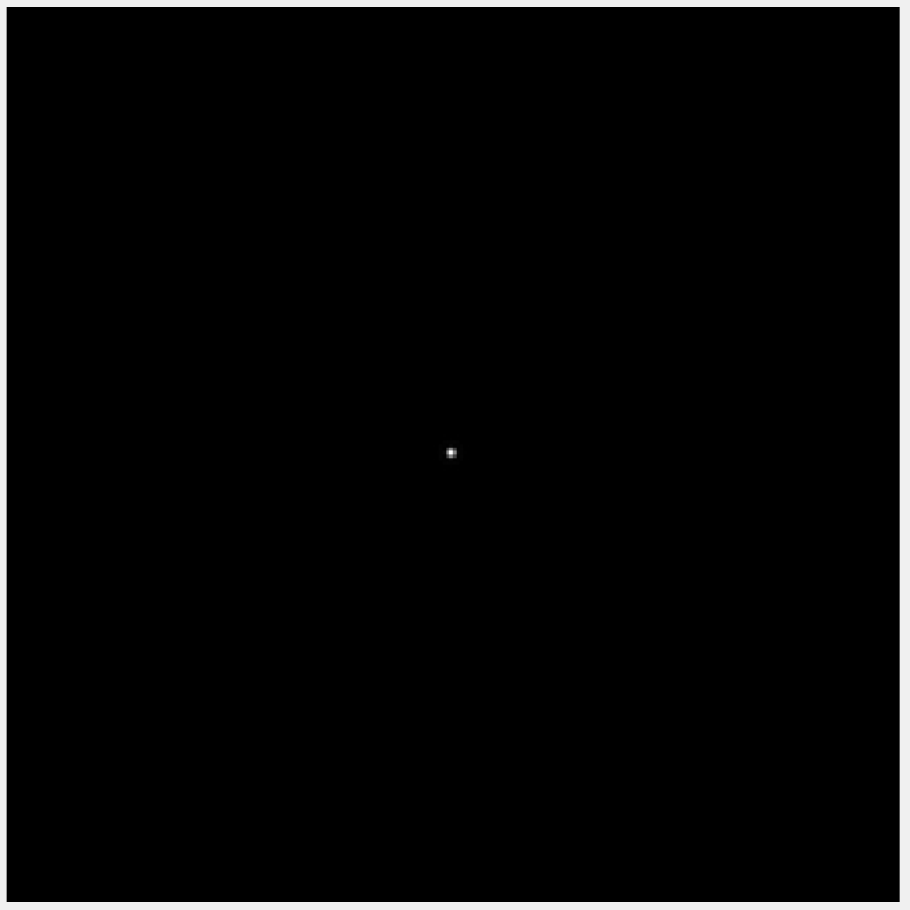}
    }\\
    \subfloat[]{\hspace{-0.0in} 
    \includegraphics[scale=0.21]{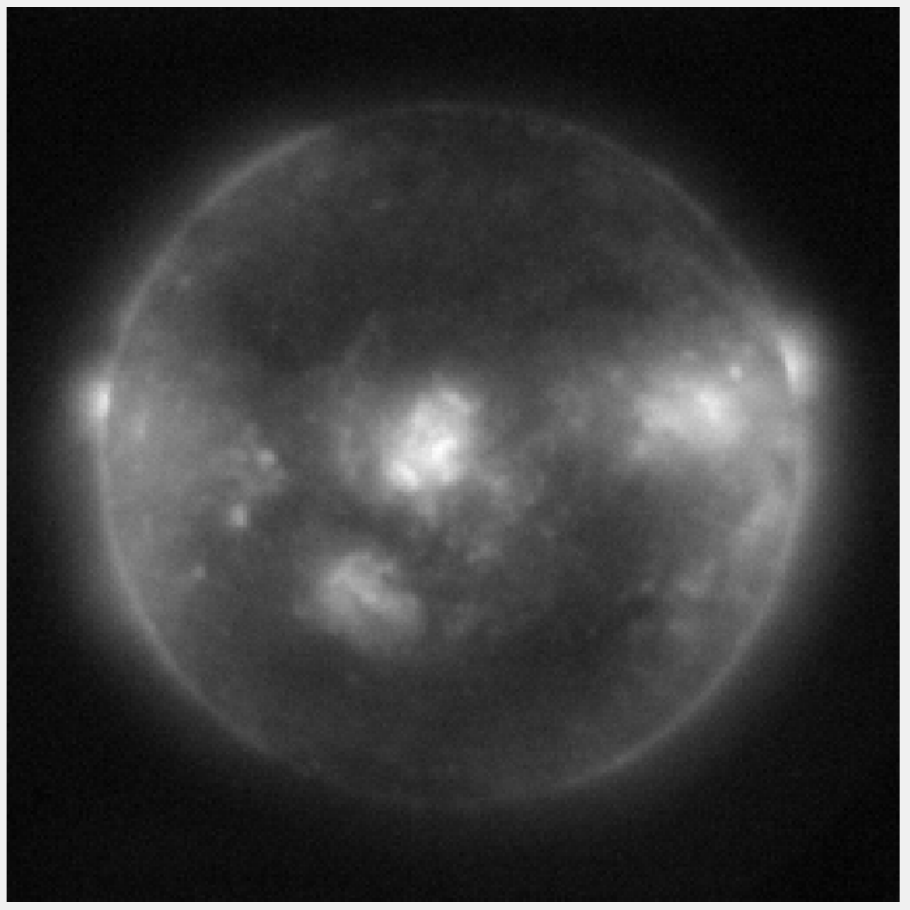}
    } 
\subfloat[]{\hspace{-0.0in}
    \includegraphics[scale=0.21]{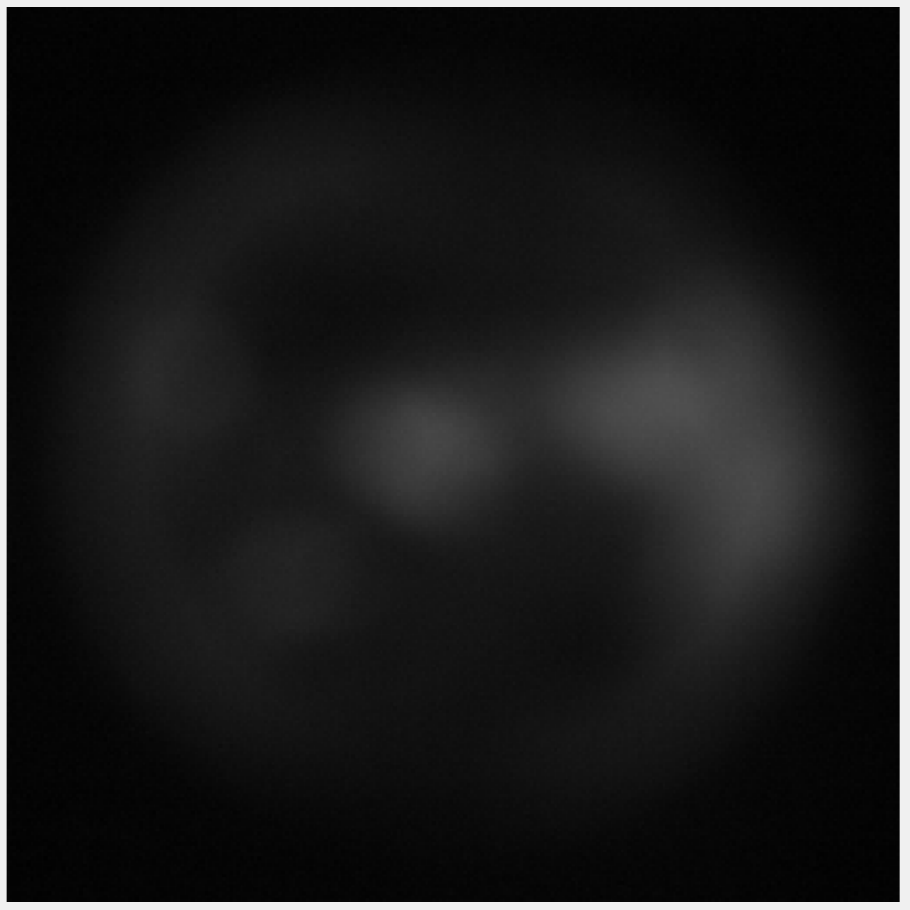}
    }
    \subfloat[]{\hspace{-0.0in}
    \includegraphics[scale=0.21]{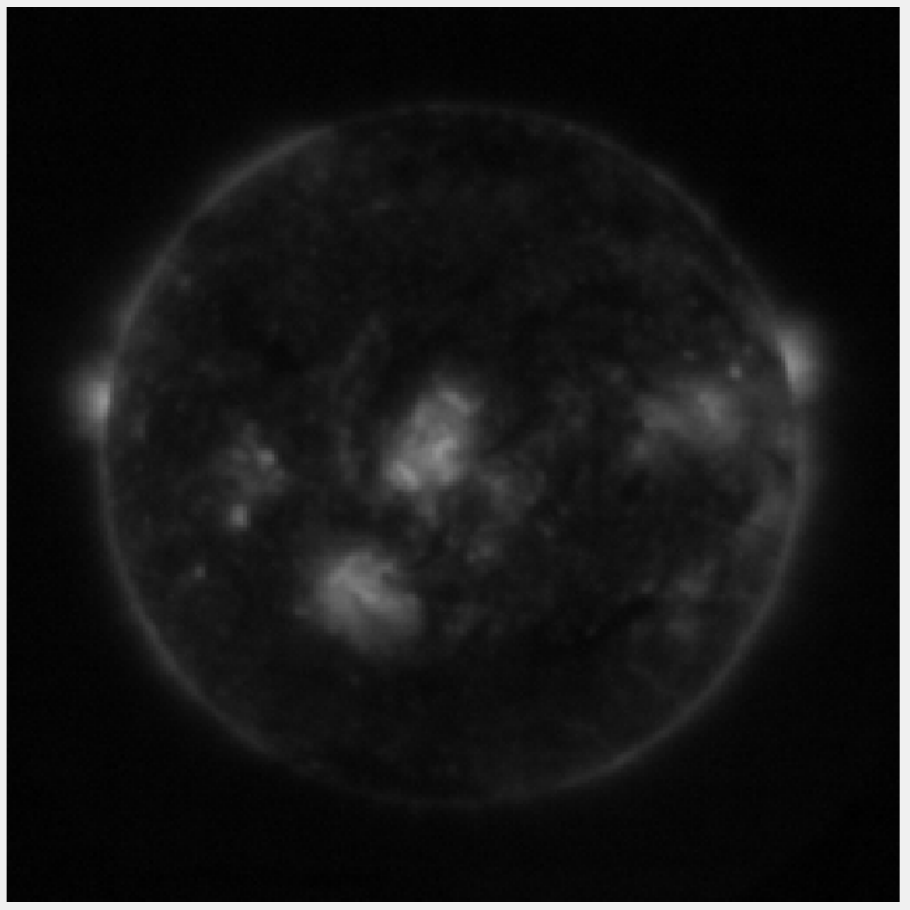}
    }
    \subfloat[]{\hspace{-0.0in}
    \includegraphics[scale=0.21]{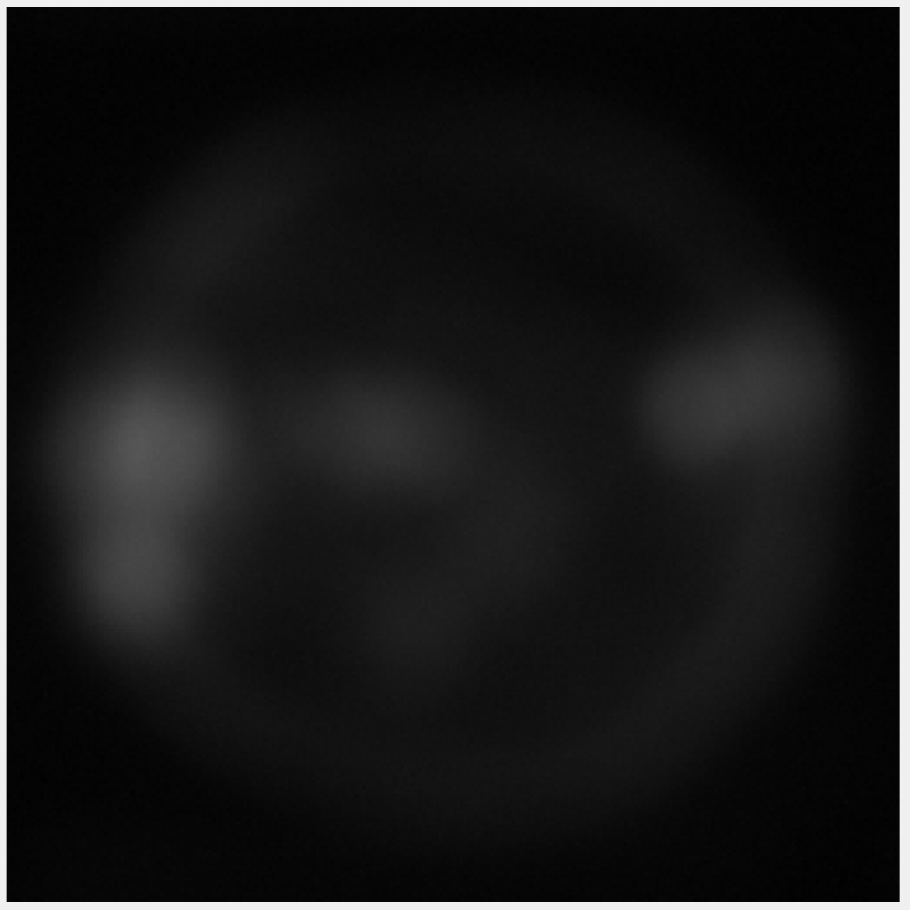}
    }
    \subfloat[]{\hspace{-0.0in}
    \includegraphics[scale=0.21]{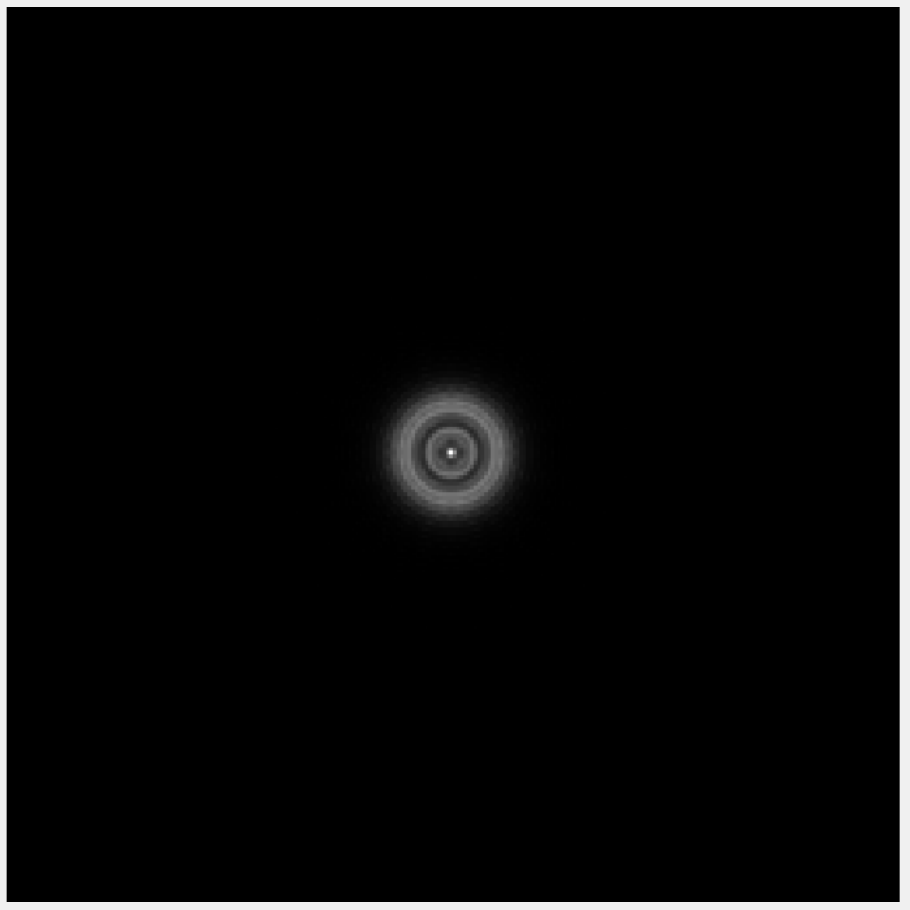}
    }\\
    \subfloat[]{\hspace{-0.0in} 
    \includegraphics[scale=0.21]{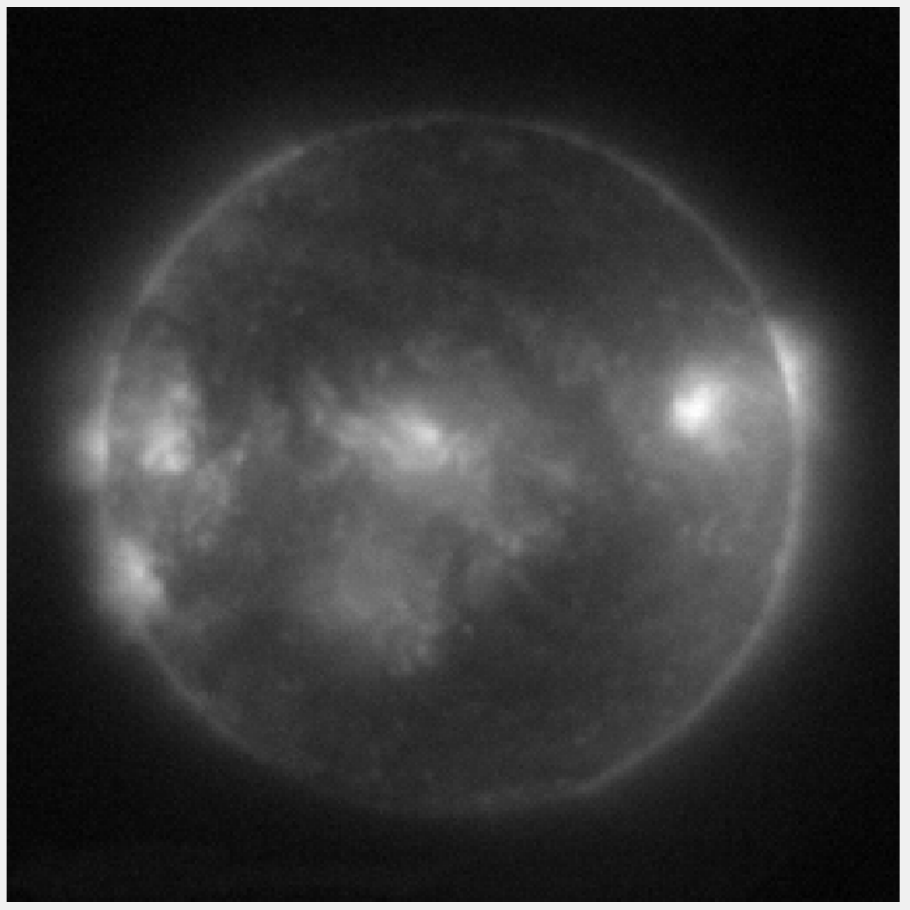}
    } 
\subfloat[]{\hspace{-0.0in}
    \includegraphics[scale=0.21]{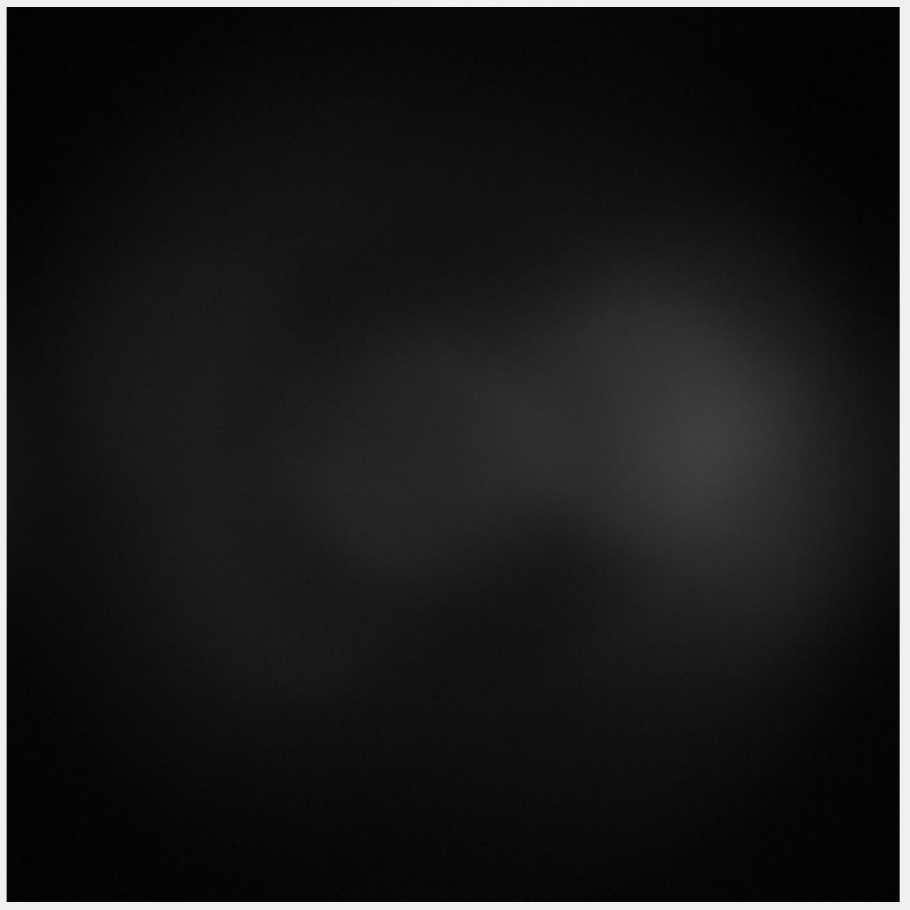}
    }
    \subfloat[]{\hspace{-0.0in}
    \includegraphics[scale=0.21]{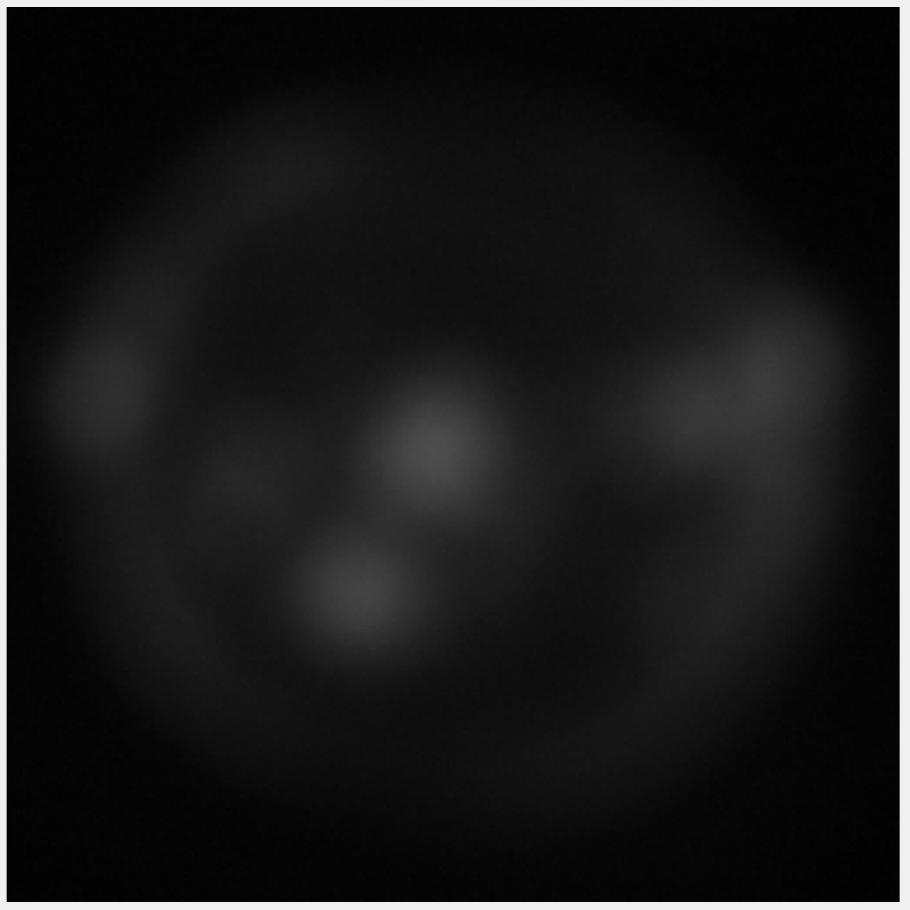}
    }
    \subfloat[]{\hspace{-0.0in}
     \includegraphics[scale=0.21]{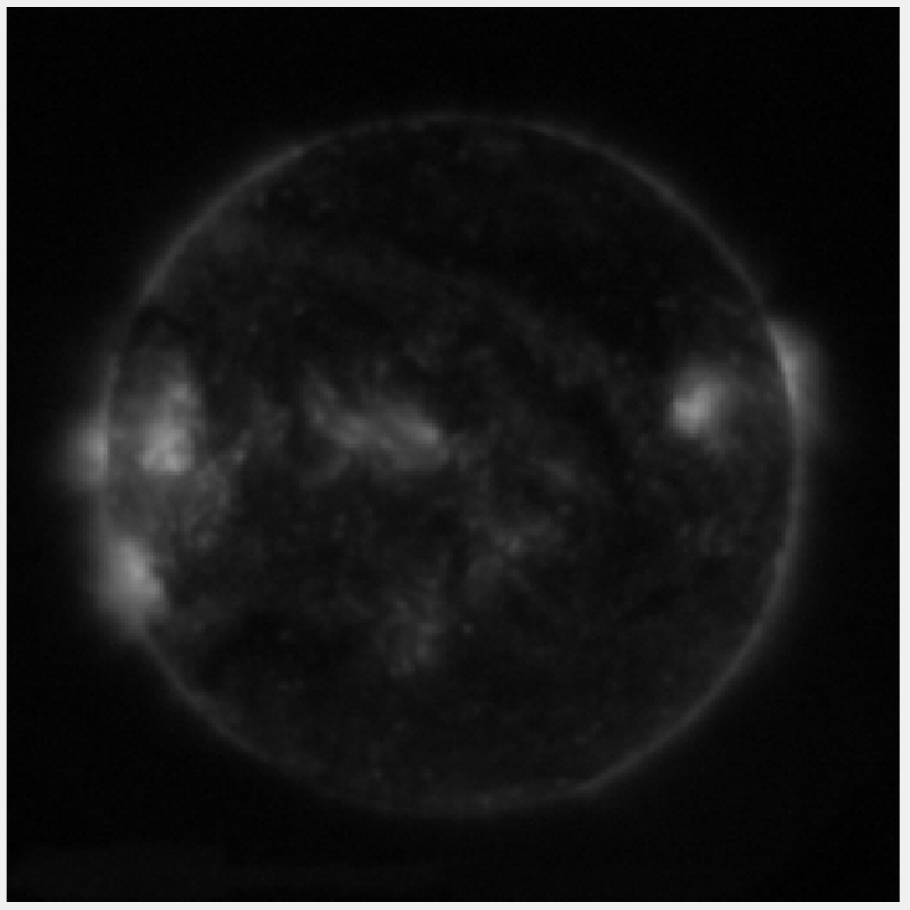}
    }
    \subfloat[]{\hspace{-0.0in}
    \includegraphics[scale=0.21]{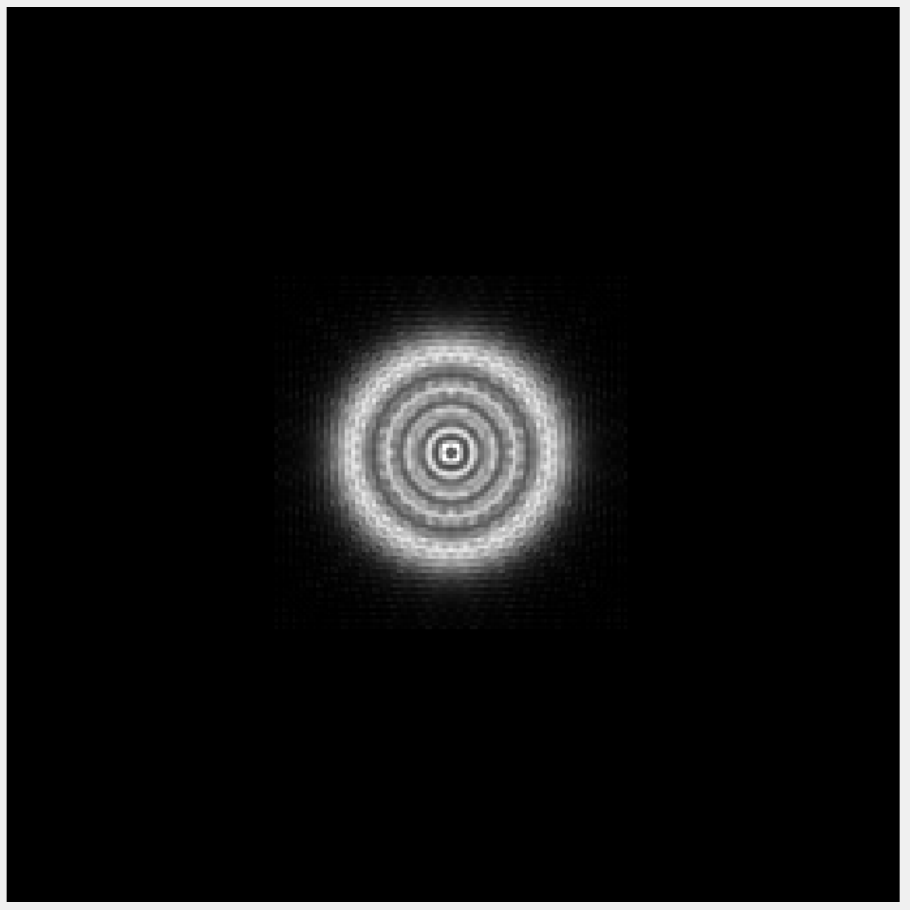}
    }
    \caption{Measured intensities for SNR=$25$ dB at the first focal plane (a), at the second focal plane (f), and at the third focal plane (k); 
    the contribution of the 1st, 2nd, and 3rd sources to the measurement at the first focal plane (b)-(d), at the second focal plane (g)-(i), at the third focal plane (l)-(n);
    sampled and zoomed PSFs of the system at the first focal plane for the 1st, 2nd, and 3rd sources, (e), (j), and (o), respectively. 
     }  
\label{solarImageMeasurements}
\end{center} 
\end{figure*}

Using the forward model in Eq.~\eqref{noisyModel}, we first simulate the measurements $\yb$ at the signal-to-noise ratio (SNR) of $25$ dB (where SNR is defined as $10 \log_{10} \frac{\sigma_t^2}{\sigma_w^2}$). 
Figure~\ref{solarImageMeasurements} illustrates the simulated measurements at the three focal planes together with the contribution from each spectral source. To also demonstrate the wavelength-dependent behavior of the system, the acting PSFs of the diffractive lens at the three 
source wavelengths are illustrated in Fig.~\ref{solarImageMeasurements}(e), (j), (o) when the measurement is taken at the 1st focal plane. These PSFs illustrate that each spectral source is exposed to a different amount of blur. As clearly seen from the shown contributions and PSFs, the first source is focused and the other two sources are defocused when the measurement is taken at the 1st focal plane. We have similar behavior for the measurements at the 2nd and 3rd focal planes. Hence the measurements involve not only the superposition of all spectral images but also substantial and varying amount of blur and noise.

We obtain the reconstructions using the algorithm in Table~\ref{tab:pssi_ps} with isotropic TV prior and the learned reconstruction method with deep prior. For TV prior, one reconstruction takes a couple of minutes while for learned reconstruction it takes a couple of seconds on a computer with 32 GB of RAM, Intel i$9-9900$K $3.60$ GHz CPU, and Nvidia GeForce RTX $2080$ Ti GPU. 

For the model-based learned reconstruction method, we train our model using solar EUV images from the NASA SDO database~\cite{lemen2011}. We use SDO images from the $1^{st}$, $5^{th}$, $10^{th}$, $15^{th}$, $25^{th}$, and $30^{th}$ days of every month for years between $2012$ and $2020$, 
to have training data diversity. After removing the images with artifacts (such as out-of-focus), 
we randomly pick $6$ images from each day. 
These images are then converted to grayscale, resized to $256 \times 256$, and normalized between $[0,1]$. Finally, we combine randomly chosen three images to form $725$ training datasets (for $P=3$ case). In a similar way, we have generated separate $105$ test datasets. 

We train four different models for signal-to-noise ratios (SNR) of $15$ dB, $20$ dB, $25$ dB, and $30$ dB. For this, we simulate the measurements $\yb$ at these SNRs 
using the forward model in Eq.~\eqref{noisyModel}. We then train our models by minimizing the $l_2$ loss between the ground truth and reconstructed images by using Adam solver ~\cite{adam2015} . We initialize the model parameters with Xavier initialization. We choose $N=5$, $L=20$, and the learning rate as $0.001$ after some empirical efforts. Codes are implemented with Tensorflow on Nvidia GeForce RTX $2080$ Ti GPU. Training takes approximately $30$ hours.  

To illustrate the performance of the developed imaging modality and the two reconstruction methods, we use the randomly constructed 105 test datasets described above. For each dataset, we perform $5$ Monte Carlo runs and compute the average PSNR and SSIM values between the diffraction-limited ground truth images and the reconstructed spectral images using TV and learned reconstruction methods. Note that here the comparison is performed with the diffraction-limited versions of the original images to check the imaging system's capability for achieving diffraction-limited spatial resolution.
Table~\ref{Table:learned-tv-recon} presents the resulting PSNR and SSIM values. The first row demonstrates the average performance over the entire test dataset as SNR changes from $15$ dB to $30$ dB. To illustrate the performance at different levels of difficulty, 5 spectral datasets are also selected from this large test dataset and their PSNR/SSIM values are also separately given. Numbered as 1 to 5, these selected datasets have increasing levels of abrupt changes along both spatial and spectral dimensions, corresponding to observations with different sizes of active and quiet regions on the sun.

As seen from Table~\ref{Table:learned-tv-recon}, the learned reconstruction method provides the best performance in this large dataset with the average PSNR changing from $29.7$ dB to $35.8$ dB and SSIM changing from $0.79$ to $0.92$ as SNR increases from $15$ to $30$ dB.
The obtained PSNR and SSIM values between the reconstructions and diffraction-limited original images illustrate the imaging system's diffraction-limited spatial resolution capability over a wide range of SNRs.
As expected, the learned reconstruction method provides significantly better performance than the TV reconstruction on the average both in terms of PSNR and SSIM. 
This illustrates the better generalization capability of the learned reconstruction method for a particular application. Moreover, as expected, the performance of the TV reconstruction decreases with the increased structure in the images (i.e. from dataset 1 to dataset 5). The performance improvement achieved with the learned reconstruction method is more significant for the datasets with significant structure (for example, for datasets 4 and 5). Note that for the case that all three spectral images are mostly smooth and have few edges (such as dataset 1) TV reconstruction performs better. This may also result from using limited number of such examples in the training phase of the learning-based method.

\begin{figure*}[tbh!]
\begin{center} 
\subfloat[]{
    \includegraphics[scale=0.29]{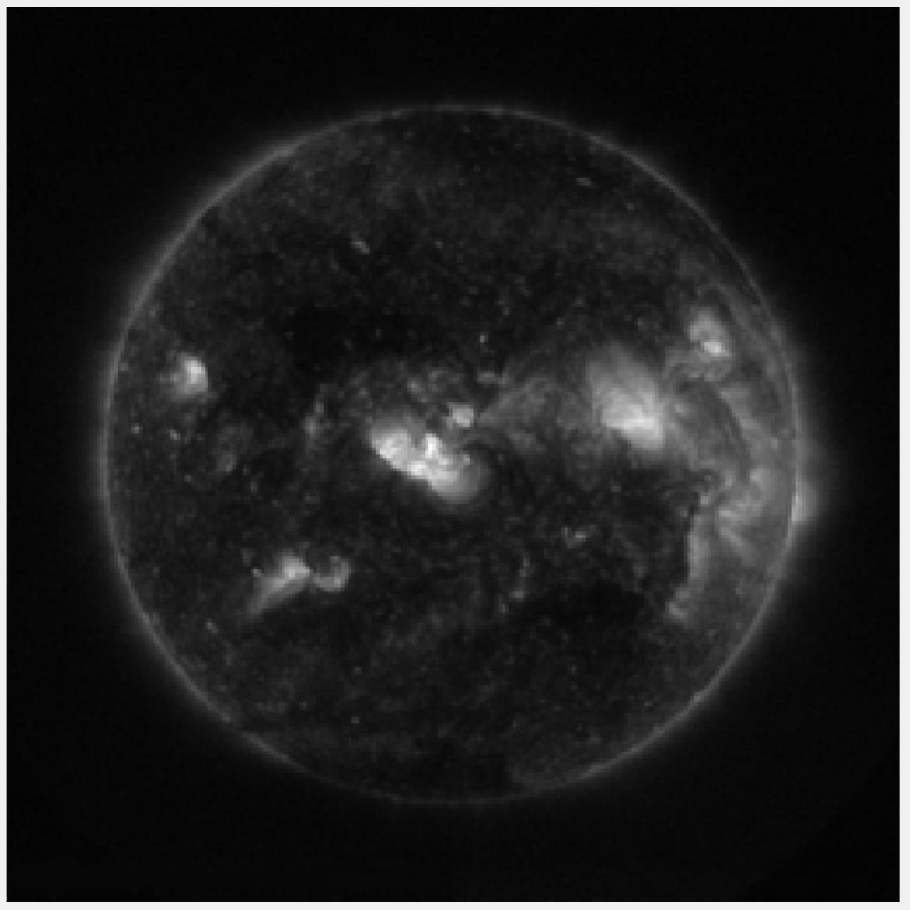}
    } 
\subfloat[]{
    \includegraphics[scale=0.29]{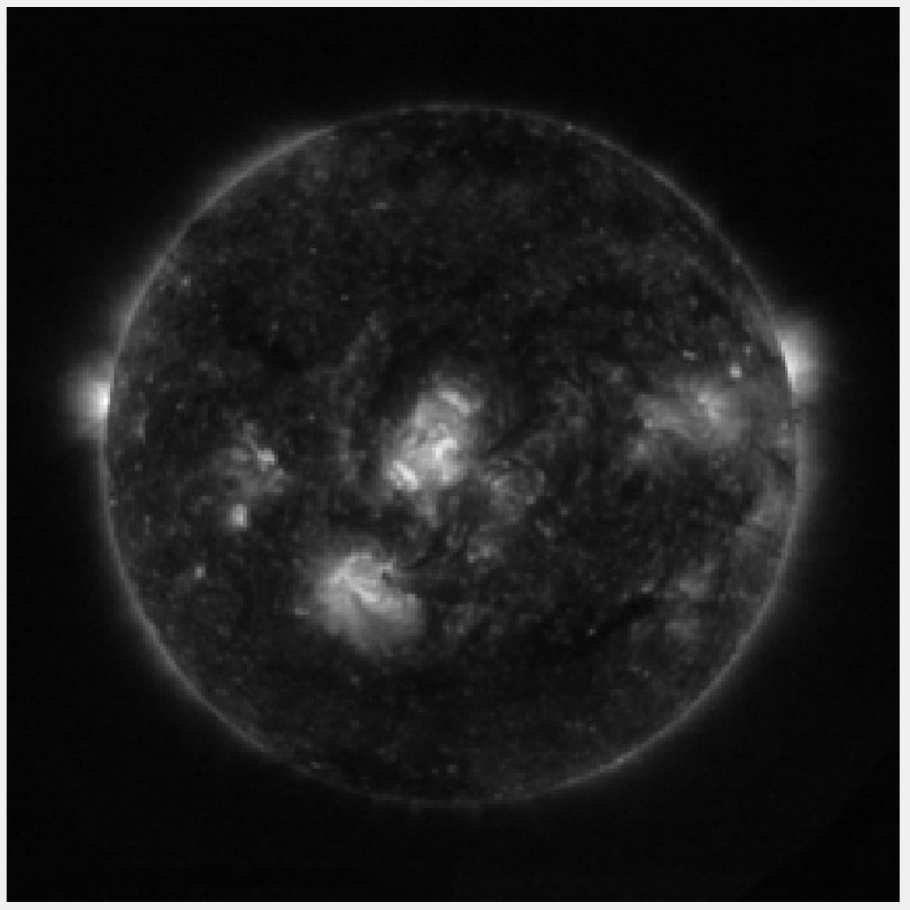}
    }
    \subfloat[]{
    \includegraphics[scale=0.29]{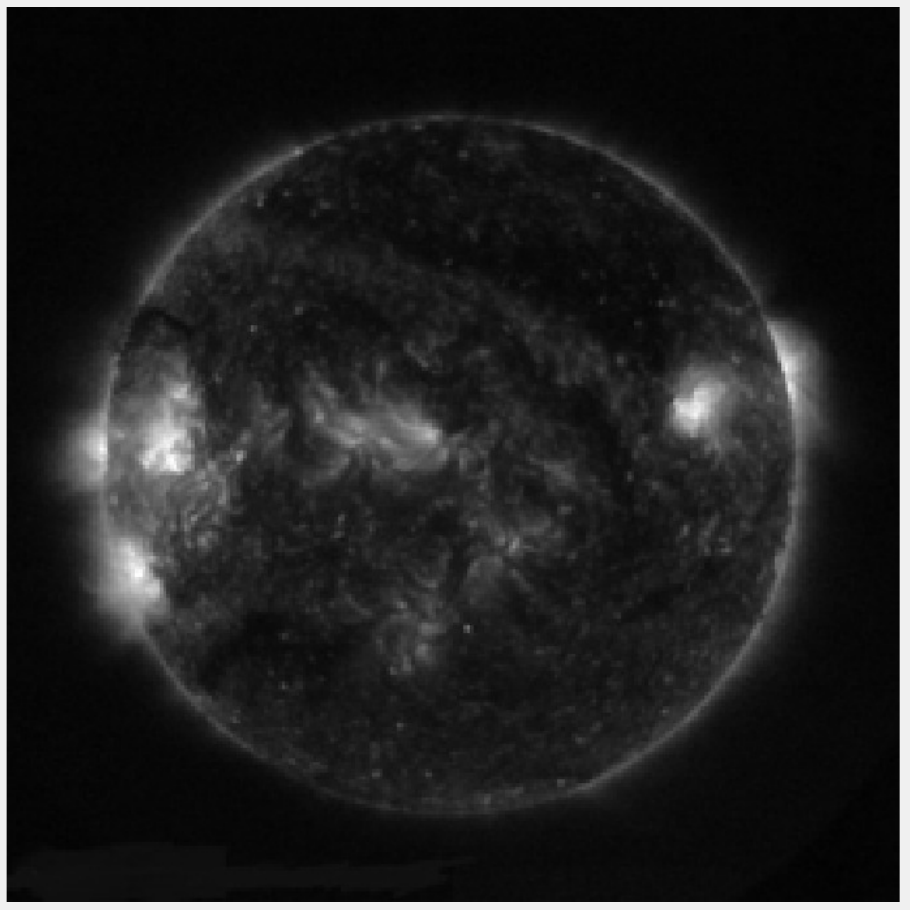}
    }
    \subfloat[]{
    \includegraphics[scale=0.29]{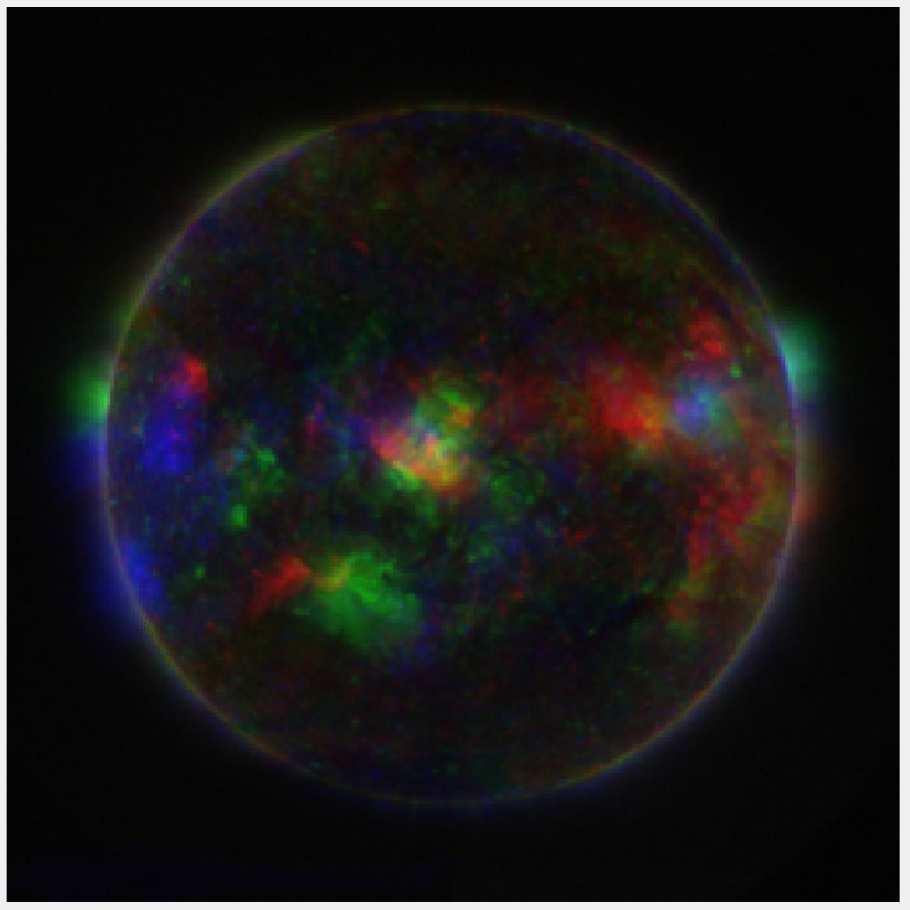}
    }
    \\
    \subfloat[]{
    \includegraphics[scale=0.29]{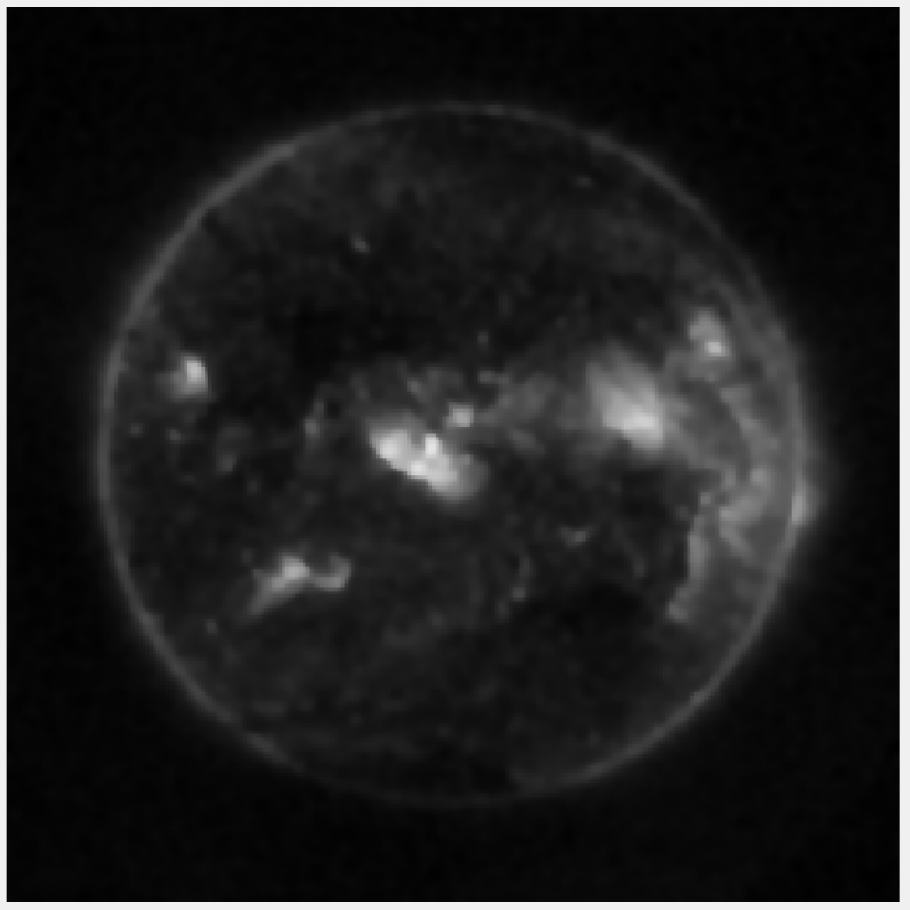}
    } 
\subfloat[]{
    \includegraphics[scale=0.29]{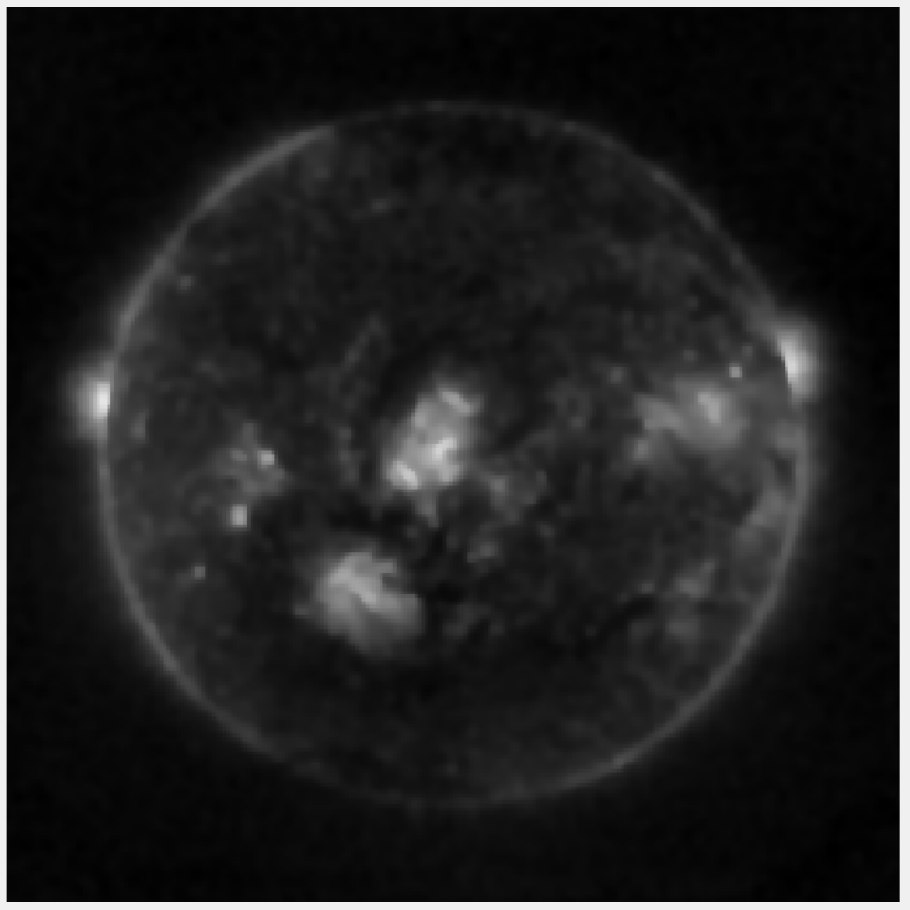}
     }
    \subfloat[]{
    \includegraphics[scale=0.29]{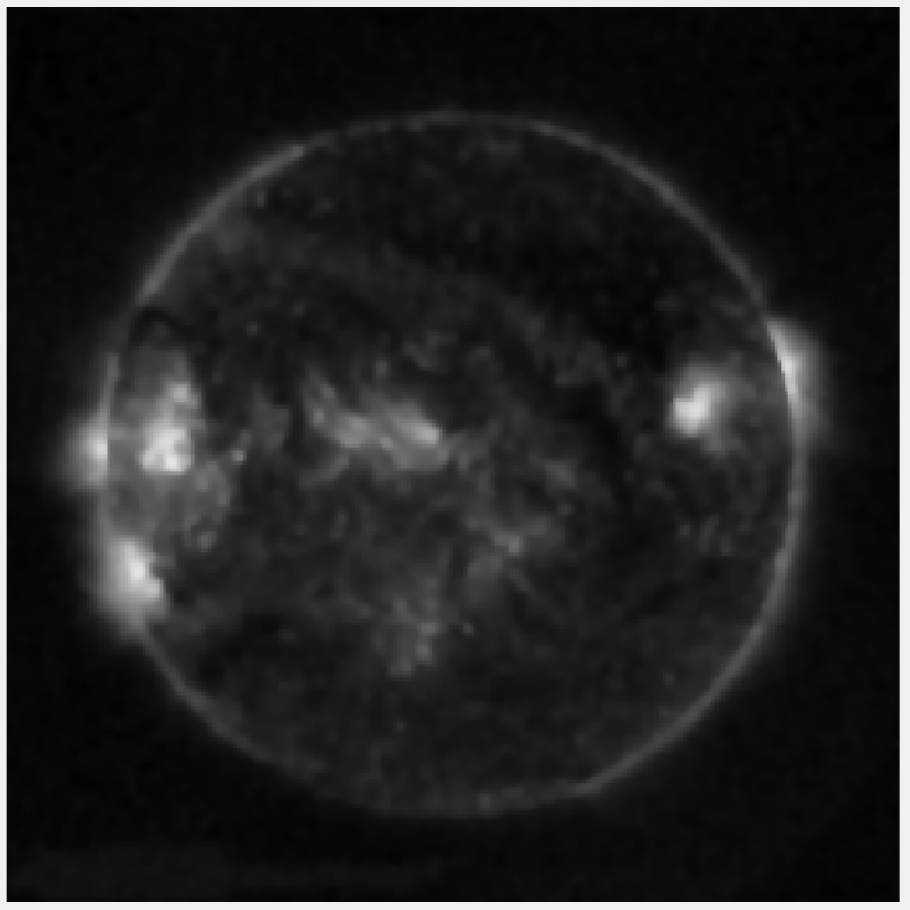}
    }
    \subfloat[]{
     \includegraphics[scale=0.29]{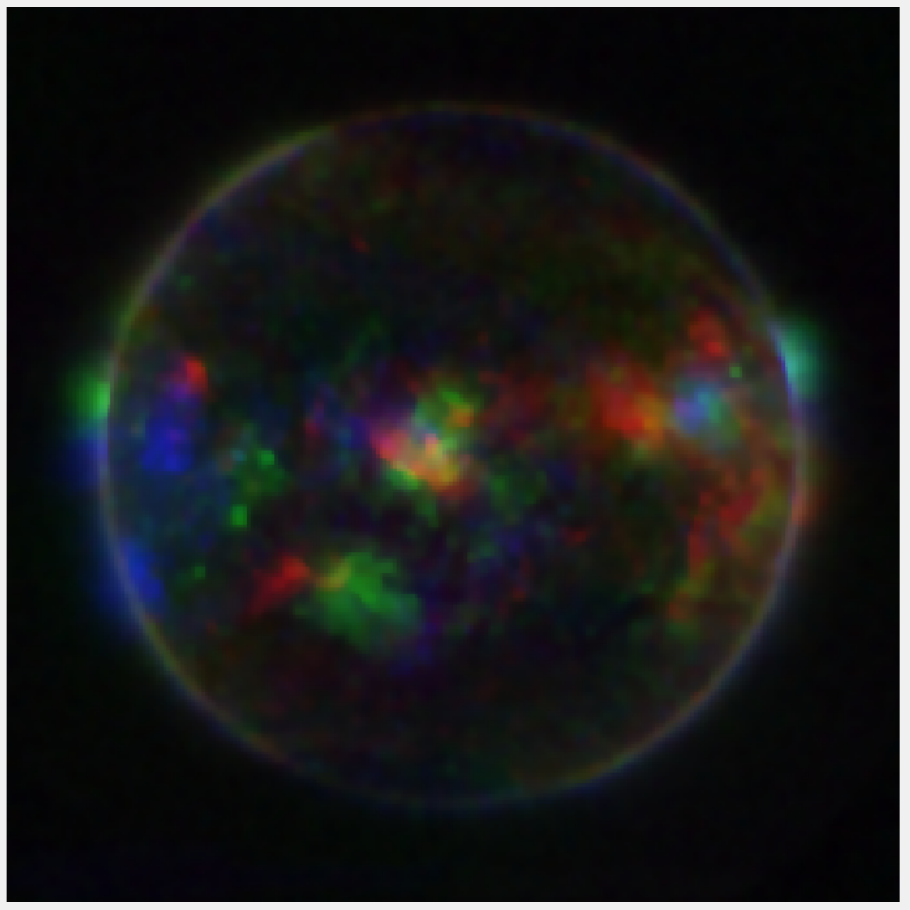}
    }
    \\
    \subfloat[]{
    \includegraphics[scale=0.29]{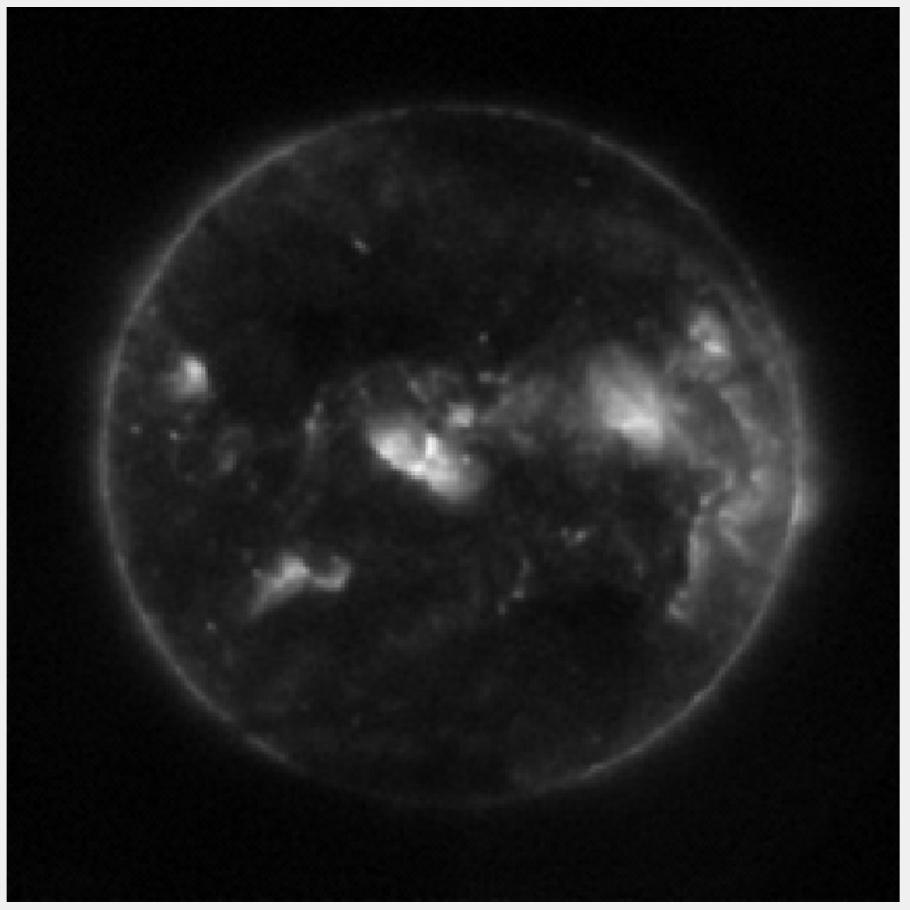}
    } 
\subfloat[]{
    \includegraphics[scale=0.29]{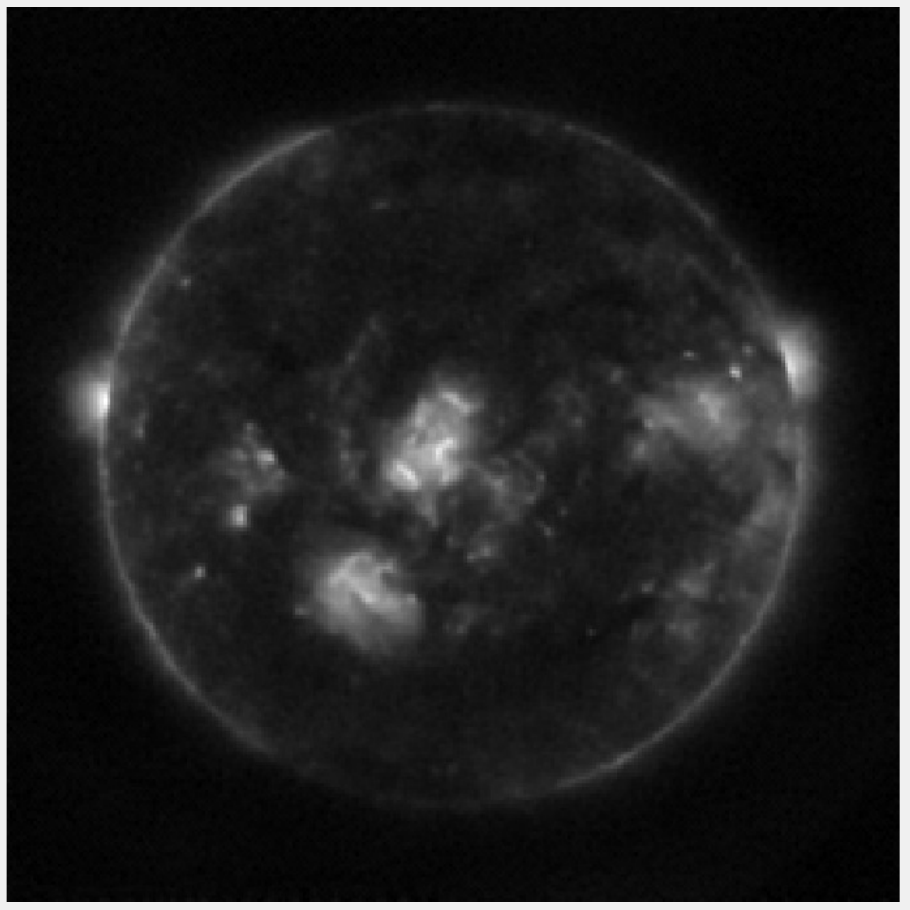}
    }
    \subfloat[]{
    \includegraphics[scale=0.29]{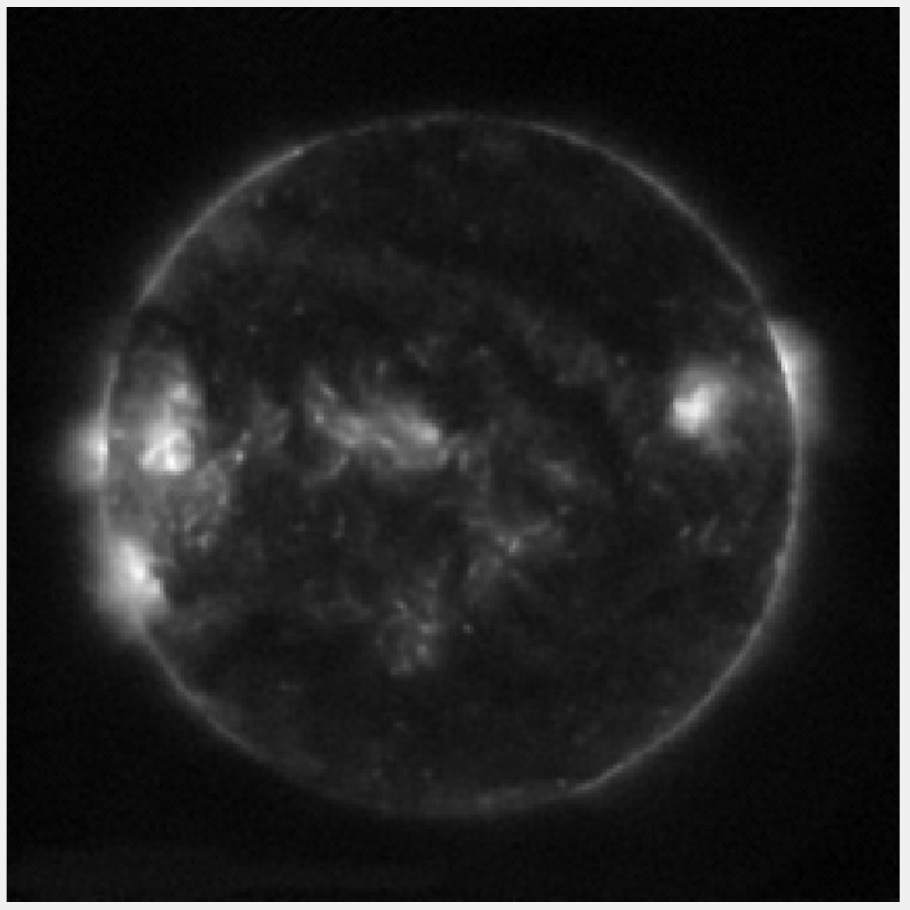}
    }
    \subfloat[]{
   \includegraphics[scale=0.29]{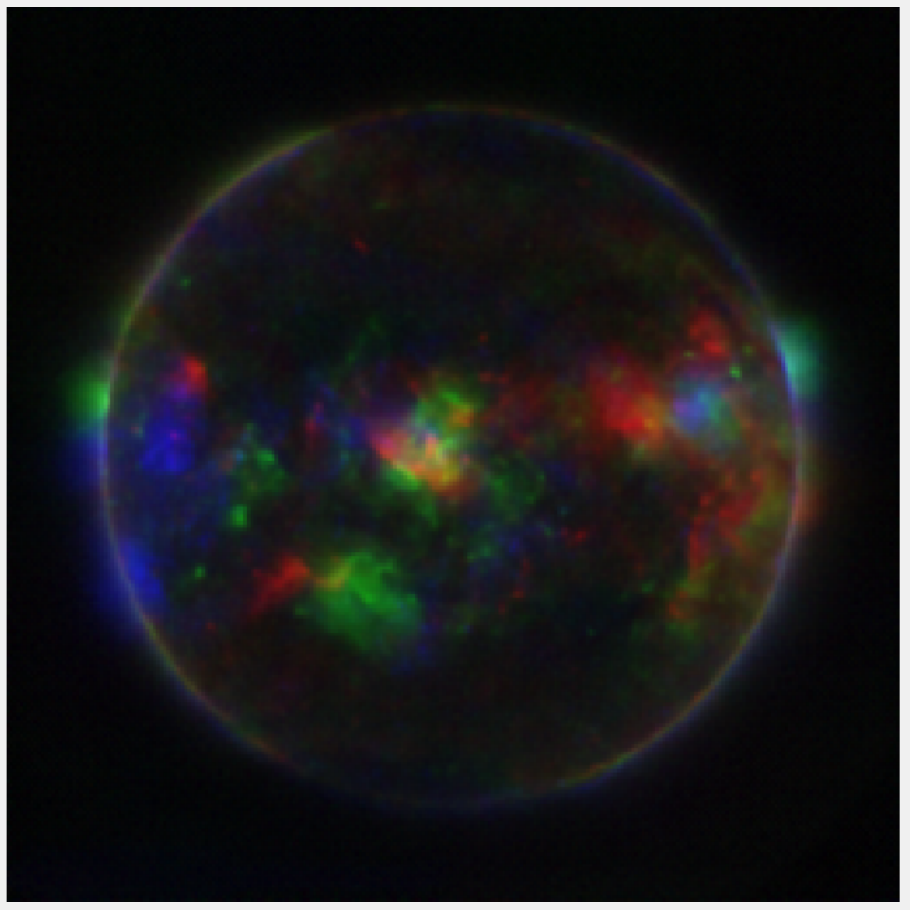}
    }

    \caption{Diffraction-limited original images (top), reconstructed spectral images with isotropic TV regularization (middle), and reconstructed spectral images with learned-reconstruction method (bottom) for dataset 2 at 25dB SNR. (Courtesy of NASA/SDO and the AIA, EVE, and HMI science teams.) For each case, false-color representations are also provided on the right.
     }  
\label{solarImageReconstructions_Solar}
\end{center} 
\end{figure*}

\begin{figure*}[tbh!]
\begin{center} 
\subfloat[]{
    \includegraphics[scale=0.29]{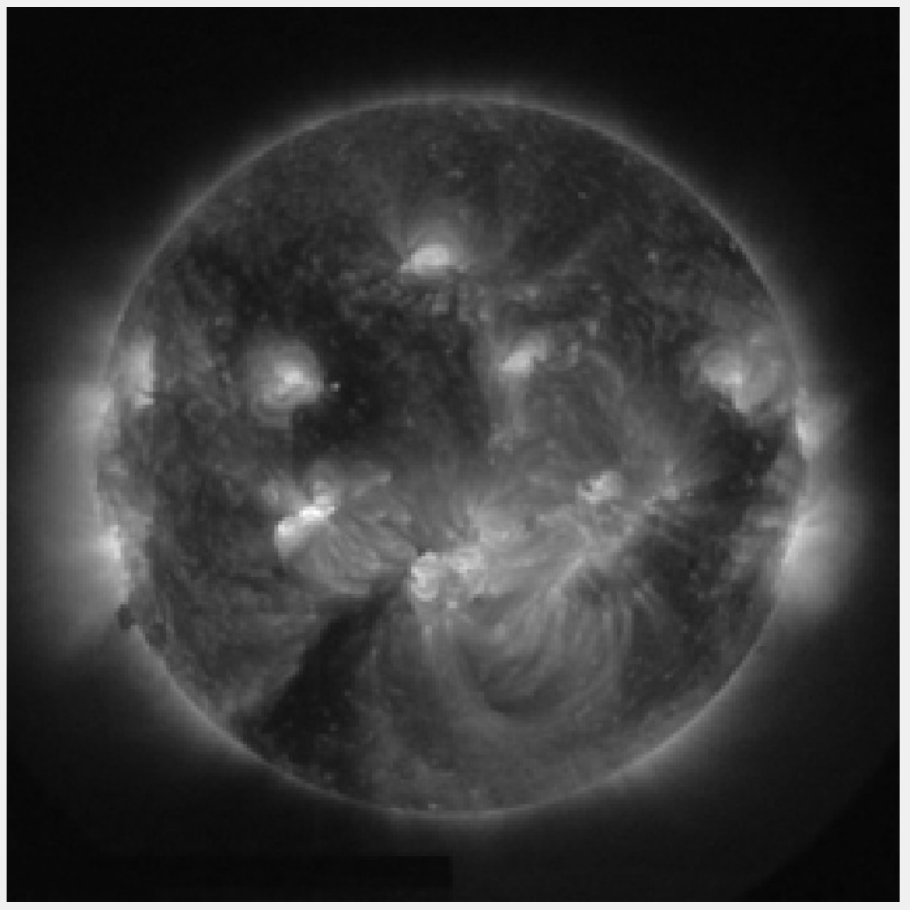}
    } 
\subfloat[]{
    \includegraphics[scale=0.29]{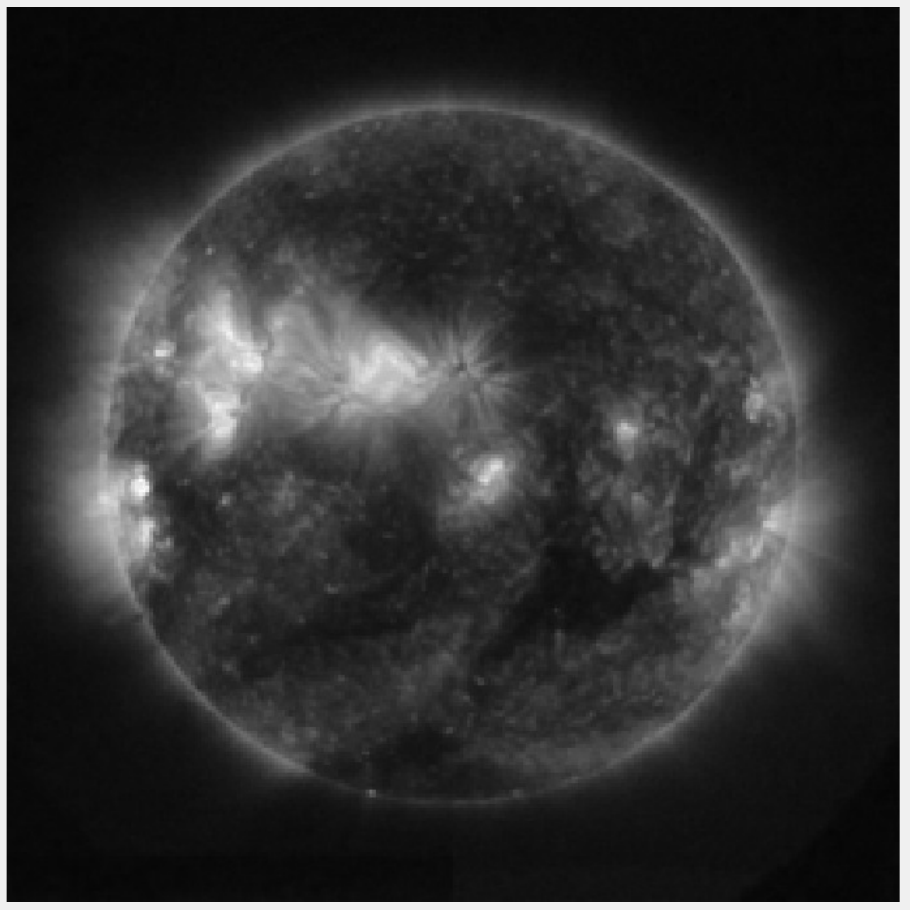}
    }
    \subfloat[]{
    \includegraphics[scale=0.29]{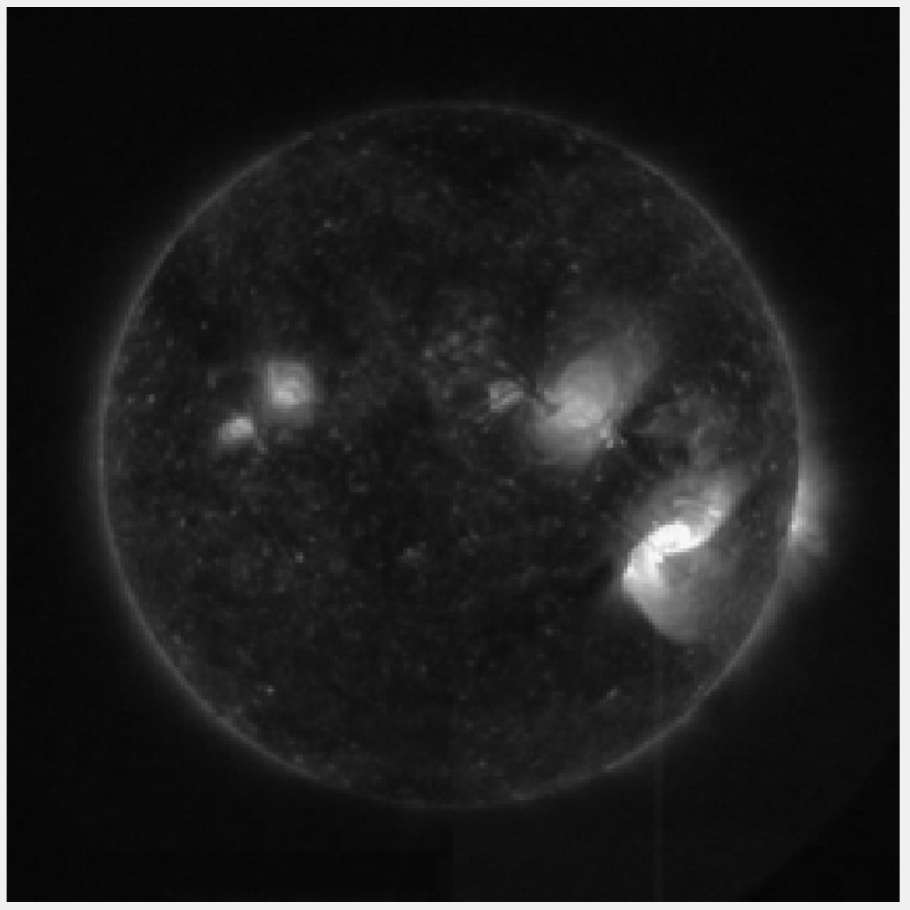}
    }
    \subfloat[]{
    \includegraphics[scale=0.29]{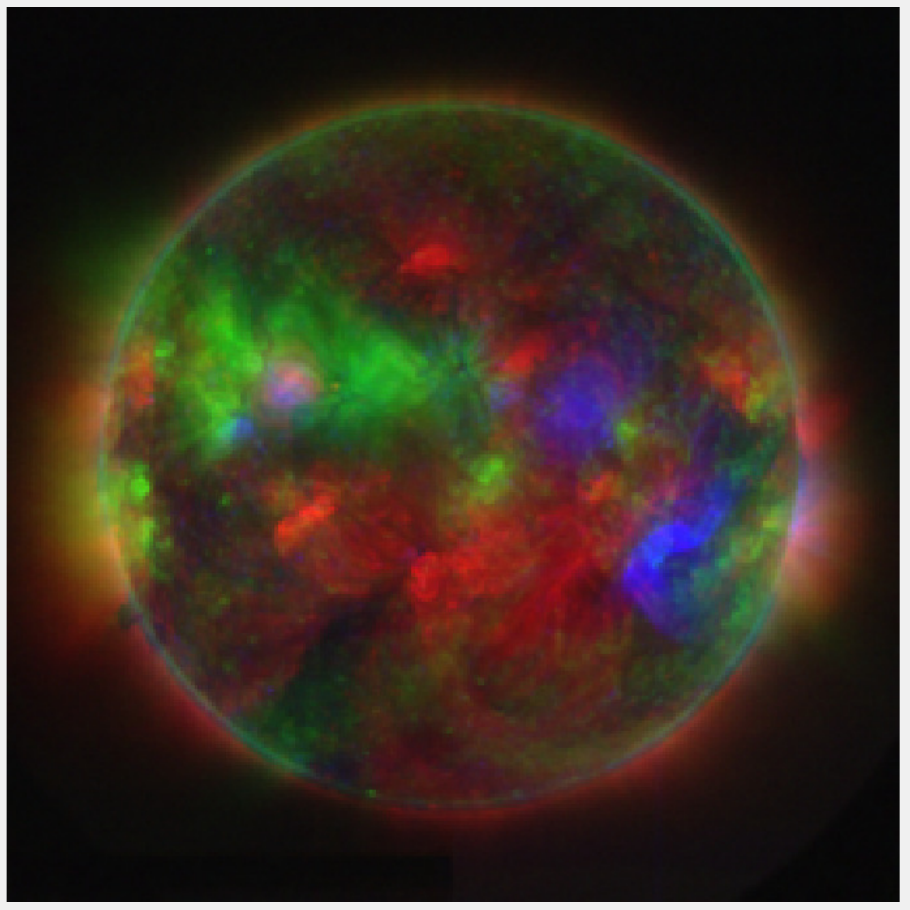}
    }
    \\
    \subfloat[]{
    \includegraphics[scale=0.29]{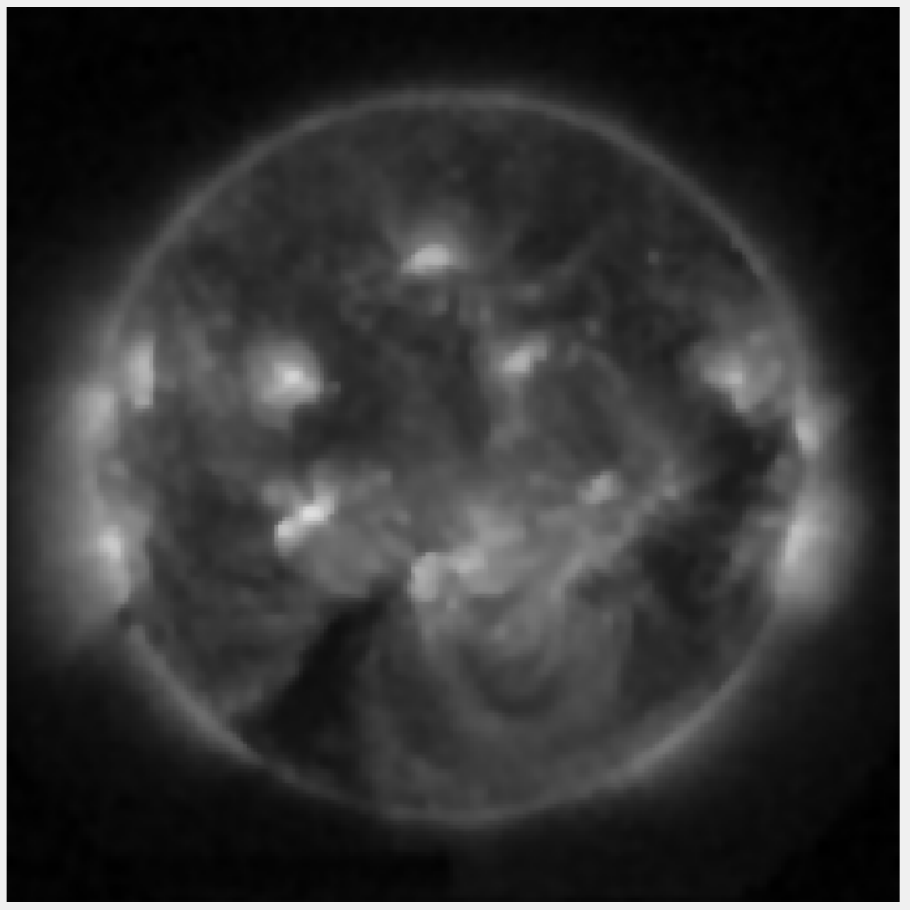}
    } 
\subfloat[]{
    \includegraphics[scale=0.29]{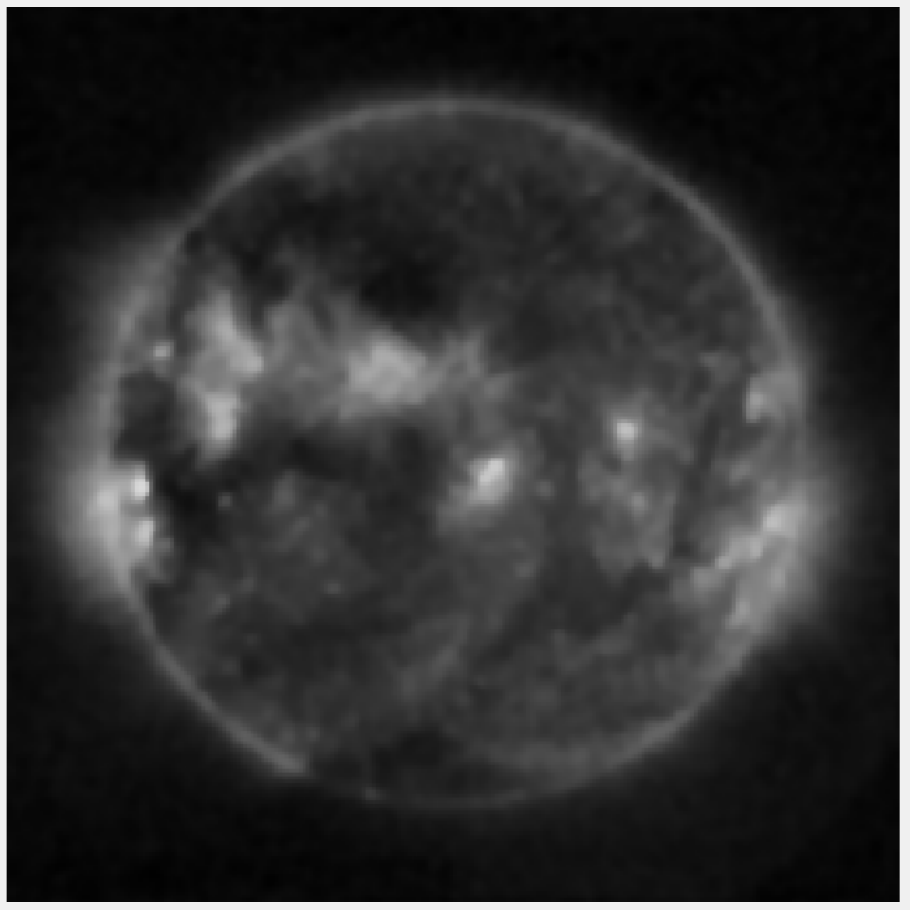}
    }
    \subfloat[]{
    \includegraphics[scale=0.29]{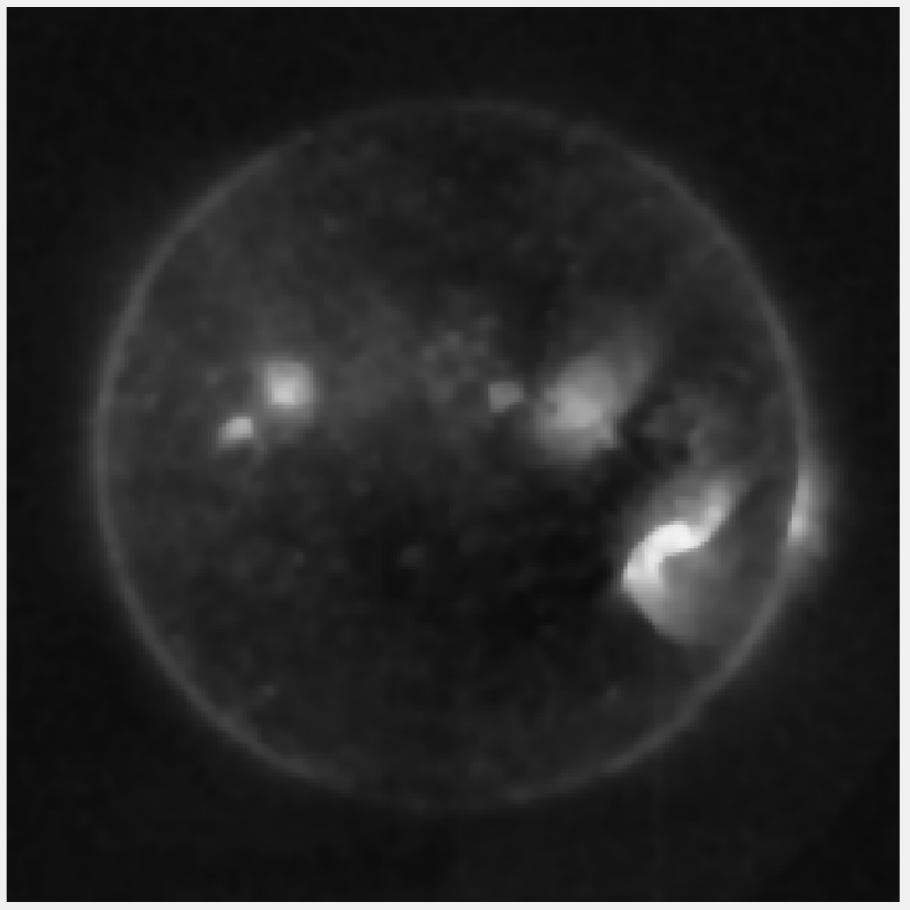}
    }
    \subfloat[]{
    \includegraphics[scale=0.29]{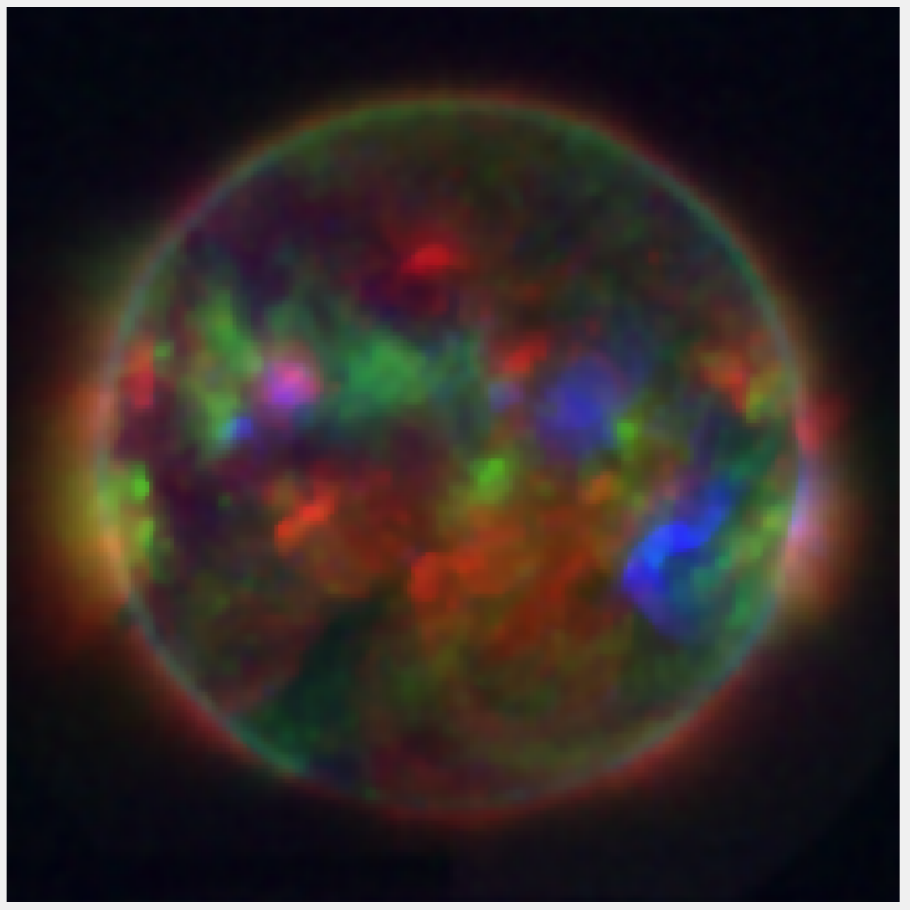}
    }
    \\
    \subfloat[]{
        \includegraphics[scale=0.29]{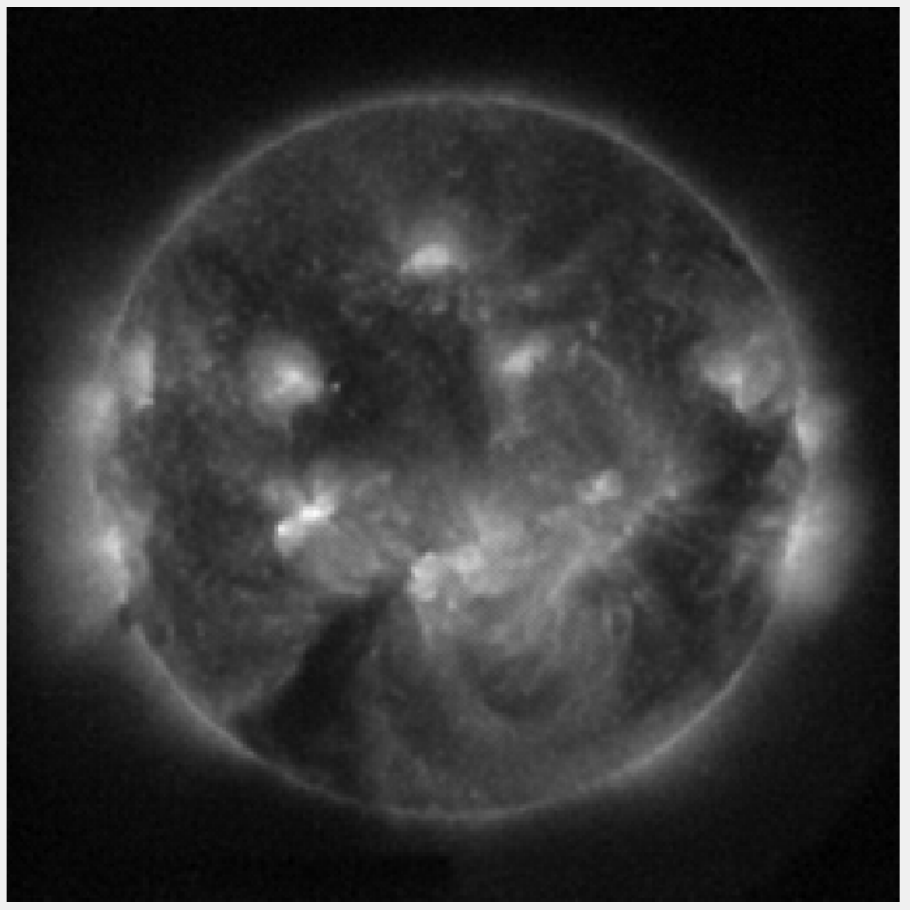}
    } 
\subfloat[]{
        \includegraphics[scale=0.29]{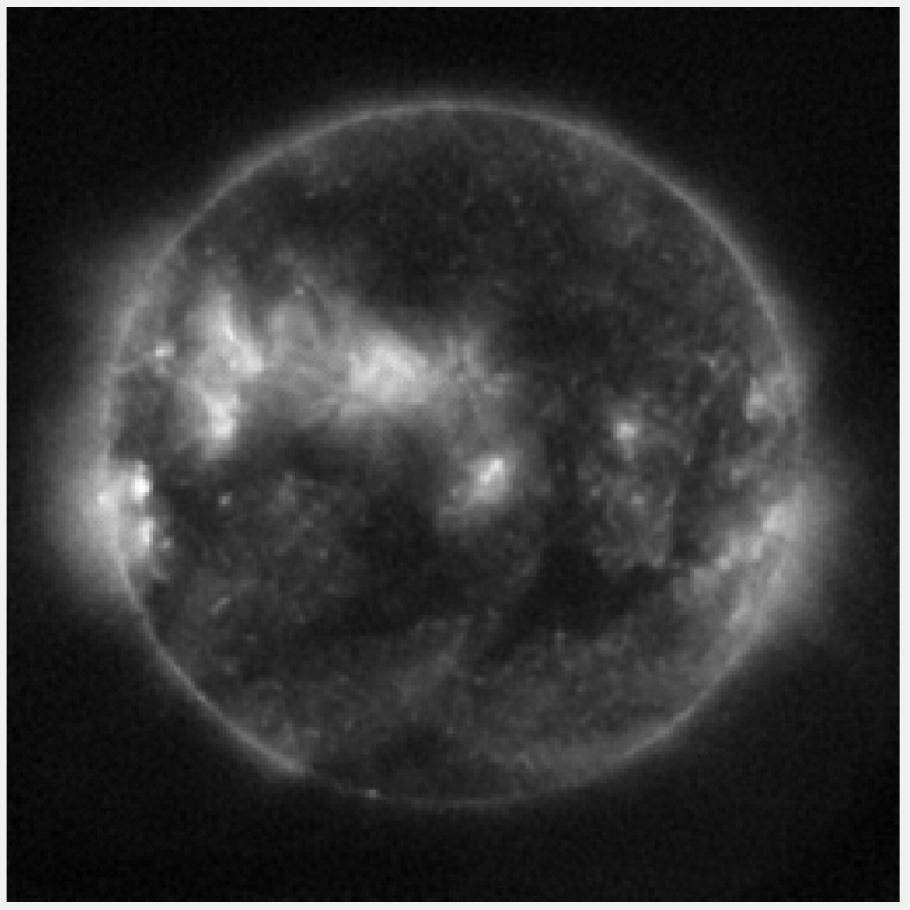}
    }
    \subfloat[]{
        \includegraphics[scale=0.29]{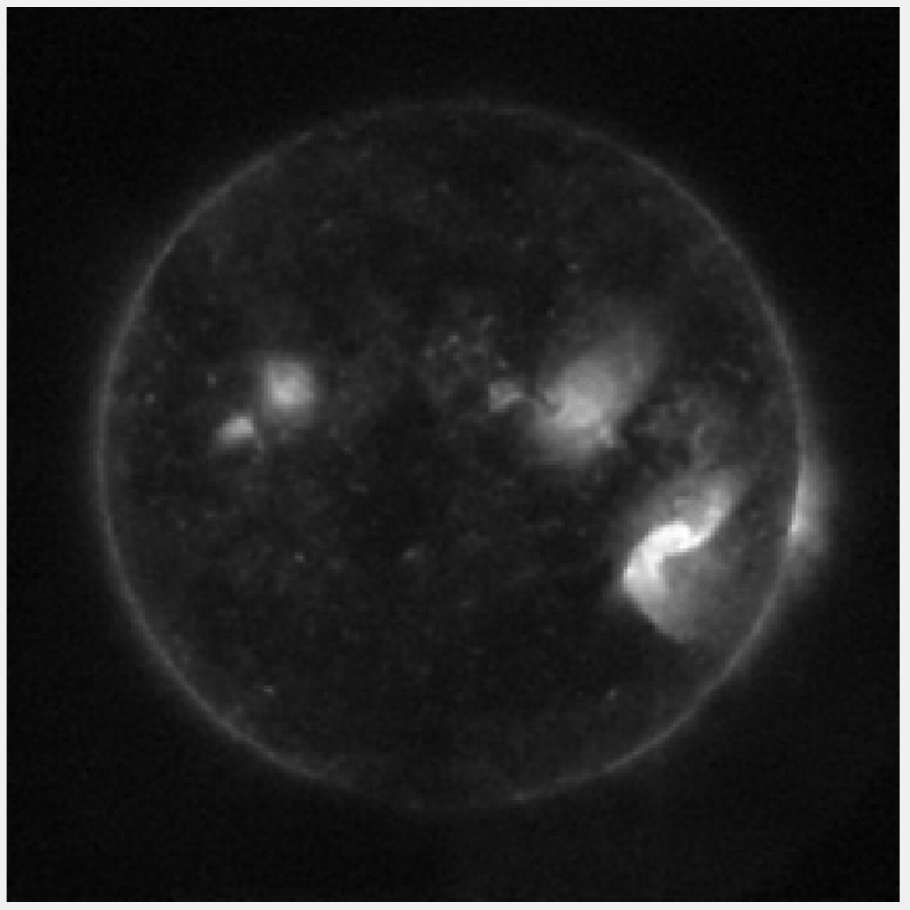}
    }
    \subfloat[]{
        \includegraphics[scale=0.29]{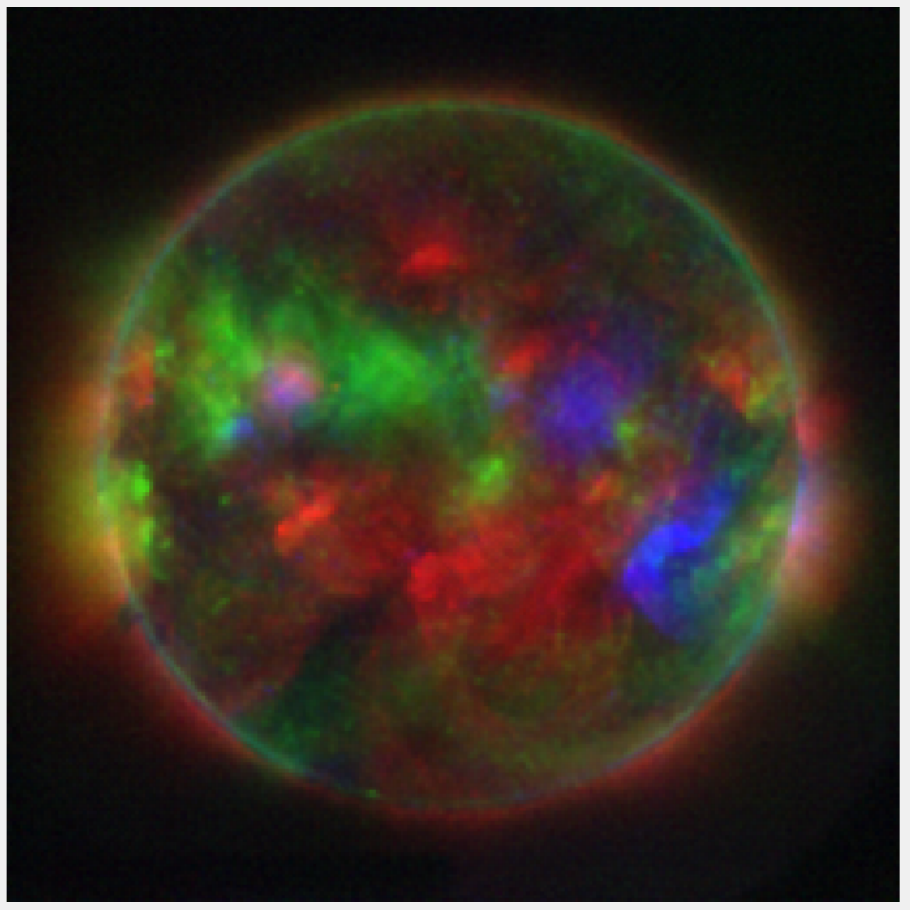}
    }

    \caption{
    Diffraction-limited original images (top), reconstructed spectral images with isotropic TV regularization (middle), and reconstructed spectral images with learned-reconstruction method (bottom) for dataset 5 at 25dB SNR. (Courtesy of NASA/SDO and the AIA, EVE, and HMI science teams.) For each case, false-color representations are also provided on the right.
     }  
\label{solarImageReconstructions_AAA}
\end{center} 
\end{figure*}

To visually evaluate the results for images with different amount of structure, we provide the reconstructed spectral images for the datasets 2 and 5 in Fig.~\ref{solarImageReconstructions_Solar} and \ref{solarImageReconstructions_AAA}, respectively. For comparison, the
diffraction-limited versions of the original images are also
given. Here SNR is taken as $25$ dB. 
These also illustrate that the developed imaging system provides near diffraction-limited resolution. Visual inspection shows that important characteristic features, such as solar
flares, are successfully recovered. The learned reconstruction method is more successful in this respect than the TV reconstruction, but the TV reconstruction performs closely to the learned reconstruction for the dataset 2.
However, even for these mostly smooth spectral images, TV reconstruction introduces some blur for the rapidly varying spatial structures.

\begin{table*}[]
	\caption{PSNR (dB) and SSIM values for TV and learned reconstruction methods at different input SNRs and for different spectral datasets.} 
	\label{Table:learned-tv-recon}
\begin{adjustbox}{width=1\textwidth}

\begin{tabular}{|c|c|c|c|c|c|c|c|c|}
\hline
SNR & \multicolumn{2}{c|}{15dB}                  & \multicolumn{2}{c|}{20dB}                  & \multicolumn{2}{c|}{25dB}                  & \multicolumn{2}{c|}{30dB}                  \\ \hline
Method    & TV Recon. & Learned Recon. & TV Recon. & Learned Recon. & TV Recon. & Learned Recon. & TV Recon. & Learned Recon. \\ \hline
Average over 105 images & 27.56/0.78         & 29.70/0.79             & 28.61/0.83         & 31.57/0.84              & 29.73/0.87         & 33.55/0.89              & 30.80/0.90          & 35.77/0.92              \\ \hline
Sample Dataset 1 & 38.56/0.93         & 37.12/0.91             & 39.99/0.95         & 37.70/0.92             & 41.82/0.97         & 40.85/0.96             & 43.54/0.98         & 43.06/0.97             \\ \hline
Sample Dataset 2 & 34.05/0.89         & 35.54/0.87             & 35.49/0.92         & 36.40/0.89             & 37.09/0.94         & 39.04/0.94             & 38.97/0.96         & 41.32/0.96             \\ \hline
Sample Dataset 3 & 32.38/0.83         & 31.20/0.82              & 33.80/0.86         & 32.62/0.86              & 35.16/0.89         & 32.26/0.90              & 36.80/0.93         & 35.88/0.93              \\ \hline
Sample Dataset 4 & 29.99/0.86          & 33.04/0.86             & 30.32/0.88         & 35.22/0.90             & 33.06/0.92         & 37.25/0.94             & 34.60/0.94         & 35.56/0.95             \\ \hline
Sample Dataset 5 & 26.44/0.83         & 29.09/0.84              & 27.48/0.86         & 31.25/0.88              & 28.83/0.90         & 35.72/0.92              & 29.98/0.93         & 36.72/0.94              \\ \hline
\end{tabular}
\end{adjustbox}
\end{table*}

Another important observation is given in Table~\ref{Table:psnr_ssim}, which illustrates that MD and FD settings provide similar reconstruction quality. Here the solar images in dataset 2 with a larger size of $512\times 512$ are used. As seen, for a given input SNR level, maximum differences in the reconstructions for the two measurement settings are less than 0.07 dB in PSNR and 0.01 in SSIM.
This suggests that one setting can be chosen over the other based on the design requirements of a particular application.

\begin{table}[tbh!] \vspace{-0.0in}
	\renewcommand{\arraystretch}{1.5}
	\fontsize{8}{10}\selectfont
	\centering
	\caption{
    	PSNR (dB) and SSIM values for different input SNRs when \textit{moving detector} (MD) and \textit{fixed detector} (FD) measurement settings are used. SNR and PSNR values are reported in decibel (dB).}
	\label{Table:psnr_ssim}
	\vspace{-0.0in}
	\begin{tabular}{|c|c|c|c|c|}
		\hline
		\textbf{Input SNR} & \textbf{MD PSNR} & \textbf{FD PSNR}& \textbf{MD SSIM} & \textbf{FD SSIM} \\
		\hline
		15&31.98 &32.00  &0.88 &0.88  \\
		\hline
		20&33.43 & 33.38 & 0.91& 0.91 \\
		\hline
		25&34.75 & 34.73 &0.93 & 0.93 \\
		\hline
		30&35.82 &35.75 & 0.94 & 0.94 \\
		\hline	
	\end{tabular} \vspace{-0.0in}
\end{table}

\begin{table}[tbh!] \vspace{-0.0in}
	\renewcommand{\arraystretch}{1.5}
	\fontsize{8}{10}\selectfont
	\centering
	\caption{
	PSNR (dB) and SSIM values for different number of sources (P) and measurements (K) for SNR=$25$ dB.}
	\label{Table:psnr_ssim_differentK}
	\vspace{-0.0in}
	\begin{tabular}{|c|c|c|}
		\hline
		\textbf{Setting} & \textbf{Average PSNR (dB)} & \textbf{Average SSIM}  \\
		\hline
		\text{K}=2\text{, P}=2&  37.55 &    0.94 \\
		\hline
		\text{K}=3\text{, P}=3& 34.75 & 0.93 
		\\
		\hline
		\text{K}=4\text{, P}=4& 31.91  & 0.91   \\

		\hline	
	\end{tabular} \vspace{-0.0in}
\end{table}

 In addition to the measurement noise, other practical challenges should also be considered in real implementation such as the dynamic range of the detector, calibration of the system, or model mismatch. Dynamic range of the chosen monochromatic detector will actually contribute to the noise level as another source of error, but this error can be kept at a desired level to achieve a particular SNR. Likewise, the calibration is also an easy task as the imaging system is shift-invariant (unlike many of the earlier spectral imaging systems). In particular, measuring the PSFs for different wavelengths is sufficient for calibration. Hence there is no need to measure the system response for each voxel in the spectral data cube. 

\begin{figure}[h]
		\subfloat[]{\hspace{-0.0in}
		\includegraphics[align=t,scale=0.18] {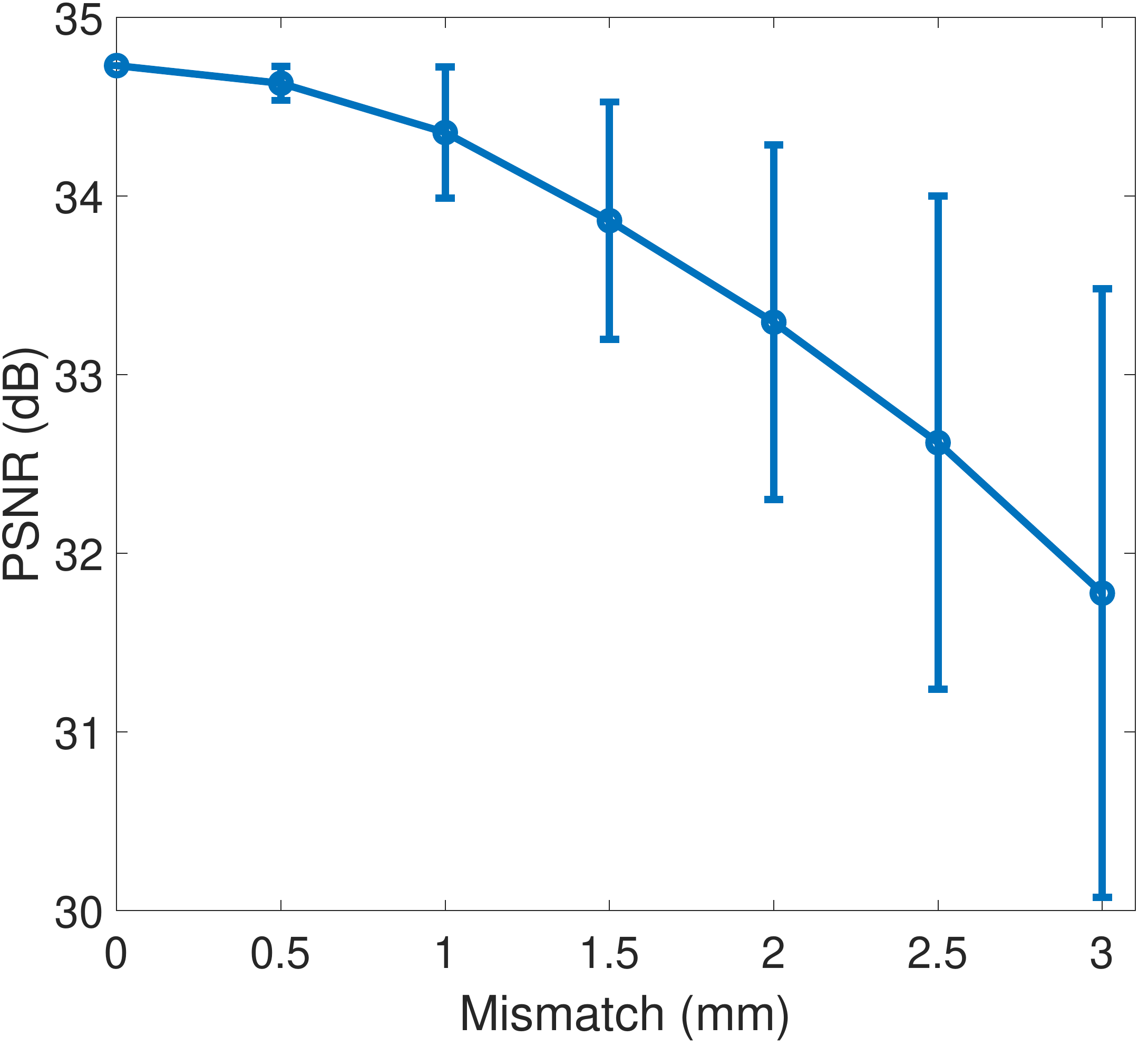}}
		\vspace{-0.00in}
		\subfloat[]{\hspace{-0.0in}
		\includegraphics[align=t,scale=0.18]{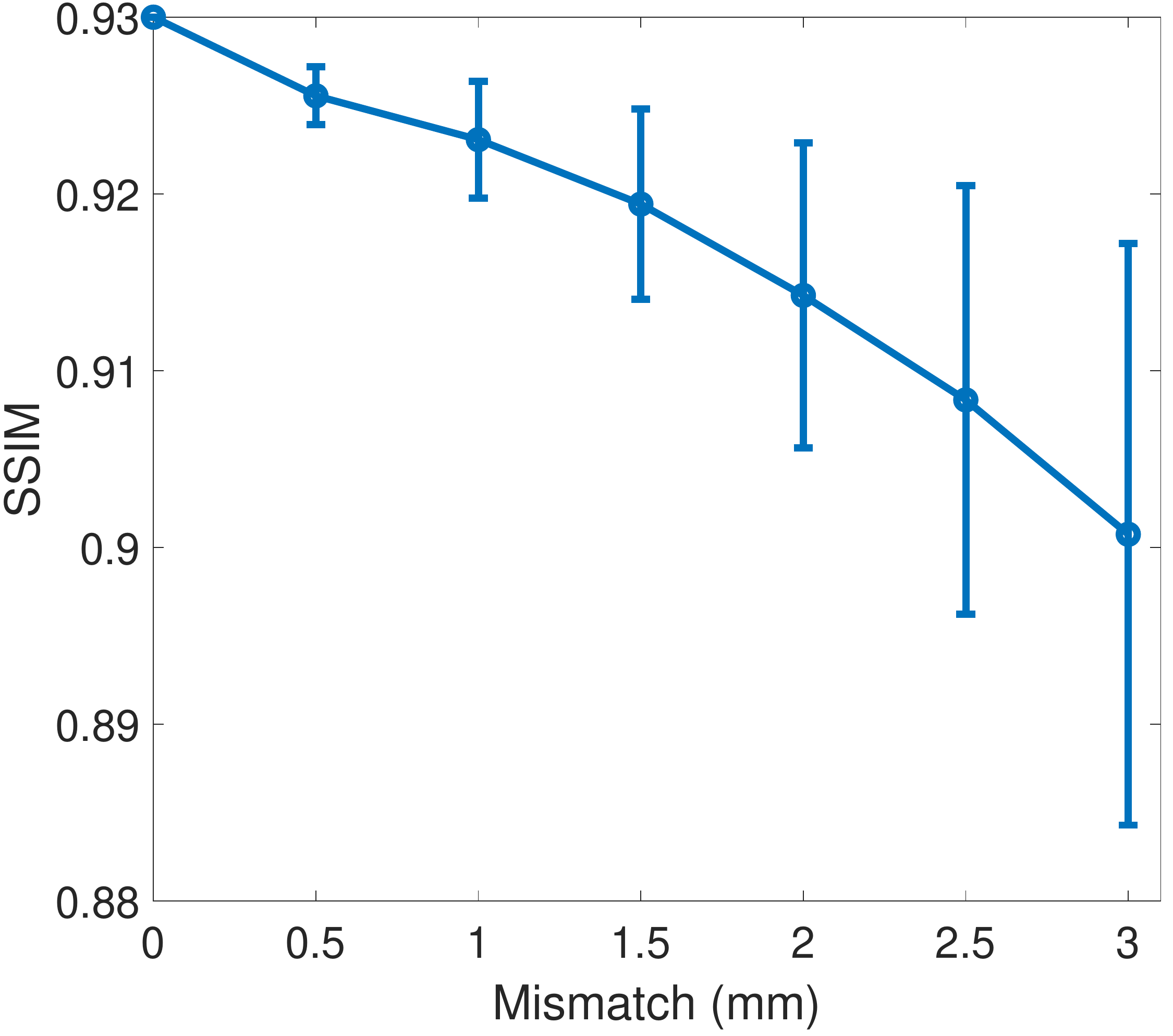}}
		\vspace{-0.00in}
		\caption{The average PSNR (a) and SSIM (b) values for the reconstructed spectral images as a function of maximum placement error $\Delta d_{\text{max}}$ in the measurement planes.
		}
		\label{fig-mismatch}
\end{figure}

Model mismatch can be, however, more critical in practice. For example, for the moving detector (MD) case, 
there can be misplacement errors due to the mechanically moving component, resulting in model errors for the assumed PSFs of different spectral bands.
Other non-ideal factors during operation (such as in-flight distortions) can also cause similar widening in the PSFs. Hence errors in the assumed PSFs will constitute the main sources of model mismatch. To illustrate the effect of such model errors on the performance, we have performed a sensitivity analysis with varying misplacement errors for the MD setting. Note that in the considered MD setting, the moving detector takes measurements from the distances varying between $d_k=3.727$ m ($f_3$) to $3.756$ m ($f_1$), covering a range of $29$ mm. In the sensitivity analysis, maximum placement error, $\Delta d_{\text{max}}$, is considered between $0$ and $3$ mm. Then the error in the measurement distance $d_k$, denoted as $\Delta d$, is modeled as a uniformly distributed random variable between $[-\Delta d_{\text{max}}, \Delta d_{\text{max}}]$ for each measurement taken. After performing a Monte-Carlo simulation at $25$ dB SNR, the average PSNR and SSIM values for the reconstructed spectral images are obtained in the presence of these measurement errors. The results are given in Fig.~\ref{fig-mismatch} as a function of $\Delta d_{\text{max}}$. As seen from this figure, the random misplacement errors up to $3$ mm degrades the performance gracefully; hence the reconstructions are mostly robust to such errors. But for larger misplacement errors, it could be better to modify the reconstruction methods to take into account such model mismatch errors (for example, through semi-blind multi-channel deconvolution).

Another type of model error can arise when the spectral sampling interval is not chosen to be smaller than the spectral bandwidth of the designed diffractive lens. To illustrate this, we have considered Eq.~\eqref{forwardModel} with a continuous spectrum consisting of three Gaussian spectral lines, each with a width slightly larger than the spectral bandwidth. 
Although the reconstruction was still successful for the three spectral images reconstructed at the central wavelengths, the performance was slightly decreased (36.57 dB PSNR and 0.92 SSIM for dataset 2 with 25 dB SNR).
To avoid such issues, one can consider to revise the design of the diffractive lens to decrease its spectral bandwidth. Other options are to use finer sampling for the spectral discretization (i.e. more terms in Eq.~\eqref{forwardModel}) and take more measurements, or train the network with the measurements obtained through a more realistic simulation. 

Lastly, we demonstrate the system performance for increased number of spectral bands, $P$. For this, we also perform simulations for $P=2$ and $P=4$ cases by taking same number of measurements as the number of spectral bands (i.e. $K=P$). 
In the first case, we consider two spectral bands at $\lambda_2 =33.42$ nm, $\lambda_3 =33.54$ nm with the measurements obtained at $f_2$ and $f_3$ for the MD setting. In the latter case, we consider four spectral bands, namely $\lambda_0 = 33.16$ nm, $\lambda_1 =33.28$ nm, $\lambda_2 =33.42$ nm, $\lambda_3 =33.54$ nm, and obtain four measurements at the corresponding focal planes, $f_0$, $f_1$, $f_2$, and $f_3$. Table~\ref{Table:psnr_ssim_differentK} shows PSNR and SSIM values for these cases together with $P=3$ case described before. The results demonstrate that the reconstruction quality gets worse with increasing $P$. This is expected since the ill-posedness of the 
inverse problem increases with the number of spectral bands, $P$. Nevertheless, the system provides high quality reconstructions even for $P=4$ case, with a PSNR of 31.91 dB and an SSIM of 0.91 at 25 dB input SNR.

\subsection{Resolution Analysis}

An important advantage of the developed technique is the unprecedented spectral resolution it enables. The technique offers the capability of separating close-by spectral components that would not otherwise be possible using conventional spectral imagers employing wavelength filters. In the experiments, the sources of interest have wavelengths $33.28$ nm, $33.42$ nm, and $33.54$ nm; hence the developed spectral imager can achieve a spectral resolution of approximately $0.1$ nm, which is nearly $0.3\%$ of the central wavelengths. 
On the other hand, the state-of-the-art EUV wavelength filters at best provide a spectral resolution of $10\%$ of the central wavelength~\cite{lemen2011}; hence such a high spectral resolution cannot be achieved with the conventional filter-based imagers. This becomes critical especially when more than one spectral line lie in this $10\%$ range as in the considered EUV spectral imaging application because, using feasible wavelength filters, it is not possible to separately resolve each line. 

The obtained PSNR and SSIM values between the reconstructions and diffraction-limited original images
also illustrate that the technique offers diffraction-limited high spatial resolution as enabled by the diffractive lenses. Here we further analyze the spatial resolution of the system to quantify 
its diffraction-limited imaging performance better. For this purpose, we first perform a conditioning-based analysis for the resolution. We also reconstruct point targets separated by various distances to verify that the resolvability of the point targets in these results agree with the diffraction-limited resolution. Each of these analysis provides additional support for the imaging system's diffraction-limited spatial resolution capability. Note that existing EUV spectral imagers cannot achieve diffraction-limited resolution due to surface roughness and figure errors of the used reflective optics~\cite{lemen2011, davila2011b}.

In the conditioning-based resolution analysis, 
we investigate the conditioning of the forward model in Eq.~\eqref{noisyModel} when the scene consists of point targets. For this, we consider different number of point sources with varying distances between them. If we assume that the locations of the point targets are known, the reconstruction quality will only depend on the columns of $\Hb$ associated with the locations of the point targets. Thus, by examining the conditioning of the relevant submatrices of $\Hb$, we will gain an understanding of the resolving capability of the system~\cite{antipa2018diffusercam}. In this analysis, we suppose an oracle tells us the exact locations of the point sources in the 3D spectral data cube, which effectively corresponds to knowing the support of the cube a priori. Then, the reconstruction task is to determine the values of these nonzero components. If this problem cannot be solved, the original problem of finding both the locations and values of the point targets will also fail. As a result, the conditioning of the resulting submatrices of $\Hb$ provides information about the best possible capability of the system for resolving point targets.

\begin{figure}[h]
	\begin{center} 
		\subfloat[]{
		\includegraphics[width=1.5in]{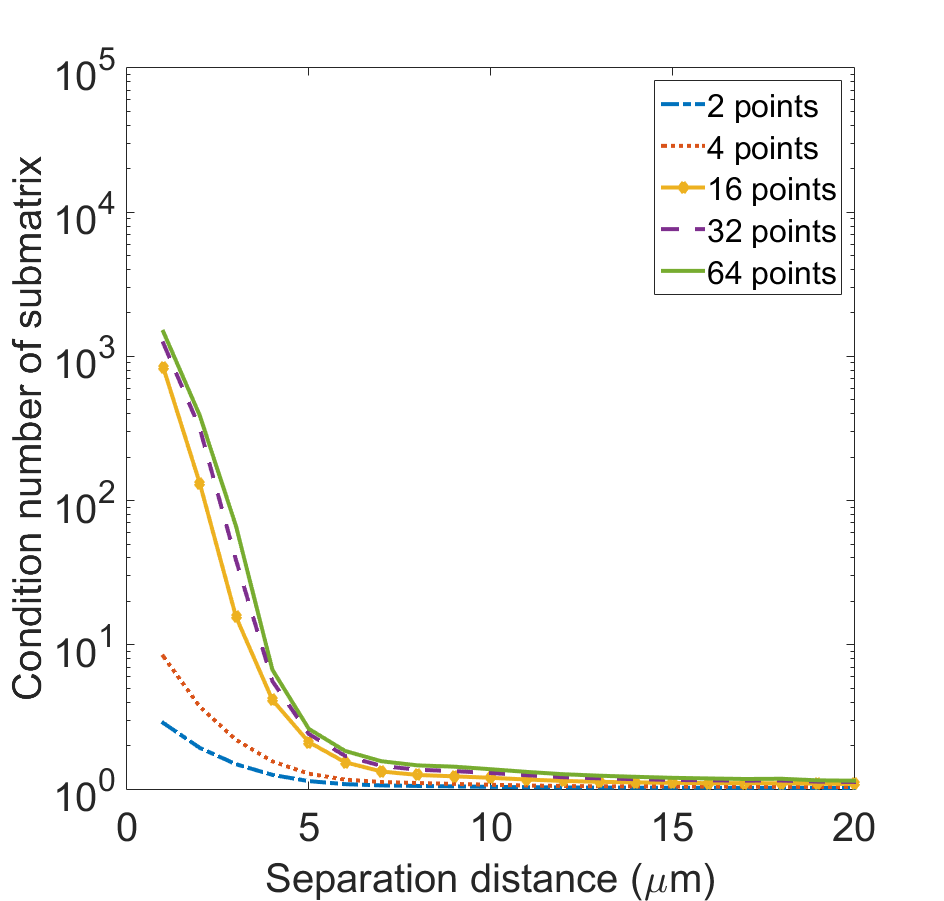}}
		\subfloat[]{
		\includegraphics[width=1.5in]{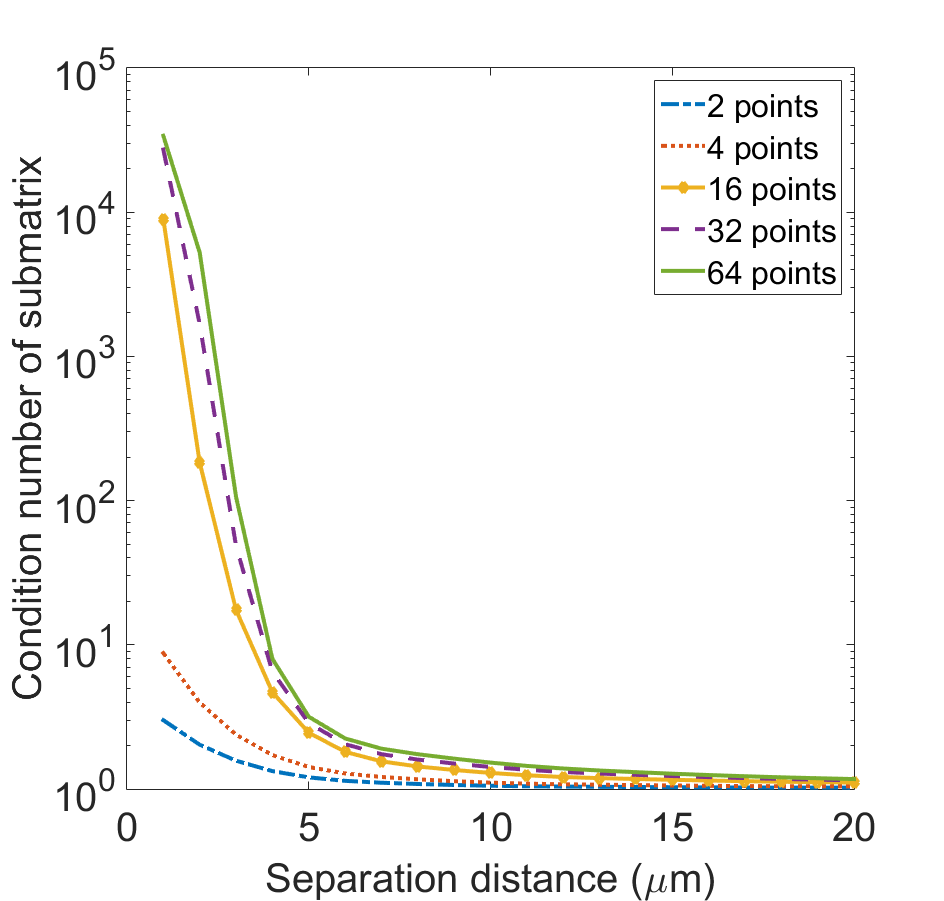}}\\
		\subfloat[]{
		\includegraphics[width=1.5in]{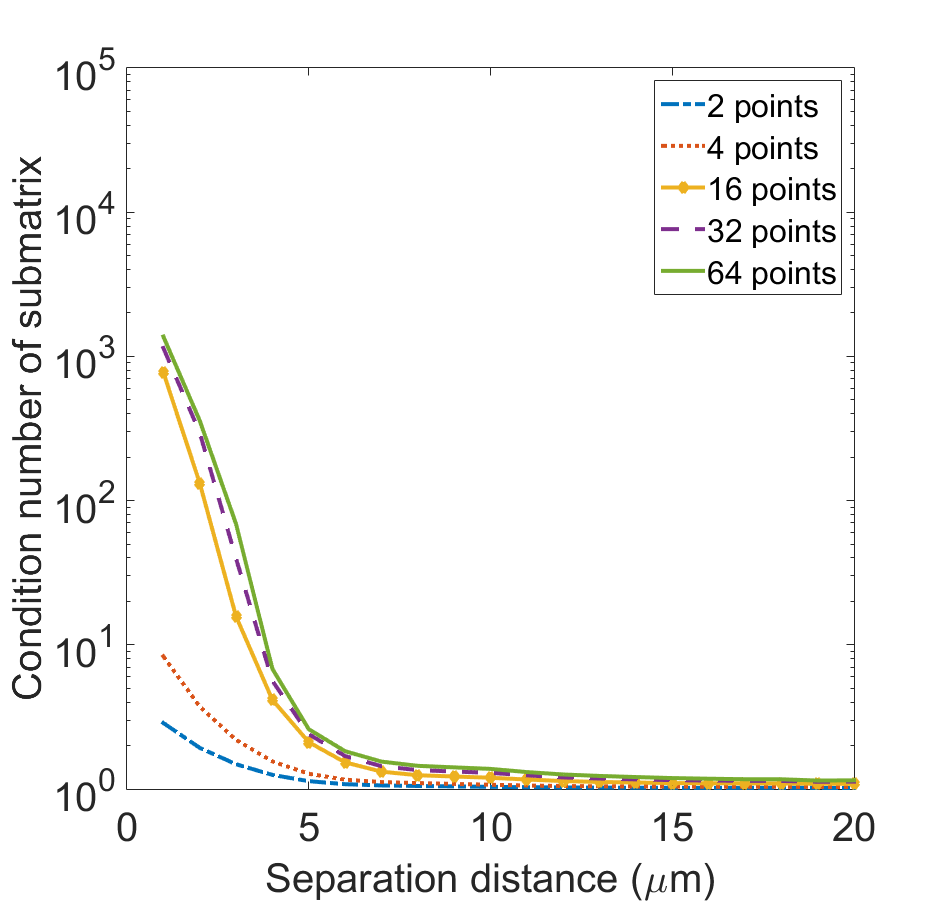}}
		\caption{Conditioning of the relevant submatrices of $\Hb$ for different number of point sources and separation distances. Conditioning results when the point sources are placed at (a) the first, (b) the second, and (c) the third spectral bands.
		}
		\vspace{-0.0in} 
		\label{fig-conditioning}
	\end{center} 
\end{figure}

Since this is not a conventional camera but a computational imaging system, two-point resolution may not reveal the system performance for more complex scenes. For this reason, we also consider a higher number of point targets than two in our analysis.
In particular, we consider $2$, $4$, $16$, $32$,
and $64$ point sources placed in a square grid. We choose the pixel size on the detector as $1$ $\mu$m for fine analysis of the resolution. We change the spacing between the point sources from $1$ $\mu$m to $20$ $\mu$m with $1$ $\mu$m steps. Then, we calculate the conditioning of the submatrix of $\Hb$ that contains the columns of $\Hb$ associated with the locations of these point sources. The results are plotted in Fig.~\ref{fig-conditioning} when all point sources are placed in the first, second or third spectral bands for $P=3$ and MD setting case. 

As seen, conditioning is similar for each spectral band and gets worse as the number of point sources increases, or the separation distance between them decreases, as expected. 
An important observation is the rapid decrease in the condition number around $5$ $\mu$m separation distance, which indeed corresponds to the theoretical diffraction-limited resolution of the imaging system for the monochromatic case. As the distance gets larger than $5$ $\mu$m, the change in conditioning becomes small. This observation agrees with the expected diffraction-limited resolution of the system.

\begin{figure}[h]
\begin{center} 
    \vspace{-0.0in}
    \subfloat[]{
    \includegraphics[scale=1.3]{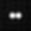}
     }
\subfloat[]{
    \includegraphics[scale=1.3]{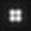}
    }
    \subfloat[]{
    \includegraphics[scale=1.3]{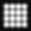}
    }
    \\
    \vspace{-0.0in}
    \subfloat[]{
    \includegraphics[scale=1.3]{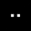}
    }
\subfloat[]{
    \includegraphics[scale=1.3]{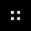}
    }
    \subfloat[]{
    \includegraphics[scale=1.3]{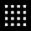}
    }\\
    \vspace{-0.0in}
    \subfloat[]{
    \includegraphics[scale=1.3]{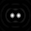}
    }
    \subfloat[]{
    \includegraphics[scale=1.3]{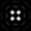}
    }
    \subfloat[]{
    \includegraphics[scale=1.3]{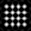}
    }\\
    \subfloat[]{
    \includegraphics[scale=1.3]{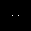}
    }
\subfloat[]{
    \includegraphics[scale=1.3]{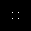}
    }
    \vspace{-0.0in}
    \subfloat[]{
    \includegraphics[scale=1.3]{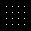}
    }
    \caption{Demonstration of $5$ $\mu$m resolution using point targets for SNR=$25$ dB. (a)-(c) Noisy measurements with 25 dB SNR at the focal planes $f_1$, $f_2$, and $f_3$, (d)-(f) reconstructed images from the noisy measurements, (g)-(i) diffraction-limited images, (j)-(l) ground truth images.
     }  
\label{fig:res_targets}
\end{center} 
\end{figure}

In the reconstruction-based resolution analysis, we simultaneously place $2$, $4$, and $16$ point sources of size $1$ $\mu$m in the first, second, and third spectral bands, respectively. The separation distance between point sources is chosen as the expected spatial resolution of $5$ $\mu$m. This data cube is reconstructed from the measurements generated using the forward model in Eq.~\eqref{noisyModel} with 25 dB SNR. Figure~\ref{fig:res_targets} shows the reconstructed and diffraction-limited images for each band, together with the ground truth and measurements. As seen, the imaging system can successfully resolve point 
sources with $5$ $\mu$m separation. In fact, the reconstructed images are even sharper than their diffraction-limited versions at this SNR level. 
This observation is certainly related to our regularization choice (isotropic TV) as well.  

By repeating this analysis for different SNRs and separation distances, we also observe that point sources of $5$ $\mu$m separation can be resolved for SNRs as low as 3 dB. This suggests that the expected theoretical resolution can be achieved even for highly noisy cases. Moreover, resolving point sources with $4$ $\mu$m spacing is also possible for SNRs as low as 5 dB. That is, the imaging system can even provide spatial resolution beyond the diffraction limit for a wide range of SNR values. However, the system can not resolve point sources even at high SNR levels when the separation distance becomes $3$ $\mu$m. That is, the system fails to resolve point sources shortly after $5$ $\mu$m separation distance, as the conditioning starts to degrade significantly. Hence, the reconstruction-based resolution analysis is in agreement with the conditioning-based analysis and also supports the high spatial resolution enabled by the developed multi-spectral imaging technique.

\section{Conclusion}

In this paper, we have developed a novel multi-spectral imaging modality that enables higher spatial and spectral resolutions than the conventional scanning-based techniques.  
Our approach exploited the use of a diffractive lens and powerful image reconstruction to overcome the resolution limits of conventional techniques. 
The proposed lightweight and low-cost imaging system enables not only unprecedented spectral resolution but also diffraction-limited high spatial resolution for a wide spectral range including UV and x-rays. Spectral imaging with these capabilities has profound impact for a variety of applications. 
Moreover, the developed learning-based method is capable of adapting the reconstruction (i.e. deep prior) to a particular application. To illustrate these capabilities, the developed computational imaging technique was applied to a potential solar imaging application in the EUV regime. The presented results for EUV spectral imaging illustrate the 
diffraction-limited spatial resolution and unprecedented spectral resolution enabled with this technique.

The developed imaging technique opens up new possibilities for high-resolution multi-spectral imaging with low-cost and simpler designs. One important application area is the observation of the spectra of space plasmas with new capabilities, as demonstrated in the paper. Lightweight and low-cost diffractive lenses provide diffraction-limited high spatial resolution for a wide spectral range including UV and x-rays. This is not possible with the imaging/collimating optics (such as mirrors and refractive lenses) prevalently used in the other spectral imaging techniques. Another important advantage of the developed technique is the unprecedented spectral resolution it enables. The technique provides the capability of resolving nearby spectral components that would not otherwise be possible with the wavelength filters used in conventional spectral imagers. 
For the considered EUV application (which was the initial targeted domain), missions are currently under development. The details of a mission concept based on a distributed solar imager that exploits diffractive lenses and two small satellites flying in formation can be found elsewhere~\cite{davila2016spd,rabin2018agu}. 

Our technique is designed to provide high resolution for general scenes including those that lack strong spectral correlation (i.e. has frequent peaks and fast transitions) or have spectra consisting of discrete spectral lines. Spectral sensing of such scenes are of interest in various applications including astrophysical imaging of space plasmas, atmospheric physics, and remote sensing~\cite{rybicki_book,hollas_book}. Note that our approach will naturally work better for spectrally correlated scenes (such as those 
encountered in the visible regime) using 3D priors to exploit spectral correlation. Sample results in visible regime can be found elsewhere~\cite{oguzhan_thesis,didem2020}.

Lastly we note that the performance of the developed technique can be further improved by taking more measurements (such as at the intermediate planes) and optimizing the measurement configuration or the diffractive lens design~\cite{widloski2020}. 
In the presented results, a conventional photon sieve has been used, but the performance may be improved with a diffractive lens specifically designed for this imaging setting. The performance of the learned reconstruction method can also be improved by using different loss functions 
and priors.


%
%

\section*{Acknowledgment}
The authors would like to thank Joseph M. Davila and Adrian Daw from NASA GSFC for many fruitful discussions about solar imaging spectroscopy and photon sieves. F.~S.~Oktem acknowledges the support and hospitality received as a Visiting Postdoctoral Researcher at the NASA GSFC,
Heliophysics Division, and the Catholic University of America.


\begin{thebibliography}{10}
\providecommand{\url}[1]{#1}
\csname url@samestyle\endcsname
\providecommand{\newblock}{\relax}
\providecommand{\bibinfo}[2]{#2}
\providecommand{\BIBentrySTDinterwordspacing}{\spaceskip=0pt\relax}
\providecommand{\BIBentryALTinterwordstretchfactor}{4}
\providecommand{\BIBentryALTinterwordspacing}{\spaceskip=\fontdimen2\font plus
\BIBentryALTinterwordstretchfactor\fontdimen3\font minus
  \fontdimen4\font\relax}
\providecommand{\BIBforeignlanguage}[2]{{%
\expandafter\ifx\csname l@#1\endcsname\relax
\typeout{** WARNING: IEEEtran.bst: No hyphenation pattern has been}%
\typeout{** loaded for the language `#1'. Using the pattern for}%
\typeout{** the default language instead.}%
\else
\language=\csname l@#1\endcsname
\fi
#2}}
\providecommand{\BIBdecl}{\relax}
\BIBdecl

\bibitem{oktem2014icip}
F.~S. Oktem, F.~Kamalabadi, and J.~M. Davila, ``High-resolution computational
  spectral imaging with photon sieves,'' in \emph{IEEE Int. Conf. on Image
  Processing (ICIP)}.\hskip 1em plus 0.5em minus 0.4em\relax IEEE, 2014, pp.
  5122--5126.

\bibitem{davila2011b}
J.~Davila, ``High-resolution solar imaging with a photon sieve,'' in \emph{SPIE
  Optical Engineering+ Applications}.\hskip 1em plus 0.5em minus 0.4em\relax
  International Society for Optics and Photonics, 2011, pp. 81\,480O--81\,480O.

\bibitem{arce2014}
G.~Arce, D.~Brady, L.~Carin, H.~Arguello, and D.~Kittle, ``Compressive coded
  aperture spectral imaging: An introduction,'' \emph{Signal Processing
  Magazine, IEEE}, vol.~31, no.~1, pp. 105--115, 2014.

\bibitem{lemen2011}
J.~R. Lemen, A.~M. Title, D.~J. Akin, P.~F. Boerner, C.~Chou, J.~F. Drake,
  D.~W. Duncan, C.~G. Edwards, F.~M. Friedlaender, G.~F. Heyman \emph{et~al.},
  ``The atmospheric imaging assembly ({AIA}) on the solar dynamics observatory
  ({SDO}),'' \emph{Solar Physics}, pp. 1--24, 2011.

\bibitem{andersen2005}
G.~Andersen, ``Large optical photon sieve,'' \emph{Optics Letters}, vol.~30,
  no.~22, pp. 2976--2978, 2005.

\bibitem{lucas2018}
A.~Lucas, M.~Iliadis, R.~Molina, and A.~K. Katsaggelos, ``Using deep neural
  networks for inverse problems in imaging: beyond analytical methods,''
  \emph{IEEE Signal Processing Magazine}, vol.~35, no.~1, pp. 20--36, 2018.

\bibitem{aggarwal2019}
H.~K. Aggarwal, M.~P. Mani, and M.~Jacob, ``Modl: Model-based deep learning
  architecture for inverse problems,'' \emph{IEEE Transactions on Medical
  Imaging}, vol.~38, no.~2, pp. 394--405, 2019.

\bibitem{ongie2020}
G.~Ongie, A.~Jalal, C.~A. M. R.~G. Baraniuk, A.~G. Dimakis, and R.~Willett,
  ``Deep learning techniques for inverse problems in imaging,'' \emph{IEEE
  Journal on Selected Areas in Information Theory}, 2020.

\bibitem{cassi1}
A.~Wagadarikar, R.~John, R.~Willett, and D.~Brady, ``Single disperser design
  for coded aperture snapshot spectral imaging,'' \emph{Applied Optics},
  vol.~47, no.~10, pp. B44--B51, 2008.

\bibitem{wu2011}
Y.~Wu, I.~O. Mirza, G.~R. Arce, and D.~W. Prather, ``Development of a
  digital-micromirror-device-based multishot snapshot spectral imaging
  system,'' \emph{Optics Letters}, vol.~36, no.~14, pp. 2692--2694, 2011.

\bibitem{cao2016computational}
X.~Cao, T.~Yue, X.~Lin, S.~Lin, X.~Yuan, Q.~Dai, L.~Carin, and D.~J. Brady,
  ``Computational snapshot multispectral cameras: toward dynamic capture of the
  spectral world,'' \emph{IEEE Signal Processing Magazine}, vol.~33, no.~5, pp.
  95--108, 2016.

\bibitem{salazar2019}
E.~Salazar, A.~Parada-Mayorga, and G.~R. Arce, ``Spectral zooming and
  resolution limits of spatial spectral compressive spectral imagers,''
  \emph{IEEE Transactions on Computational Imaging}, vol.~5, no.~2, pp.
  165--179, 2019.
  
\bibitem{osa2018}
O.~F. Kar, U.~Kamaci, F.~C. Akyon, and F.~S. Oktem, ``Compressive photon-sieve
  spectral imaging,'' in \emph{Computational Optical Sensing and
  Imaging}.\hskip 1em plus 0.5em minus 0.4em\relax Optical Society of America,
  2018, pp. CTu5D--8.

\bibitem{kar2019}
O.~F. Kar and F.~S. Oktem, ``Compressive spectral imaging with diffractive
  lenses,'' \emph{Optics Letters}, vol.~44, no.~18, pp. 4582--4585, 2019.

\bibitem{rybicki_book}
G.~B. Rybicki and A.~P. Lightman, \emph{Radiative processes in
  astrophysics}.\hskip 1em plus 0.5em minus 0.4em\relax John Wiley \& Sons,
  2008.

\bibitem{hollas_book}
J.~M. Hollas, \emph{Modern spectroscopy}.\hskip 1em plus 0.5em minus
  0.4em\relax Wiley. com, 2004.

\bibitem{attwood_book}
D.~Attwood, \emph{Soft x-rays and extreme ultraviolet radiation: principles and
  applications}.\hskip 1em plus 0.5em minus 0.4em\relax Cambridge University
  Press, Cambridge, 2000.

\bibitem{kipp2001}
L.~Kipp, M.~Skibowski, R.~Johnson, R.~Berndt, R.~Adelung, S.~Harm, and
  R.~Seemann, ``Sharper images by focusing soft x-rays with photon sieves,''
  \emph{Nature}, vol. 414, no. 6860, pp. 184--188, 2001.

\bibitem{menon2005}
R.~Menon, D.~Gil, G.~Barbastathis, and H.~I. Smith, ``Photon-sieve
  lithography,'' \emph{J. Opt. Soc. Am. A}, vol.~22, no.~2, pp. 342--345, 2005.

\bibitem{kim2012}
G.~Kim, J.~A. Dom{\'\i}nguez-Caballero, and R.~Menon, ``Design and analysis of
  multi-wavelength diffractive optics,'' \emph{Optics Express}, vol.~20, no.~3,
  pp. 2814--2823, 2012.

\bibitem{oktem2018analytical}
F.~S. Oktem, F.~Kamalabadi, and J.~M. Davila, ``Analytical {Fresnel} imaging
  models for photon sieves,'' \emph{Optics Express}, vol.~26, no.~24, pp.
  32\,259--32\,279, 2018.

\bibitem{andersen2007}
G.~Andersen and D.~Tullson, ``Broadband antihole photon sieve telescope,''
  \emph{Applied Optics}, vol.~46, no.~18, pp. 3706--3708, 2007.

\bibitem{zhou2009}
C.~Zhou, X.~Dong, L.~Shi, C.~Wang, and C.~Du, ``Experimental study of a
  multiwavelength photon sieve designed by random-area-divided approach,''
  \emph{Applied Optics}, vol.~48, no.~8, pp. 1619--1623, 2009.

\bibitem{wang2003}
Y.~Wang, W.~Yun, and C.~Jacobsen, ``Achromatic {Fresnel} optics for wideband
  extreme-ultraviolet and x-ray imaging,'' \emph{Nature}, vol. 424, no. 6944,
  pp. 50--53, 2003.

\bibitem{zhao2015}
X.~Zhao, F.~Xu, J.~Hu, and C.~Wang, ``Broadband photon sieves imaging with
  wavefront coding,'' \emph{Optics Express}, vol.~23, no.~13, pp.
  16\,812--16\,822, 2015.

\bibitem{wang2016}
P.~Wang, N.~Mohammad, and R.~Menon, ``Chromatic-aberration-corrected
  diffractive lenses for ultra-broadband focusing,'' \emph{Scientific Reports},
  vol.~6, p. 21545, 2016.

\bibitem{mohammad2018}
N.~Mohammad, M.~Meem, B.~Shen, P.~Wang, and R.~Menon, ``Broadband imaging with
  one planar diffractive lens,'' \emph{Scientific Reports}, vol.~8, no.~1, pp.
  1--6, 2018.

\bibitem{li2018}
Y.~Li, C.~Wang, X.~Zhao, F.~Xu, and C.~Wang, ``Multispectral and large
  bandwidth achromatic imaging with a single diffractive photon sieve,''
  \emph{Optics Express}, vol.~26, no.~16, pp. 21\,141--21\,152, 2018.

\bibitem{dun2020}
X.~Dun, H.~Ikoma, G.~Wetzstein, Z.~Wang, X.~Cheng, and Y.~Peng, ``Learned
  rotationally symmetric diffractive achromat for full-spectrum computational
  imaging,'' \emph{Optica}, vol.~7, no.~8, pp. 913--922, 2020.

\bibitem{davila2016spd}
J.~M. {Davila}, F.~S. {Oktem}, F.~{Kamalabadi}, J.~{O'Neill}, A.-M.
  {Novo-Gradac}, A.~N. {Daw}, and D.~M. {Rabin}, ``{Milli-Arcsecond (MAS)
  Imaging of the Solar Corona},'' in \emph{AAS/Solar Physics Division Abstracts
  \#47}, ser. AAS/Solar Physics Division Meeting, vol.~47, May 2016, p. 3.10.

\bibitem{osa2019}
O.~F. Kar and F.~S. Oktem, ``Fast computational spectral imaging with a
  programmable diffractive lens,'' in \emph{Computational Optical Sensing and
  Imaging}.\hskip 1em plus 0.5em minus 0.4em\relax Optical Society of America,
  2019, pp. CTh3A--3.

\bibitem{hallada2017}
F.~D. Hallada, A.~L. Franz, and M.~R. Hawks, ``Fresnel zone plate light field
  spectral imaging,'' \emph{Optical Engineering}, vol.~56, no.~8, p. 081811,
  2017.

\bibitem{wang2018}
P.~Wang and R.~Menon, ``Computational multispectral video imaging,'' \emph{JOSA
  A}, vol.~35, no.~1, pp. 189--199, 2018.

\bibitem{nimmer2018}
M.~Nimmer, G.~Steidl, R.~Riesenberg, and A.~Wuttig, ``Spectral imaging based on
  {2D} diffraction patterns and a regularization model,'' \emph{Optics
  Express}, vol.~26, no.~22, pp. 28\,335--28\,348, 2018.

\bibitem{jeon2019}
D.~S. Jeon, S.-H. Baek, S.~Yi, Q.~Fu, X.~Dun, W.~Heidrich, and M.~H. Kim,
  ``Compact snapshot hyperspectral imaging with diffracted rotation,''
  \emph{ACM Transactions on Graphics (Proc. SIGGRAPH 2019)}, vol.~38, no.~4,
  pp. 117:1--13, 2019.

\bibitem{le2017}
R.~Le~Goff, H.~Krol, C.~Grezes-Besset, M.~Lequime, T.~Begou, C.~Hecquet,
  M.~Zerrad, B.~Badoil, G.~Montay, and K.~Gasc, ``Multispectral filters
  assemblies for {Earth} remote sensing imagers,'' in \emph{International
  Conference on Space Optics 2014}, vol. 10563.\hskip 1em plus 0.5em
  minus 0.4em\relax International Society for Optics and Photonics, 2017, p.
  1056305.

\bibitem{schowengerdt2006}
R.~A. Schowengerdt, \emph{Remote sensing: models and methods for image
  processing}.\hskip 1em plus 0.5em minus 0.4em\relax Elsevier, 2006.

\bibitem{yi2019}
C.~Yi, Y.-Q. Zhao, and J.~C.-W. Chan, ``Spectral super-resolution for
  multispectral image based on spectral improvement strategy and spatial
  preservation strategy,'' \emph{IEEE Transactions on Geoscience and Remote
  Sensing}, vol.~57, no.~11, pp. 9010--9024, 2019.

\bibitem{geelen2013}
B.~Geelen, N.~Tack, and A.~Lambrechts, ``A snapshot multispectral imager with
  integrated tiled filters and optical duplication,'' in \emph{Advanced
  Fabrication Technologies for Micro/Nano Optics and Photonics VI}, vol.
  8613.\hskip 1em plus 0.5em minus 0.4em\relax International Society for Optics
  and Photonics, 2013, p. 861314.

\bibitem{lapray2014}
P.-J. Lapray, X.~Wang, J.-B. Thomas, and P.~Gouton, ``Multispectral filter
  arrays: Recent advances and practical implementation,'' \emph{Sensors},
  vol.~14, no.~11, pp. 21\,626--21\,659, 2014.

\bibitem{wang2014}
Z.~Wang and Z.~Yu, ``Spectral analysis based on compressive sensing in
  nanophotonic structures,'' \emph{Optics Express}, vol.~22, no.~21, pp.
  25\,608--25\,614, 2014.

\bibitem{wang2019}
Z.~Wang, S.~Yi, A.~Chen, M.~Zhou, T.~S. Luk, A.~James, J.~Nogan, W.~Ross,
  G.~Joe, A.~Shahsafi \emph{et~al.}, ``Single-shot on-chip spectral sensors
  based on photonic crystal slabs,'' \emph{Nature communications}, vol.~10,
  no.~1, pp. 1--6, 2019.

\bibitem{hahn2020}
R.~Hahn, F.-E. H{\"a}mmerling, T.~Haist, D.~Fleischle, O.~Schwanke, O.~Hauler,
  K.~Rebner, M.~Brecht, and W.~Osten, ``Detailed characterization of a mosaic
  based hyperspectral snapshot imager,'' \emph{Optical Engineering}, vol.~59,
  no.~12, p. 125102, 2020.

\bibitem{book_chapter}
F.~S. Oktem, L.~Gao, and F.~Kamalabadi, ``Computational spectral and ultrafast
  imaging via convex optimization,'' in \emph{Handbook of Convex Optimization
  Methods in Imaging Science}.\hskip 1em plus 0.5em minus 0.4em\relax Springer,
  2017, pp. 105--127.

\bibitem{okamoto1991}
T.~Okamoto and I.~Yamaguchi, ``Simultaneous acquisition of spectral image
  information,'' \emph{Optics Letters}, vol.~16, no.~16, pp. 1277--1279, 1991.

\bibitem{descour1995}
M.~Descour and E.~Dereniak, ``Computed-tomography imaging spectrometer:
  experimental calibration and reconstruction results,'' \emph{Applied Optics},
  vol.~34, no.~22, pp. 4817--4826, 1995.

\bibitem{ford2001}
B.~K. Ford, C.~E. Volin, S.~M. Murphy, R.~M. Lynch, and M.~R. Descour,
  ``Computed tomography-based spectral imaging for fluorescence microscopy,''
  \emph{Biophysical Journal}, vol.~80, no.~2, pp. 986--993, 2001.

\bibitem{oktem2014}
F.~S. Oktem, F.~Kamalabadi, and J.~M. Davila, ``A parametric estimation
  approach to instantaneous spectral imaging,'' \emph{IEEE Transactions on
  Image Processing}, vol.~23, no.~12, pp. 5707--5721, 2014.

\bibitem{august2013}
Y.~August, C.~Vachman, Y.~Rivenson, and A.~Stern, ``Compressive hyperspectral
  imaging by random separable projections in both the spatial and the spectral
  domains,'' \emph{Applied Optics}, vol.~52, no.~10, pp. D46--D54, 2013.

\bibitem{baek2020}
S.-H. Baek, H.~Ikoma, D.~S. Jeon, Y.~Li, W.~Heidrich, G.~Wetzstein, and M.~H.
  Kim, ``End-to-end hyperspectral-depth imaging with learned diffractive
  optics,'' \emph{arXiv preprint arXiv:2009.00463}, 2020.

\bibitem{baek2017}
S.-H. Baek, I.~Kim, D.~Gutierrez, and M.~H. Kim, ``Compact single-shot
  hyperspectral imaging using a prism,'' \emph{ACM Transactions on Graphics
  (TOG)}, vol.~36, no.~6, pp. 1--12, 2017.

\bibitem{monakhova2020}
K.~Monakhova, K.~Yanny, N.~Aggarwal, and L.~Waller, ``Spectral diffusercam:
  Lensless snapshot hyperspectral imaging with a spectral filter array,''
  \emph{Optica}, vol.~7, no.~10, pp. 1298--1307, 2020.

\bibitem{parada2016}
A.~Parada-Mayorga and G.~R. Arce, ``Spectral super-resolution in colored coded
  aperture spectral imaging,'' \emph{IEEE Transactions on Computational
  Imaging}, vol.~2, no.~4, pp. 440--455, 2016.

\bibitem{arguello2012}
H.~Arguello, H.~F. Rueda, and G.~R. Arce, ``Spatial super-resolution in code
  aperture spectral imaging,'' in \emph{Compressive Sensing}, vol. 8365.\hskip
  1em plus 0.5em minus 0.4em\relax International Society for Optics and
  Photonics, 2012, p. 83650A.

\bibitem{oguzhan_thesis}
O.~F. Kar, ``Computational spectral imaging techniques using diffractive lenses
  and compressive sensing,'' Master's thesis, METU, Turkey, 2019.

\bibitem{oktem2014agu}
F.~S. {Oktem}, F.~{Kamalabadi}, and J.~M. {Davila}, ``{High-Resolution solar
  imaging with photon sieves},'' in \emph{AGU Fall Meeting Abstracts}, vol.
  2014, Dec. 2014, pp. SH53B--4219.

\bibitem{chung2008dual}
H.-H. Chung, N.~Bradman, M.~R. Davidson, and P.~H. Holloway, ``Dual wavelength
  photon sieves,'' \emph{Optical Engineering}, vol.~47, no.~11, p. 118001,
  2008.

\bibitem{yontem2018imaging}
A.~{\"O}. Y{\"o}ntem, J.~Li, and D.~Chu, ``Imaging through a projection screen
  using bi-stable switchable diffusive photon sieves,'' \emph{Optics Express},
  vol.~26, no.~8, pp. 10\,162--10\,170, 2018.

\bibitem{ayazgok2019}
S.~Ayazgok and F.~S. Oktem, ``Snapshot spectral imaging with generalized photon
  sieves,'' in \emph{Imaging Systems and Applications}.\hskip 1em plus 0.5em
  minus 0.4em\relax Optical Society of America, 2019, pp. JW2A--8.

\bibitem{ayazgok2020}
------, ``Efficient computation of {2D} point-spread functions for diffractive
  lenses,'' \emph{Applied Optics}, vol.~59, no.~2, pp. 445--451, 2020.

\bibitem{blahut_book}
R.~E. Blahut, \emph{Theory of remote image formation}.\hskip 1em plus 0.5em
  minus 0.4em\relax Cambridge University Press, Cambridge, 2004.

\bibitem{hansen_book}
P.~C. Hansen, \emph{Discrete Inverse Problems: Insight and Algorithms}.\hskip
  1em plus 0.5em minus 0.4em\relax SIAM, 2010, vol.~7.

\bibitem{vogel2002computational}
C.~R. Vogel, \emph{Computational methods for inverse problems}.\hskip 1em plus
  0.5em minus 0.4em\relax SIAM, 2002, vol.~23.

\bibitem{geman1987stochastic}
S.~Geman and D.~Geman, ``Stochastic relaxation, {Gibbs} distributions, and the
  {Bayesian} restoration of images,'' in \emph{Readings in Computer
  Vision}.\hskip 1em plus 0.5em minus 0.4em\relax Elsevier, 1987, pp. 564--584.

\bibitem{c-salsa}
M.~V. Afonso, J.~M. Bioucas-Dias, and M.~A.~T. Figueiredo, ``An augmented
  {Lagrangian} approach to the constrained optimization formulation of imaging
  inverse problems,'' \emph{IEEE Transactions on Image Processing}, vol.~20,
  no.~3, pp. 681--695, 2011.

\bibitem{rudin1992nonlinear}
L.~I. Rudin, S.~Osher, and E.~Fatemi, ``Nonlinear total variation based noise
  removal algorithms,'' \emph{Physica D: Nonlinear Phenomena}, vol.~60, no.
  1-4, pp. 259--268, 1992.

\bibitem{nocedal2006numerical}
J.~Nocedal and S.~Wright, \emph{Numerical optimization}.\hskip 1em plus 0.5em
  minus 0.4em\relax Springer Science \& Business Media, 2006.

\bibitem{ng2010solving}
M.~K. Ng, P.~Weiss, and X.~Yuan, ``Solving constrained total-variation image
  restoration and reconstruction problems via alternating direction methods,''
  \emph{SIAM Journal on Scientific Computing}, vol.~32, no.~5, pp. 2710--2736,
  2010.

\bibitem{kamaci2017efficient}
U.~Kamaci, F.~C. Akyon, T.~Alkanat, and F.~S. Oktem, ``Efficient sparsity-based
  inversion for photon-sieve spectral imagers with transform learning,'' in
  \emph{2017 IEEE Global Conference on Signal and Information Processing
  (GlobalSIP)}.\hskip 1em plus 0.5em minus 0.4em\relax IEEE, 2017, pp.
  1225--1229.

\bibitem{noble1988}
B.~Noble and J.~W. Daniel, \emph{Applied Linear Algebra}.\hskip 1em plus 0.5em
  minus 0.4em\relax Prentice-Hall New Jersey, 1988, vol.~3.

\bibitem{chambolle}
A.~Chambolle, ``An algorithm for total variation minimization and
  applications,'' \emph{Journal of Mathematical Imaging and Vision}, vol.~20,
  no. 1-2, pp. 89--97, 2004.

\bibitem{brekke2000extreme}
P.~Brekke, W.~Thompson, T.~Woods, and F.~Eparvier, ``The extreme-ultraviolet
  solar irradiance spectrum observed with the {Coronal Diagnostic Spectrometer
  (CDS)} on {SOHO},'' \emph{The Astrophysical Journal}, vol. 536, no.~2, p.
  959, 2000.

\bibitem{young2007euv}
P.~R. Young, D.~G. Zanna, H.~E. Mason, K.~P. Dere, E.~Landi, M.~Landini, G.~A.
  Doschek, C.~M. Brown, L.~Culhane, L.~K. Harra \emph{et~al.}, ``{EUV} emission
  lines and diagnostics observed with {Hinode/EIS},'' \emph{Publications of the
  Astronomical Society of Japan}, vol.~59, no. sp3, pp. S857--S864, 2007.

\bibitem{adam2015}
D.~P. Kingma and J.~Ba, ``Adam: A method for stochastic optimization,'' in
  \emph{Proc. Int. Conf. Learn. Represent.}, 2015, pp. 1--15.

\bibitem{antipa2018diffusercam}
N.~Antipa, G.~Kuo, R.~Heckel, B.~Mildenhall, E.~Bostan, R.~Ng, and L.~Waller,
  ``Diffusercam: lensless single-exposure {3D} imaging,'' \emph{Optica},
  vol.~5, no.~1, pp. 1--9, 2018.

\bibitem{rabin2018agu}
D.~Rabin, A.~N. Daw, P.~Calhoun \emph{et~al.}, ``A mission to achieve
  ultrahigh-resolution imaging of the solar corona,'' \emph{AGUFM}, vol. 2018,
  pp. A44G--07, 2018.

\bibitem{didem2020}
D.~Dogan and F.~S. Oktem, ``Efficient algorithms for convolutional inverse
  problems in multidimensional imaging,'' \emph{arXiv preprint
  arXiv:2006.08303}, 2020.

\bibitem{widloski2020}
E.~Widloski, U.~Kamaci, and F.~Kamalabadi, ``Optimal measurement configuration
  in computational diffractive imaging,'' in \emph{2020 IEEE International
  Conference on Image Processing (ICIP)}.\hskip 1em plus 0.5em minus
  0.4em\relax IEEE, 2020, pp. 281--285.

\end{thebibliography}
\end{document}